\documentclass[preprint,prd,showpacs,preprintnumbers,amsmath,amssymb,eqsecnum,nofootinbib]{revtex4}

\usepackage{graphicx}
\usepackage{dcolumn}
\usepackage{bm}

\newcommand{\bea}{\begin{eqnarray}}
\newcommand{\eea}{\end{eqnarray}}

\begin{document}

\preprint{YNU-HEPTh-07-101} 
\preprint{KUNS-2067}

\title{The  Virtual Photon Structure to the Next-to-next-to-leading Order
 in QCD}

\author{Takahiro UEDA}
\email{t-ueda@phys.ynu.ac.jp}
\author{Ken SASAKI}%
 \email{sasaki@phys.ynu.ac.jp}
\affiliation{
Dept. of Physics,  
Faculty of Engineering, Yokohama National University, 
 Yokohama 240-8501, JAPAN
}%

\author{Tsuneo UEMATSU}
\email{uematsu@scphys.kyoto-u.ac.jp}
\affiliation{Dept. of Physics, Graduate School of Science,  Kyoto University,  
 Yoshida, Kyoto 606-8501, JAPAN 
\vspace{2cm}}%


\begin{abstract}
We investigate the unpolarized virtual photon structure
functions $F_2^\gamma(x,Q^2,P^2)$ and $F_L^\gamma(x,Q^2,P^2)$ in perturbative QCD  
for the kinematical region $\Lambda^2 \ll P^2 \ll Q^2$, where $-Q^2(-P^2)$  is the mass squared 
of the probe (target) photon and $\Lambda$ is the
QCD scale  parameter.
Using the framework of the operator product expansion supplemented by the renormalization group  method, 
we derive the definite predictions for the moments of $F_2^\gamma(x,Q^2,P^2)$
up to the next-to-next-to-leading order (NNLO) (the order $\alpha\alpha_s$) and for the 
moments of $F_L^\gamma(x,Q^2,P^2)$  up to the next-to-leading order (NLO) (the order $\alpha\alpha_s$).
The NNLO corrections to the sum rule of  $F_2^\gamma(x,Q^2,P^2)$ are negative and 
found to be   $7\% \sim10\%$ of the sum of the LO and NLO contributions, when
 $P^2\!=\!1 {\rm GeV}^2$ and  $Q^2\!=\!30\sim 100{\rm GeV}^2$ or 
$P^2\!=\!3 {\rm GeV}^2$ and  $Q^2\!=\!100{\rm GeV}^2$,  and the number of active quark flavors $n_f$ is 
three or four. The NLO corrections to $F_L^\gamma$ are also negative. 
The moments are inverted numerically to obtain the predictions for $F_2^\gamma(x,Q^2,P^2)$ and 
$F_L^\gamma(x,Q^2,P^2)$ as functions of $x$.
\end{abstract}

\maketitle

\section{Introduction}

\begin{figure}
\begin{center}
\includegraphics[scale=1.0]{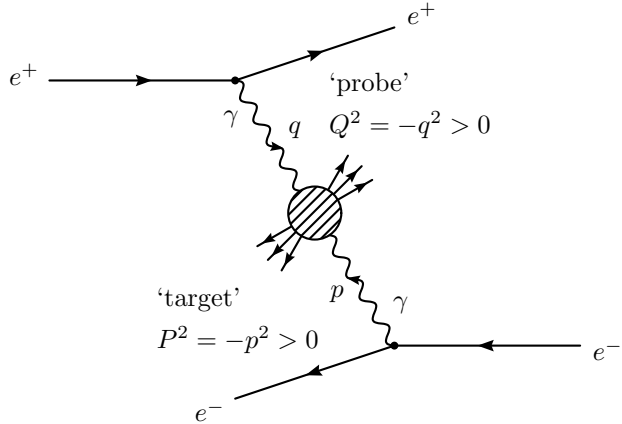}
\vspace{-0.5cm}
\caption{\label{Two Photon Process} Deep inelastic scattering on a virtual photon in the $e^+~e^-$
collider experiments.}
\end{center}
\end{figure}

The experiments at the Large Hadron Collider (LHC) will be started shortly and it is much 
anticipated that signals for the new physics beyond the Standard Model (SM) will be discovered~\cite{LHC}. 
Once these signals are observed, they would be examined more closely in a 
proposed $e^+e^-$  collider machine called the International Linear Collider (ILC)~\cite{ILC}.
In analyzing these signals for the new physics, the knowledge of the SM, especially, of QCD will be more
important than ever before. 
It is well known that, in $e^+e^-$ collision experiments, the cross section for the two-photon 
processes $e^+e^-\rightarrow e^+e^- + {\rm hadrons}$~ shown in 
Fig.\ref{Two Photon Process}  dominates  at high 
energies over other processes such as the 
annihilation process $e^+e^-\rightarrow \gamma^* \rightarrow {\rm hadrons}$. 
Here we consider the two-photon processes in the double-tag events, where 
both of the outgoing $e^+$ and $e^-$ are detected.
In particular, we investigate the case 
in which one of the virtual photon is very far off shell (large $Q^2\equiv -q^2$), while the other 
is close to the mass shell (small $P^2\equiv -p^2$). 
This process can be viewed as a deep-inelastic electron-photon
scattering where the target is a photon rather than a nucleon~\cite{WalshBKT}.
In the deep-inelastic scattering off a photon target, we can study the photon structure 
functions, which are the analogs of the nucleon structure functions. The photon structure functions 
are defined in the lowest order of the QED coupling  constant $\alpha=e^2/4\pi$ and, 
in this paper, they are of order $\alpha$.

The unpolarized (spin-averaged) photon structure functions $F_2^\gamma(x, Q^2)$ and $F_L^\gamma(x, Q^2)$ 
of the real photon ($P^2\!=\!0$) were first studied in the parton-model (PM)~\cite{WalshZerwas} and 
then investigated in perturbative QCD. A pioneering work was done by Witten~\cite{Witten} 
in which he derived the leading order (LO) QCD contributions to 
$F_2^\gamma$ and $F_L^\gamma$. A few years later the next-to-leading order (NLO) corrections to 
$F_2^\gamma$ were calculated~\cite{BB}. These results were obtained in the framework 
based on the operator product expansion (OPE)~\cite{CHM} supplemented by the renormalization group (RG)
method. The same results were rederived by the QCD improved PM powered by  the parton 
evolution equations~\cite{Dewitt,GR}. 
Recently, the lowest six even-integer Mellin moments of the photon-parton splitting functions were
caluculated to the  next-to-next-to-leading order 
(NNLO) and the parton distributions of  real photon and the sturucture function 
$F_2^\gamma$  were analyzed~\cite{MVV}. The same authors later gave the compact and accurate
parameterization of the  photon-parton splitting functions up to the NNLO in Ref.\cite{MVV3}.

When  polarized  beams are used in $e^+e^-$ collision experiments, we can 
get  information on  the spin structure of the photon. 
The QCD analysis of  
the polarized structure function $g_1^\gamma(x,Q^2)$ for the real photon target 
was performed in the LO~\cite{KS} and in the NLO~\cite{SV,GRS}.
For more information on the theoretical and experimental investigation of 
both unpolarized and polarized photon structure, see Ref.\cite{Krawczyk}.

A unique and interesting feature  of the photon structure functions is
that, in contrast with the nucleon case, the target mass squared $-P^2$ 
is not fixed but can take various values and that the structure functions show 
different behaviors depending on the values of $P^2$. The photon has two characters: 
The photon couples directly to quarks (pointlike nature) and also
it behaves as vector bosons (hadronic nature)~\cite{JJS}. Thus the structure function 
$F_2^\gamma(x,Q^2)$ of the real photon ($P^2\!=\!0$) may be decomposed as 
\begin{equation}
F_2^\gamma(x,Q^2)=F_2^\gamma(x,Q^2)|_{\rm pointlike}+F_2^\gamma(x,Q^2)|_{\rm hadronic}~.
\end{equation}
The first term, 
a pointlike piece, can be calculated in principle in a perturbative method. 
On the other hand, the second term, a hadronic piece, can only be computed by some nonperturbative methods
like  lattice QCD, or estimated, for example,  by the vector meson dominance model~\cite{JJS}. 

The moments of $F_2^\gamma(x,Q^2)|_{\rm pointlike}$ and $F_2^\gamma(x,Q^2)|_{\rm hadronic}$ 
for even $n$ may be written,  respectively, as
\bea
\int_0^1 dx x^{n-2}F_2^\gamma(x,Q^2)|_{\rm pointlike}&=&\alpha\Bigl\{ \frac{1}{\alpha_s(Q^2)}a_n 
+b_n+{\cal O}(\alpha_s(Q^2)) \Bigr\}~, \label{pointlike}\\
\int_0^1 dx x^{n-2}F_2^\gamma(x,Q^2)|_{\rm hadronic}&=&\alpha h_n(\alpha_s(Q^2))~,\label{hadronic}
\eea
where  $x$ is the Bjorken variable   and 
$\alpha_s(Q^2)=g^2(Q^2)/4\pi$ is the  QCD running coupling constant. Since $1/\alpha_s(Q^2)$ behaves as
$\ln (Q^2/\Lambda^2)$ at large-$Q^2$, where $\Lambda$  is the QCD scale parameter, the first term
$a_n/\alpha_s(Q^2)$ dominates over the 
$b_n$ term and also over the hadronic term $h_n(\alpha_s(Q^2))$.
The definite prediction for the  LO contributions $a_n$ was given in  Ref.\cite{Witten}. Meanwhile, 
the NLO corrections $b_n$ were calculated only for $n\!>\!2$ in Ref.\cite{BB}. For $n\!>\!2$, 
the hadronic moments $h_n(\alpha_s(Q^2))$   vanish in the large-$Q^2$ limit and 
the $b_n$ terms give finite contributions. However, at $n\!=\!2$, 
the hadronic energy-momentum tensor operator comes into play. Due to the conservation of this operator,  
$b_n$ shows a singularity at $n\!=\!2$ and $h_{n=2}(\alpha_s(Q^2))$ does not vanish at large-$Q^2$.
Actually, $h_{n}(\alpha_s(Q^2))$ also develops a singularity at $n\!=\!2$  which cancels out the one
of $b_n$, and $h_{n}(\alpha_s(Q^2))$ and $b_n$ in combination give a finite but 
perturbatively uncalculable contribution at $n\!=\!2$~\cite{UW2}. The fact that a definite  information 
on the NLO second moment is missing prevents us to fully predict the shape and magnitude of the 
structure function of $F_2^\gamma(x,Q^2)$ up to the order ${\cal O}(\alpha)$.

It was then pointed out in Ref.\cite{UW2} that the situation changes significantly when we
analyze the  structure function of a virtual photon  with $P^2$ much larger than the QCD parameter
$\Lambda^2$. More specifically, we consider the  following kinematical region,
\begin{equation}
\Lambda^2 \ll P^2 \ll Q^2 \label{Kinematical}~.
\end{equation}
In this region, the hadronic component of the photon can also be dealt with {\it perturbatively} 
and thus a definite prediction of the whole structure function, its shape and magnitude,  may become
possible. In fact,   
the virtual photon structure function $F_2^\gamma(x,Q^2,P^2)$ in the kinematical region
(\ref{Kinematical}) was calculated  in the LO (the order $\alpha/\alpha_s$) \cite{UW1} and in the NLO 
(the order $\alpha$)~\cite{UW2,Rossi}, and the longitudinal  
structure function $F_L^\gamma(x,Q^2,P^2)$ in the LO (the order $\alpha$)~\cite{UW2} 
without any unknown parameters.
It is notable that  the pathology of singularity, which appeared at $n\!=\!2$ in the term $b_n$ of
Eq.(\ref{pointlike}) for the real photon target, disappeared from the moments of
$F_2^\gamma(x,Q^2,P^2)$. The parton contents of the virtual photon for the case (\ref{Kinematical}) 
were studied in Refs.\cite{DG,GRStratmann,Fontannaz}.

In the same kinematical region (\ref{Kinematical}), the polarized virtual structure function 
$g_1^\gamma(x,Q^2,P^2)$ was investigated up to the NLO in QCD in Ref.\cite{SU1} and in the second 
paper of \cite{GRS}. Moreover, the polarized parton distributions inside the virtual photon 
were analyzed in various factorization schemes~\cite{SU2}. Quite recently the first moment of 
$g_1^\gamma(x,Q^2,P^2)$ was calculated up to the NNLO~\cite{SUU}. 

In this paper we investigate the unpolarized virtual photon structure
functions $F_2^\gamma(x,Q^2,P^2)$ and $F_L^\gamma(x,Q^2,P^2)$ 
in the kinematical region (\ref{Kinematical}) in QCD. 
Here we neglect all the power corrections of the form $(P^2/Q^2)^k$ ($k=1,2,\cdots$) 
which may arise from target mass and  higher-twist effects.
We present definite predictions for $F_2^\gamma(x,Q^2,P^2)$
up to the NNLO (the order $\alpha\alpha_s$) and for $F_L^\gamma(x,Q^2,P^2)$ up to the 
NLO  (the order $\alpha\alpha_s$). 
The recent calculations of the three-loop anomalous dimensions for the quark and gluon 
operators~\cite{MVV1,MVV2} and  
of the three-loop photon-quark and photon-gluon splitting functions~\cite{MVV3} 
have paved the way for this investigation.
Using the framework of the OPE supplemented by the RG method, we give, 
in the next section, an  expression for the moments of 
$F_2^\gamma(x,Q^2,P^2)$ up to the NNLO corrections. In Sec.\ref{Parameters}
we enumerate all the necessary QCD parameters to evaluate the NNLO corrections.
In Sec.\ref{SectionSecondMoment} the second moment of  
$F_2^\gamma(x,Q^2,P^2)$ will be evaluated up to the NNLO. 
The numerical analysis of $F_2^\gamma(x,Q^2,P^2)$ as a function of $x$ will be given in 
Sec.\ref{NumericalAnalysis}. In Sec.\ref{Longitudinal} the longitudinal virtual photon 
structure function $F_L^\gamma(x,Q^2,P^2)$ will be analyzed up to the NLO. 
The final section is devoted to the conclusions.

\section{Theoretical framework based on the OPE and the NNLO corrections to $F_2^\gamma(x,Q^2,P^2)$
\label{Framework}}

In this article we analyze the virtual photon structure functions 
$F_2^{\gamma}(x,Q^2,P^2)$ and $F_L^{\gamma}(x,Q^2,P^2)$ using the
theoretical framework based on the OPE and RG method. 
Unless otherwise stated, we will follow the notation of Ref.\cite{BB}.
\begin{figure}
\begin{center}
\includegraphics[scale=0.4]{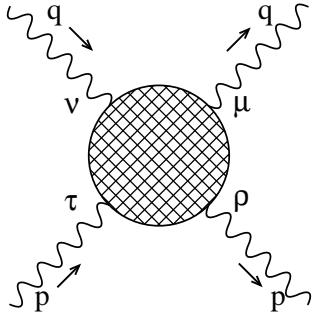}
\vspace{-0.5cm}
\caption{\label{FourPhotons} Forward scattering of a virtual photon with momentum $q$ and 
another virtual photon with momentum $p$. The Lorentz indices are denoted by $\mu, \nu, \rho, \tau$.}
\end{center}
\end{figure}
Let us consider the forward virtual photon scattering amplitude for 
$\gamma(q)+\gamma(p)\rightarrow \gamma(q)+\gamma(p)$ illustrated in Fig.\ref{FourPhotons},
\begin{equation}
T_{\mu\nu\rho\tau}(p,q)=i\int d^4 x d^4 y d^4 z e^{iq\cdot x}e^{ip\cdot (y-z)}
\langle 0|T(J_\mu(x) J_\nu(0) J_\rho(y) J_\tau(z))|0\rangle~,
\end{equation}
where $J_\mu$ is the electromagnetic current.
Its absorptive part is related to the structure tensor 
$W_{\mu\nu\rho\tau}(p,q)$ for the target photon with mass squared 
$p^2=-P^2$ probed by the photon with $q^2=-Q^2$:
\begin{equation}
W_{\mu\nu\rho\tau}(p,q)=\frac{1}{\pi}{\rm Im}T_{\mu\nu\rho\tau}(p,q)~.
\end{equation}
Taking a spin average for the target photon, we get
\bea
W_{\mu\nu}^\gamma(p,q)&=&\frac{1}{2}\sum_\lambda\epsilon^{\rho*}_{(\lambda)}(p)
W_{\mu\nu\rho\tau}(p,q)\epsilon^{\tau}_{(\lambda)}(p)~\nonumber\\
&=&-\frac{1}{2}g^{\rho\tau}W_{\mu\nu\rho\tau}(p,q)\nonumber\\
&=&\frac{1}{2}\int d^4 x e^{iqx}\langle \gamma (p)|J_\mu(x) J_\nu(0)|\gamma (p)\rangle_{\rm spin\  av}.
\label{Wmunu}
\eea
Now $W_{\mu\nu}^\gamma(p,q)$ is expressed in terms of two independent structure functions 
$F_L^\gamma(x,Q^2,P^2)$ and $F_2^\gamma(x,Q^2,P^2)$:
\bea
W_{\mu\nu}^\gamma(p,q)&=&\Bigl\{g_{\mu\nu}-\frac{q_\mu q_\nu}{q^2}\Bigr\}\frac{1}{x}
F_L^\gamma(x,Q^2,P^2)
\nonumber\\ &&+\Bigl\{ -g_{\mu\nu}+\frac{q^{\mu}p^{\nu}+p^{\mu}q^{\nu}}{p\cdot q}
-\frac{p_\mu p_\nu}{(p\cdot q)^2}q^2  \Bigr\}\frac{1}{x}F_2^\gamma(x,Q^2,P^2)~, 
\label{DefF2FL}
\eea
where $x=Q^2/2p\cdot q$. 

Applying OPE for the product of two electromagnetic currents at short distance we get
\bea
i\int d^4x e^{iqx}T(J_\mu(x)J_\nu(0))&=&
\left[g_{\mu\nu}-\frac{q_\mu q_\nu}{q^2}  \right]
\sum_{n=0\atop n={\rm even}}\left(\frac{2}{Q^2}\right)^n \hspace{-0.3cm}
q_{\mu_1}\cdots q_{\mu_{n}}
\sum_i C^i_{L,n} O_i^{\mu_1\cdots \mu_{n}} \nonumber\\
&+& \Bigl[-g_{\mu\lambda} g_{\nu\sigma}q^2+g_{\mu\lambda}q_\nu q_\sigma
+g_{\nu\sigma}q_\mu q_\lambda-g_{\mu\nu}q_\lambda q_\sigma \Bigr] \nonumber \\
&&\times \sum_{n=2 \atop n={\rm even}} 
\left(\frac{2}{Q^2}\right)^n \hspace{-0.3cm}
q_{\mu_1}\cdots q_{\mu_{n-2}}
\sum_i C^i_{2,n} O_i^{\lambda\sigma\mu_1\cdots \mu_{{n-2}}}\nonumber \\
&+& \cdots~, 
\eea
where $C^i_{L,n}$ and $C^i_{2,n}$ are the coefficient functions 
which contribute to the structure functions $F_L^\gamma$ and $F_2^\gamma$, 
respectively, and   
$O_i^{\mu_1\cdots \mu_{n}}$ and $O_i^{\lambda\sigma\mu_1\cdots \mu_{{n-2}}}$ are 
spin-$n$ twist-2 operators  (hereafter we often refer 
to $O_i^{\mu_1\cdots \mu_{n}}$ as $O_i^n$).  
The sum on $i$ runs over  the possible twist-2 operators and 
$\cdots$ represents other terms with irrelevant coefficient functions and operators. In fact, the
relevant 
$O_i^n$ are singlet quark ($\psi$), gluon ($G$), nonsinglet quark ($NS$) and 
photon ($\gamma$) operators as follows: 
\begin{subequations}
\bea
  O_{\psi}^{\mu_{1}\cdots \mu_{n}} &=&
         i^{n-1}  \overline{\psi}
       \gamma^{\{\mu_1}D^{\mu_2} \cdots D^{\mu_{n}\}}~\bm{1}\psi -
            {\rm trace \ terms}
~, \label{Singletquark}\\
  O_G^{\mu_{1}\cdots \mu_{n}} &=& \frac{1}{2}~i^{n-2}
       {G_\alpha}^{\{\mu_1}
               D^{\mu_2} \cdots D^{\mu_{n-1}}G^{\alpha\mu_n\}} -
            {\rm trace \ terms}~,
     \label{gluon}\\
  O_{NS}^{\mu_{1}\cdots \mu_{n}} &=&
         i^{n-1}  \overline{\psi}
       \gamma^{\{\mu_1}D^{\mu_2} \cdots D^{\mu_{n}\}}
             (Q^2_{ch}-\langle e^2 \rangle \bm{1})\psi  -
            {\rm trace \ terms}~,\label{NSquark} \\
 O_{\gamma}^{\mu_{1}\cdots \mu_{n}} &=& \frac{1}{2}~i^{n-2}
        {F_\alpha}^{\{\mu_1} \partial^{\mu_2} \cdots
\partial^{\mu_{n-1}}F^{\alpha\mu_n\}}
                - {\rm trace \ terms}~,
     \label{photon}
\eea
\end{subequations}
where $\{\ \ \  \}$ means complete symmetrization over the Lorentz indices
$\mu_{1}\cdots \mu_{n}$ and $D^{\mu}$ denotes covariant derivative. 
In quark operators $O_{\psi}^n$ and $O_{NS}^n$ given in Eqs.(\ref{Singletquark}) and 
(\ref{NSquark}),  
$\bm{1}$ is an $n_f \times n_f$ unit matrix, $Q^2_{ch}$ is the square of the
$n_f \times n_f$ quark-charge matrix, with $n_f$ being the number of active quark 
 (i.e., the massless quark) flavors,  and $\langle e^2
\rangle=(\sum_i^{n_f}e^2_i)/n_f$ is  the average charge squared where $e_i$ is 
the electromagnetic charge of the active quark  with flavor $i$ in the unit of proton charge. 
It is noted  that we have a relation ${\rm Tr}(Q^2_{ch}-\langle e^2 \rangle \bm{1})=0$.
The essential feature in the analysis of the photon structure functions, in contrast to the 
case of the nucleon counterparts,  is the
appearance  of photon operators $O_\gamma^n$ in addition to the  familiar hadronic operators 
$O_\psi^n$, $O_G^n$ and $O_{NS}^n$~\cite{Witten}.

The spin-averaged matrix elements of these operators sandwiched 
by the photon state with momentum $p$ are expressed as
\begin{equation}
 \langle \gamma (p)|O_i^{\mu_1\cdots \mu_{n}}|\gamma (p)\rangle_{\rm spin\  av}=
 A_n^{i}(\mu^2, P^2)~\{p^{\mu_1}\cdots p^{\mu_n}~-\ {\rm trace\ terms} \}
\end{equation}
with $i=\psi,G,NS,\gamma$,
and $\mu$ is the renormalization point. Then the moment sum rules for 
$F_2^\gamma$ and $F_L^\gamma$ are given  as follows~\cite{CHM}:
\begin{subequations}
\bea
\int_0^1dx x^{n-2} F_2^\gamma(x,Q^2,P^2)&=&
\sum_{i=\psi,G,NS,\gamma} C_{2,n}^i(Q^2/\mu^2,{\bar g}(\mu^2),\alpha)~A_n^{i}(\mu^2, P^2) ~,
\label{F2moment}\\
\int_0^1dx x^{n-2} F_L^\gamma(x,Q^2,P^2)&=&
\sum_{i=\psi,G,NS,\gamma} C_{L,n}^i(Q^2/\mu^2,{\bar g}(\mu^2),\alpha)~A_n^{i}(\mu^2, P^2)~, 
\label{FLmoment}
\eea
\end{subequations}
with ${\bar g}(\mu^2)$ being effective running QCD coupling constant at $\mu^2$. 
Recall that in this article the photon structure functions 
are defined to be  of order $\alpha$. 
Since the coefficient functions $C_{2,n}^\gamma$ and $C_{L,n}^\gamma$ are ${\cal O}(\alpha)$, 
it is sufficient to evaluate $A^\gamma_{n}$ at ${\cal O}(1)$.  Thus we have
\begin{equation}
A^\gamma_{n}(\mu^2, P^2)=1~. \label{PhotonMat}
\end{equation}
On the other hand, the matrix elements $A^i_n$ $(i=\psi, G, NS)$ for the hadronic operators 
start at ${\cal O}(\alpha)$.
For $-p^2=P^2\gg \Lambda^2$, we can calculate $A^i_{n}$ $(i=\psi, G, NS)$ perturbatively 
in each power of $g^2$. 
When $\mu^2$ is chosen at $P^2$, they are expressed in the form as 
\begin{equation}
A^i_{n}(\mu^2, P^2)|_{\mu^2=P^2}=\frac{\alpha}{4\pi}{\widetilde A}^i_{n}({\bar g}(P^2))~, \qquad {\rm
for}
\quad i=\psi,G,NS~. \label{HadronicMat}
\end{equation}

Let us first analyze the structure function $F_2^\gamma(x,Q^2,P^2)$. 
We will evaluate its  moment sum rule up to the NNLO. The $Q^2$-dependence of 
the coefficient functions $C_{2,n}^i(Q^2/\mu^2,{\bar g}(\mu^2),\alpha)$ in (\ref{F2moment}) 
is governed by the RG equation. Putting $\mu^2=-p^2=P^2$, its solution is given by 
\begin{equation}
C_{2,n}^i(Q^2/P^2,{\bar g}(P^2),\alpha)
=\biggl(T\exp \biggl[ \int_{{\bar g}(Q^2)}^{{\bar g}(P^2)} dg
\frac{\gamma_n (g,\alpha)}{\beta(g)} \biggr]\biggr)_{ij}
 C^j_{2,n}(1,{\bar g}(Q^2),\alpha)~, \label{RGequation}
\end{equation}
with $i,j=\psi, G, NS$ and  $\gamma$~.
Here $\beta(g)$ is the beta function and $\gamma_n (g^2,\alpha)$ is the 
anomalous dimension matrix. To the lowest order in $\alpha$, this matrix has 
the following form:
\begin{equation}
\gamma_{n}(g, \alpha)=\biggl(
\begin{array}{c|c}
{\hat {\gamma}}_{n}(g) & \bm{0} \\
 \hline
 \bm{K}_{n}(g, \alpha) & {0}
 \end{array}
 \biggr)~,
\end{equation}
where ${\widehat \gamma}_{n}(g^2)$ is the usual $3\times 3$ anomalous
dimension matrix in the hadronic sector
\begin{equation}
  {\widehat \gamma}_{n}(g)=\begin{pmatrix}
           \gamma^n_{\psi\psi}(g)&\gamma^n_{G\psi}(g) &0\\
           \gamma^n_{\psi G}(g)&\gamma^n_{GG}(g)&0\\
            0&0&\gamma^n_{NS}(g) \end{pmatrix}~, \label{HadronicAnomalousDim}
\end{equation}
and $\bm{K}_{n}(g,\alpha)$ is the three-component row vector
\begin{equation}
   \bm{K}_{n}(g,\alpha)=\Bigl( K^n_{\psi}(g,\alpha)~,~ 
    K^n_{G}(g,\alpha)~,~ K^n_{NS}(g,\alpha)\Bigr)~,
\end{equation}
which represents the mixing between photon operator and remaining three hadronic operators.
Then the evolution factor in (\ref{RGequation}) is expressed in the form as~\cite{BB},
\begin{equation}
 T\exp\biggl[\int_{{\bar g}(Q^2)}^{{\bar
g}(P^2)}dg\frac{\gamma_{n}(g,\alpha)}{\beta(g)}
 \biggr]=
\left(
\begin{array}{c|c}
 M_n & {\bf 0} \\
 \hline
 \bm{X}_n & { 1}
 \end{array}
 \right)~,\label{Tanomalous}
\end{equation}
where
 \bea
 M_n(Q^2/P^2,{\bar g}(P^2))&=&
 T\exp\left[\int_{{\bar g}(Q^2)}^{{\bar
 g}(P^2)}dg\frac{{\widehat{\gamma}}_{n}(g)}{\beta(g)}
 \right]~,\label{M_n}
 \\
 \bm{X}_n(Q^2/P^2,{\bar g}(P^2),\alpha)&=&
 \int_{{\bar g}(Q^2)}^{{\bar g}(P^2)}dg\frac{\bm{K}_{n}(g,\alpha)}{\beta(g)}
 T\exp\left[\int_{{\bar
g}(Q^2)}^{g}dg'\frac{{\hat{\gamma}}_{n}({g'})}{\beta(g')}
 \right]~.\label{X_n}
 \eea
Thus using (\ref{PhotonMat})-(\ref{RGequation}) and 
(\ref{Tanomalous})-(\ref{X_n}), we get
\bea
\int_0^1 dx ~ x^{n-2}F_2^\gamma(x,Q^2,P^2)
&=&\frac{\alpha}{4\pi}{\widetilde{\bm{A}}}_n({\bar g}(P^2))\cdot M_n(Q^2/P^2,{\bar g}(P^2)) \cdot
\bm{C}_{2,n}(1,{\bar g}(Q^2)) \nonumber\\
&&+\bm{X}_n(Q^2/P^2,{\bar g}(P^2),\alpha) \cdot \bm{C}_{2,n}(1,{\bar g}(Q^2)) \nonumber\\
&&+C_{2,n}^\gamma(1,{\bar g}(Q^2),\alpha)~, \label{F2momb}
\eea
with
\begin{equation}
{\widetilde{\bm{A}}}_n({\bar g})=({\widetilde A}_n^\psi({\bar g}),
{\widetilde A}_n^G({\bar g}),
{\widetilde A}_n^{NS}({\bar g}))~, \label{PhotonHadronicMat}
\end{equation}
and 
\begin{equation}
\bm{C}_{2,n}(1,{\bar g})=
\left(
\begin{array}{c}
C_{2,n}^\psi(1,{\bar g}) \\
C_{2,n}^G(1,{\bar g}) \\
C_{2,n}^{NS}(1,{\bar g}) \label{HadronicCoefficient}
\end{array}
\right)~.
\end{equation}

In order to evaluate $M_n(Q^2/P^2,{\bar g}(P^2))$ in (\ref{M_n}) up to the NNLO, 
we first expand ${\widehat \gamma}_{n}(g)$   in powers of $g^2$ up to the three-loop level as
\bea
   {\widehat \gamma}_{n}(g)&=&{\widehat \gamma}_{n}^{(0)}(g)+
{\widehat \gamma}_{n}^{(1)}(g) +
{\widehat \gamma}_{n}^{(2)}(g)+ \cdots\nonumber\\
&=&\frac{g^2}{16\pi^2}{\widehat \gamma}_{n}^{(0)}+
\frac{g^4}{(16\pi^2)^2}{\widehat \gamma}_{n}^{(1)} +
\frac{g^6}{(16\pi^2)^3}{\widehat \gamma}_{n}^{(2)}+ \cdots~. \label{GammaExpansion}
\eea
Then putting  ${\bar g}_1={\bar g}(P^2)$ and ${\bar g}_2={\bar g}(Q^2)$, we find that 
$M_n(Q^2/P^2,{\bar g}(P^2))$  is expanded as
\begin{eqnarray}
&&M_n(Q^2/P^2,{\bar g}(P^2))= T\exp\left[\int_{{\bar
g}_2}^{{\bar g}_1}dg\frac{{\hat{\bm{\gamma}}}_{n}(g)}{\beta(g)}
 \right] \nonumber\\
&&\qquad =\exp\left[\int_{{\bar g}_2}^{{\bar
 g}_1}dg\frac{{\hat{\bm{\gamma}}}_{n}^{(0)}(g)}{\beta(g)}
 \right] \nonumber\\
&&\quad\qquad +\int_{{\bar g}_2}^{{\bar g}_1}dg
\exp\left[\int_{g}^{{\bar g}_1}dg'
\frac{{\hat{\bm{\gamma}}}_{n}^{(0)}(g')}{\beta(g')} \right]
\frac{{\hat{\bm{\gamma}}}_{n}^{(1)}(g)}{\beta(g)}
\exp\left[\int_{{\bar g}_2}^{g}dg''
\frac{{\hat{\bm{\gamma}}}_{n}^{(0)}(g'')}{\beta(g'')} \right]\nonumber\\
&&\quad\qquad +\int_{{\bar g}_2}^{{\bar g}_1}dg
\exp\left[\int_{g}^{{\bar g}_1}dg'
\frac{{\hat{\bm{\gamma}}}_{n}^{(0)}(g')}{\beta(g')} \right]
\frac{{\hat{\bm{\gamma}}}_{n}^{(2)}(g)}{\beta(g)}
\exp\left[\int_{{\bar g}_2}^{g}dg''
\frac{{\hat{\bm{\gamma}}}_{n}^{(0)}(g'')}{\beta(g'')} \right]\nonumber\\
&&\quad\qquad +\int_{{\bar g}_2}^{{\bar g}_1}dg_a
\exp\left[\int_{g_a}^{{\bar g}_1}dg'
\frac{{\hat{\bm{\gamma}}}_{n}^{(0)}(g')}{\beta(g')} \right]
\frac{{\hat{\bm{\gamma}}}_{n}^{(1)}(g_a)}{\beta(g_a)}
\int_{{\bar g}_2}^{g_a}dg_b
\exp\left[\int_{g_b}^{g_a}dg''
\frac{{\hat{\bm{\gamma}}}_{n}^{(0)}(g'')}{\beta(g'')} \right]\nonumber\\
&&\qquad \qquad \qquad\times 
\frac{{\hat{\bm{\gamma}}}_{n}^{(1)}(g_b)}{\beta(g_b)}
\exp\left[\int_{{\bar g}_2}^{g_b}dg'''
\frac{{\hat{\bm{\gamma}}}_{n}^{(0)}(g''')}{\beta(g''')} \right]+\cdots~.
\label{M_nExpansion}
\end{eqnarray}
To evaluate the integrals,  we make a full use of the projection operators obtained from 
the one-loop anomalous dimension matrix ${\widehat \gamma}^{(0)}_{n}$ in
(\ref{GammaExpansion})~\cite{BB}:
\begin{equation}
  {\widehat \gamma}^{(0)}_{n}=\sum_{i=+, -, NS} \lambda^n_i~P^n_i~, \label{EigenProjector}
\end{equation}
where $\lambda^n_i~ (i=+, -, NS)$ and $P^n_i$ are eigenvalues of ${\widehat \gamma}^{(0)}_{n}$ 
and the corresponding projection operators, respectively.  The explicit forms of $\lambda^n_i$ and 
$P^n_i$ are  given in Appendix A. Expanding $\beta(g)$ in powers of $g^2$ up to the three-loop level as 
\begin{equation}
\beta(g)=-\frac{g^3}{16\pi^2}\beta_0-\frac{g^5}{(16\pi^2)^2}\beta_1
-\frac{g^7}{(16\pi^2)^3}\beta_2 + \cdots~,
\label{Beta}
\end{equation}
we perform integration in (\ref{M_nExpansion}). The final form of $M_n(Q^2/P^2,{\bar g}(P^2))$
up to the NNLO is given in (\ref{M_nExpression}) in Appendix A.

Similarly, expanding ${\bm K}_n(g,\alpha)$ in powers of $g^2$ up to the three-loop level as
\begin{equation}
   {\bm K}_n(g,\alpha)=-\frac{e^2}{16\pi^2}{\bm K}_n^{(0)}-
\frac{e^2g^2}{(16\pi^2)^2}{\bm K}_n^{(1)} -
\frac{e^2g^4}{(16\pi^2)^3}{\bm K}_n^{(2)}+\cdots~,
\label{ExpandK}
\end{equation}
we can evaluate $\bm{X}_n(Q^2/P^2,{\bar g}(P^2),\alpha)$  in (\ref{X_n}) up to the NNLO.
The result is given in (\ref{X_nExpression}) in Appendix A.

Finally,  expansions are made for the photon matrix elements of hadronic operators
${\widetilde{\bm{A}}}_n({\bar g}(P^2))$ in (\ref{PhotonHadronicMat}) 
as well as the coefficient functions ${\bm C}_{2,n}(1,{\bar g}(Q^2))$ in (\ref{HadronicCoefficient})
and  $C_{2,n}^\gamma(1,{\bar g}(Q^2),\alpha)$ in (\ref{F2momb}) 
up to the two-loop level  as follows:
\bea
  {\widetilde{\bm{A}}}_n({\bar g}(P^2))&=&{\widetilde{\bm{A}}}_n^{(1)}+ 
\frac{{\bar g}^2(P^2)}{16\pi^2}{\widetilde{\bm{A}}}_n^{(2)} +
 \cdots~,\label{ExpandA} \\
{\bm C}_{2,n}(1,{\bar g}(Q^2))&=&{\bm C}_{2,n}^{(0)}+\frac{{\bar
g}^2(Q^2)}{16\pi^2}{\bm C}_{2,n}^{(1)}+
\frac{{\bar g}^4(Q^2)}{(16\pi^2)^2}{\bm C}_{2,n}^{(2)}+\cdots \label{ExpandChadronic}\\
C_{2,n}^\gamma(1,{\bar g}(Q^2),\alpha)&=&\frac{e^2}{16\pi^2}C_{2,n}^{\gamma(1)}+\frac{e^2{\bar
g}^2(Q^2)}{(16\pi^2)^2}C_{2,n}^{\gamma(2)}+\cdots ~.\label{ExpandCgamma}
\eea
Then putting (\ref{M_nExpression}), (\ref{X_nExpression}), (\ref{ExpandA}), (\ref{ExpandChadronic})
and (\ref{ExpandCgamma}) into (\ref{F2momb}), we obtain the expression for the 
moment sum rule of $F_2^\gamma(x,Q^2,P^2)$ up to the NNLO ($\alpha\alpha_s$) corrections as follows:
\bea
&&\hspace{-1.5cm}\int_0^1 dx x^{n-2}F_2^\gamma(x,Q^2,P^2)  \nonumber\\
&=&\frac{\alpha}{4\pi}\frac{1}{2\beta_0}
\Biggl\{\frac{4\pi}{\alpha_s(Q^2)}\sum_{i}{\cal L}^n_i
\left[1-\left(\frac{\alpha_s(Q^2)}{\alpha_s(P^2)}\right)^{d_i^n+1}
\right]\nonumber\\
&&\qquad+\sum_{i}{\cal
A}_i^n\left[1-\left(\frac{\alpha_s(Q^2)}{\alpha_s(P^2)}\right)^{d_i^n}\right]
+\sum_{i}{\cal
B}_i^n\left[1-\left(\frac{\alpha_s(Q^2)}{\alpha_s(P^2)}\right)^{d_i^n+1}\right]
+{\cal C}^n \nonumber\\
&&\qquad+\frac{\alpha_s(Q^2)}{4\pi}\Biggl(\sum_{i}{\cal
D}_i^n\left[1-\left(\frac{\alpha_s(Q^2)}{\alpha_s(P^2)}\right)^{d_i^n-1}\right]
+\sum_{i}{\cal
E}_i^n\left[1-\left(\frac{\alpha_s(Q^2)}{\alpha_s(P^2)}\right)^{d_i^n}\right]
\nonumber\\
&&\qquad \qquad \qquad+\sum_{i}{\cal
F}_i^n\left[1-\left(\frac{\alpha_s(Q^2)}{\alpha_s(P^2)}\right)^{d_i^n+1}\right]
+{\cal G}^n \Biggr) \nonumber\\
&&\qquad+{\cal O}(\alpha_s^2)
 \Biggr\}~,\hspace{2cm} {\rm with}\quad i=+, -, NS~,
\label{master1}
\eea
where $d_i^n=\frac{\lambda_i^n}{2\beta_0}$~.
The coefficients ${\cal L}^n_i$, ${\cal A}^n_i$, ${\cal B}^n_i$, ${\cal C}^n$, ${\cal D}^n_i$, 
${\cal E}^n_i$, ${\cal F}^n_i$ and ${\cal G}^n$ are given by
\bea
{\cal L}^n_i&=&\bm{K}_n^{(0)}P_i^n\mbox{\boldmath
$C$}_{2,n}^{(0)}\frac{1}{d_i^n+1}~,\label{ExpressionL}\\
&&\nonumber\\
{\cal A}^n_i&=&-\bm{K}_n^{(0)} \sum_j \frac{P^n_j \hat \gamma_n^{(1)}
P^n_i}
             {\lambda^n_j - \lambda^n_i+2\beta_0 } {\bm{C}}_{2,n}^{(0)}
     \frac{1}{d_i^n}
            -\bm{K}_n^{(0)}P^n_i {\bm{C}}_{2,n}^{(0)}\frac{\beta_1}{\beta_0}
      \frac{1-d_i^n}{d_i^n} \nonumber  \\
    & & + \bm{K}_n^{(1)} P^n_i {\bm{C}}_{2,n}^{(0)} \frac{1}{d_i^n}
      - 2\beta_0 {\widetilde{\bm{A}}}_n^{(1)} P^n_i {\bm{C}}_{2,n}^{(0)}~,\label{ExpressionA}\\
&&\nonumber\\
{\cal B}^n_i&=& \bm{K}_n^{(0)} \sum_j \frac{P^n_i \hat \gamma_n^{(1)}
P^n_j}
             {\lambda^n_i - \lambda^n_j +2\beta_0} {\bm{C}}_{2,n}^{(0)}
     \frac{1}{1+d_i^n} \nonumber  \\
   &+& \bm{K}_n^{(0)} P^n_i{\bm{C}}_{2,n}^{(1)}
    \frac{1}{1+d_i^n} 
     -\bm{K}_n^{(0)}   P^n_i {\bm{C}}_{2,n}^{(0)}\frac{\beta_1}{\beta_0}
      \frac{d_i^n}{1+d_i^n}~, \label{ExpressionB} \\
&&\nonumber\\
{\cal C}^n&=&2\beta_0 (C_{2,n}^{\gamma (1)} +
         {\widetilde{\bm{A}}}_n^{(1)} \cdot{\bm{C}}_{2,n}^{(0)} )~, \label{ExpressionC}\\
&&\nonumber\\
{\cal D}^n_i&=&-\bm{K}_n^{(0)}P^n_i {\bm{C}}_{2,n}^{(0)}
\left(\frac{\beta^2_1}{\beta^2_0}-\frac{\beta_2}{\beta_0}\frac{1}{1-d_i^n}\right)
     \left(1- \frac{d_i^n}{2}\right)\nonumber\\
&& -\bm{K}_n^{(0)} \sum_j\frac{P^n_j\hat \gamma_n^{(1)} P^n_i}
             {\lambda^n_j - \lambda^n_i +2\beta_0} {\bm{C}}_{2,n}^{(0)}
    \frac{\beta_1}{\beta_0}\frac{1-d_j^n}{1-d_i^n} \nonumber\\
&& -\bm{K}_n^{(0)} \sum_j\frac{P^n_j\hat \gamma_n^{(1)} P^n_i}
             {\lambda^n_j - \lambda^n_i +4\beta_0} {\bm{C}}_{2,n}^{(0)}
    \frac{\beta_1}{\beta_0}\Bigl( \frac{1-d_i^n+d_j^n}{1-d_i^n} \Bigr)\nonumber\\
&& +\bm{K}_n^{(0)} \sum_j\frac{P^n_j\hat \gamma_n^{(2)} P^n_i}
             {\lambda^n_j - \lambda^n_i +4\beta_0} {\bm{C}}_{2,n}^{(0)}
   \frac{1}{1-d_i^n} \nonumber\\
&& -\bm{K}_n^{(0)} \sum_{j, k}
\frac{P^n_k\hat \gamma_n^{(1)} P^n_j\hat \gamma_n^{(1)} P^n_i}
             {(\lambda^n_j - \lambda^n_i +2\beta_0)(\lambda^n_k - \lambda^n_i
+4\beta_0)} {\bm{C}}_{2,n}^{(0)}
   \frac{1}{1-d_i^n} \nonumber\\
&&+\bm{K}_n^{(1)}P^n_i {\bm{C}}_{2,n}^{(0)} \frac{\beta_1}{\beta_0}
+\bm{K}_n^{(1)} \sum_j\frac{P^n_j\hat \gamma_n^{(1)} P^n_i}
             {\lambda^n_j - \lambda^n_i +2\beta_0} {\bm{C}}_{2,n}^{(0)}
   \frac{1}{1-d_i^n} \nonumber\\
&&-\bm{K}_n^{(2)}P^n_i {\bm{C}}_{2,n}^{(0)} \frac{1}{1-d_i^n}
+2\beta_0{\widetilde{\bm{A}}}_n^{(1)} \sum_j\frac{P^n_j\hat \gamma_n^{(1)} P^n_i}
             {\lambda^n_j - \lambda^n_i +2\beta_0} {\bm{C}}_{2,n}^{(0)}  \nonumber\\
&&-2\beta_0{\widetilde{\bm{A}}}_n^{(1)}P^n_i {\bm{C}}_{2,n}^{(0)}
\frac{\beta_1}{\beta_0}d_i^n -2\beta_0{\widetilde{\bm{A}}}_n^{(2)}P^n_i
{\bm{C}}_{2,n}^{(0)}~,
\eea
\bea
{\cal E}^n_i&=&-\bm{K}_n^{(0)}P^n_i {\bm{C}}_{2,n}^{(1)}
\frac{\beta_1}{\beta_0}\frac{1-d_i^n}{d_i^n}
     -\bm{K}_n^{(0)} \sum_j\frac{P^n_j\hat \gamma_n^{(1)} P^n_i}
             {\lambda^n_j - \lambda^n_i +2\beta_0} {\bm{C}}_{2,n}^{(1)}
   \frac{1}{d_i^n} \nonumber\\
&&+\bm{K}_n^{(1)}P^n_i {\bm{C}}_{2,n}^{(1)}\frac{1}{d_i^n}
+\bm{K}_n^{(0)}P^n_i {\bm{C}}_{2,n}^{(0)}
\frac{\beta^2_1}{\beta^2_0}(1-d_i^n)\nonumber\\
&& -\bm{K}_n^{(0)} \sum_j\frac{P^n_i\hat \gamma_n^{(1)} P^n_j}
             {\lambda^n_i - \lambda^n_j +2\beta_0} {\bm{C}}_{2,n}^{(0)}
    \frac{\beta_1}{\beta_0} \frac{1-d_i^n}{d_i^n}
 +\bm{K}_n^{(0)} \sum_j\frac{P^n_j\hat \gamma_n^{(1)} P^n_i}
             {\lambda^n_j - \lambda^n_i +2\beta_0} {\bm{C}}_{2,n}^{(0)}
    \frac{\beta_1}{\beta_0} \nonumber\\
&& -\bm{K}_n^{(0)} \sum_{j, k}
\frac{P^n_j\hat \gamma_n^{(1)} P^n_i\hat \gamma_n^{(1)} P^n_k}
             {(\lambda^n_i - \lambda^n_k +2\beta_0)(\lambda^n_j - \lambda^n_i
+2\beta_0)} {\bm{C}}_{2,n}^{(0)}
   \frac{1}{d_i^n} \nonumber\\
&&-\bm{K}_n^{(1)}P^n_i {\bm{C}}_{2,n}^{(0)} \frac{\beta_1}{\beta_0}
+\bm{K}_n^{(1)} \sum_j\frac{P^n_i\hat \gamma_n^{(1)} P^n_j}
             {\lambda^n_i - \lambda^n_j +2\beta_0} {\bm{C}}_{2,n}^{(0)}
   \frac{1}{d_i^n} \nonumber\\
&&
-2\beta_0{\widetilde{\bm{A}}}_n^{(1)} \sum_j\frac{P^n_i\hat \gamma_n^{(1)} P^n_j}
             {\lambda^n_i - \lambda^n_j +2\beta_0} {\bm{C}}_{2,n}^{(0)}  
+2\beta_0{\widetilde{\bm{A}}}_n^{(1)}P^n_i {\bm{C}}_{2,n}^{(0)}
\frac{\beta_1}{\beta_0}d_i^n -2\beta_0{\widetilde{\bm{A}}}_n^{(1)}P^n_i
{\bm{C}}_{2,n}^{(1)}~,\nonumber\\
\eea
\bea
{\cal F}^n_i&=&\bm{K}_n^{(0)}P^n_i {\bm{C}}_{2,n}^{(2)}
\frac{1}{1+d_i^n}-\bm{K}_n^{(0)}P^n_i {\bm{C}}_{2,n}^{(1)}
\frac{\beta_1}{\beta_0}\frac{d_i^n}{1+d_i^n}\nonumber\\
&&    +\bm{K}_n^{(0)} \sum_j\frac{P^n_i\hat \gamma_n^{(1)} P^n_j}
             {\lambda^n_i - \lambda^n_j +2\beta_0} {\bm{C}}_{2,n}^{(1)}
   \frac{1}{1+d_i^n} \nonumber\\
&&+\bm{K}_n^{(0)}P^n_i {\bm{C}}_{2,n}^{(0)}
\left(\frac{\beta^2_1}{\beta^2_0}-\frac{\beta_2}{\beta_0}\frac{1}{1+d_i^n}\right)
      \frac{d_i^n}{2}\nonumber\\
&& -\bm{K}_n^{(0)} \sum_j\frac{P^n_i\hat \gamma_n^{(1)} P^n_j}
             {\lambda^n_i - \lambda^n_j +2\beta_0} {\bm{C}}_{2,n}^{(0)}
    \frac{\beta_1}{\beta_0} \frac{d_j^n}{1+d_i^n} \nonumber\\
&& -\bm{K}_n^{(0)} \sum_j\frac{P^n_i\hat \gamma_n^{(1)} P^n_j}
             {\lambda^n_i - \lambda^n_j +4\beta_0} {\bm{C}}_{2,n}^{(0)}
    \frac{\beta_1}{\beta_0}  \frac{1+d_i^n-d_j^n}{1+d_i^n} \nonumber\\
&& +\bm{K}_n^{(0)} \sum_j\frac{P^n_i\hat \gamma_n^{(2)} P^n_j}
             {\lambda^n_i - \lambda^n_j +4\beta_0} {\bm{C}}_{2,n}^{(0)}
     \frac{1}{1+d_i^n} \nonumber\\
&& +\bm{K}_n^{(0)} \sum_{j, k}
\frac{P^n_i\hat \gamma_n^{(1)} P^n_j\hat \gamma_n^{(1)} P^n_k}
             {(\lambda^n_j - \lambda^n_k +2\beta_0)} {\bm{C}}_{2,n}^{(0)}
  \Bigl(\frac{1}{\lambda^n_i - \lambda^n_j
+2\beta_0}-\frac{1}{\lambda^n_i - \lambda^n_k
+4\beta_0} \Bigr)\frac{1}{1+d_i^n}~, \nonumber\\
&& \\
{\cal G}^n&=&2\beta_0 (C_{2,n}^{\gamma (2)} +
         {\widetilde{\bm{A}}}_n^{(1)} \cdot{\bm{C}}_{2,n}^{(1)} +{\widetilde{\bm{A}}}_n^{(2)}
\cdot{\bm{C}}_{2,n}^{(0)})~, \label{CoeffiG}
\eea
with $i,j,k=+, -, NS$. The LO term ${\cal L}^n_i$ was obtained by Witten~\cite{Witten}. 
The NLO ($\alpha$) corrections ${\cal A}^n_i$, ${\cal B}^n_i$ and ${\cal C}^n$ without
terms with ${\widetilde{\bm{A}}}_n^{(1)}$  were  first derived by Bardeen and Buras~\cite{BB} for the
case of the {\it real} photon target  (i.e. $P^2=0$). Later
authors in Ref.\cite{UW2} analyzed the NLO ($\alpha$) corrections for the 
case of the {\it virtual} photon target  ($P^2\gg\Lambda^2$) and the terms with 
${\widetilde{\bm{A}}}_n^{(1)}$ were added to ${\cal A}^n_i$ and ${\cal C}^n$.
The coefficients ${\cal D}^n_i$, ${\cal E}^n_i$, ${\cal F}^n_i$ and ${\cal G}^n$ are the NNLO 
($\alpha\alpha_s$) corrections and new.

For $n=2$, one of the eigenvalues, $\lambda_-^{n=2}$, in Eq.(\ref{EigenProjector})
vanishes and we have $d_-^{n=2}=0$. 
This is due to the fact that the corresponding operator is the hadronic 
energy-momentum tensor and is, therefore,  conserved with a null anomalous dimension~\cite{BB}. 
The coefficients ${\cal A}^{n=2}_-$ and ${\cal E}^{n=2}_-$ have terms which are proportional to 
$\frac{1}{d_-^{n=2}}$ and thus diverge. However, we see from (\ref{master1}) that these coefficients
are multiplied by a factor $\Bigl[1-\Bigl(\alpha_s(Q^2)/\alpha_s(P^2)\Bigr)^{d_-^{n=2}}\Bigr]$ which 
vanishes. In the end, the coefficients ${\cal A}^{n=2}_-$ and ${\cal E}^{n=2}_-$ multiplied by this factor
remain finite~\cite{UW2}.

\section{Parameters in the $\overline{\rm MS}$ scheme  \label{Parameters}}

All the quantities necessary to evaluate the NNLO ($\alpha\alpha_s$) corrections to the moments of 
$F_2^\gamma(x,Q^2,P^2)$ have been calculated and most of them are presented in the literature, 
except for the two-loop photon matrix elements of hadronic operators 
${\widetilde A}_n^{(2)\psi}$, ${\widetilde A}_n^{(2)G}$ and
${\widetilde A}_n^{(2)NS}$.  Also for the three-loop anomalous dimensions 
$K_\psi^{{(2),n}}$, $ K_G^{{(2),n}}$ and $K_{NS}^{{(2),n}}$, we only have approximate expressions 
in the form of photon-quark and photon-gluon splitting functions. In the following we will enumerate 
all these necessary parameters. The expressions  are the ones
calculated  in the modified minimal subtraction ($\overline{\rm MS}$) scheme~\cite{BBDM}. 

\subsection{Quark-charge factors and $\beta$ function parameters}
The following quark-charge factors are often used below:
\begin{equation}
\delta_\psi=\langle e^2 \rangle=\sum_{i=1}^{n_f} e_i^2/n_f~, \qquad \delta_{NS}=1~, \qquad
\delta_\gamma=3n_f \langle e^4\rangle=3\sum_{i=1}^{n_f} e_i^4~. 
\end{equation}
The $\beta$ function parameters $\beta_0$, $\beta_1$ and $\beta_2$~\cite{TVZ} are given by 
\bea
\beta_0&=&\frac{11}{3}C_A- \frac{2}{3}n_f~, \label{beta0}\\
\beta_1&=&\frac{34}{3}C_A^2- \frac{10}{3}C_A n_f -2C_F n_f~, \label{beta1}\\
\beta_2&=&\frac{2857}{54}C_A^3- \frac{1415}{54}C_A^2 n_f- \frac{205}{18}C_A C_F n_f
+\frac{79}{54}C_A n_f^2+C_F^2 n_f  +\frac{11}{9}C_F n_f^2~,\label{beta2}
\eea
with $C_A=3$ and $C_F=\frac{4}{3}$ in QCD.

\subsection{Coefficient functions}
As shown in (\ref{ExpandChadronic}) and (\ref{ExpandCgamma}), 
we need the hadronic coefficient functions $C_{2,n}^i(1,{\bar g}(Q^2))$ with $i=\psi, G$ and 
$NS$, and the photon coefficient function $C_{2,n}^\gamma(1,{\bar g}(Q^2),\alpha)$ up to
the  two-loop level. At the tree-level, we have
\begin{equation}
C_{2,n}^{\psi(0)}=\delta_\psi~, \quad C_{2,n}^{G(0)}=0~, \quad C_{2,n}^{NS(0)}=\delta_{NS}
~, \quad C_{2,n}^{\gamma(0)}=0~.
\end{equation}
The one-loop coefficient functions were  calculated  in the MS scheme in Refs.\cite{BBDM} and 
\cite{FRS}. 
The $\overline{\rm MS}$ results are written in the form as
\begin{equation}
C_{2,n}^{\psi(1)}=\delta_\psi {\overline B}^n_\psi, \quad C_{2,n}^{G(1)}=\delta_\psi {\overline B}^n_G~, 
\quad C_{2,n}^{NS(1)}=\delta_{NS}{\overline B}^n_{NS} ~, \quad
C_{2,n}^{\gamma(1)}=\delta_{\gamma}{\overline B}^n_\gamma~, \label{OneLoopCoefficient}
\end{equation}
where ${\overline B}^n_\psi={\overline B}^n_{NS}$ and ${\overline B}^n_G$ are obtained, 
for example,  
from  the MS-scheme results for $B^n_\psi=B^n_{NS}$ and $B^n_G$ given 
in Eqs.(4.10) and (4.11) of Ref.\cite{BB} by discarding the terms proportional to 
${\rm ln}(4\pi-\gamma_E)$.  ${\overline B}^n_\gamma$ is related to ${\overline B}^n_G$ 
by ${\overline B}^n_\gamma=(2/n_f){\overline B}^n_G$~.

The two-loop coefficient functions corresponding to the hadronic operators 
were calculated in the $\overline{\rm MS}$ scheme in Ref.\cite{vNZ,ZvN1}. 
They were expressed in fractional momentum space as functions $x$. The results in Mellin space 
as functions of $n$ are found, for example, in Ref.~\cite{MochVermaseren}:
\bea
C_{2,n}^{\psi(2)}&=&\delta_\psi \{c_{2,q}^{(2),+{\rm ns}}(n)+c_{2,q}^{(2),-{\rm ns}}(n) + 
c_{2,q}^{(2),{\rm ps}}(n)\}, \\
C_{2,n}^{G(2)}&=&\delta_\psi c_{2,g}^{(2)}(n), \label{GluonCoeffi2}\\
\quad C_{2,n}^{NS(2)}&=&\delta_{NS}\{c_{2,q}^{(2),+{\rm ns}}(n)+c_{2,q}^{(2),-{\rm ns}}(n)\}~. ~
\eea
where $c_{2,q}^{(2),+{\rm ns}}(n)$, $c_{2,q}^{(2),-{\rm ns}}(n)$, $c_{2,q}^{(2),{\rm
ps}}(n)$ and $c_{2,g}^{(2)}(n)$ are given in Eqs.(197), (198), (201) and (202)  
in Appendix B of Ref.\cite{MochVermaseren}, respectively,  with $N$ being replaced by $n$.
The two-loop photon coefficient function
$C_{2,n}^{\gamma(2)}$ is expressed as 
\begin{equation}
C_{2,n}^{\gamma(2)}=\delta_\gamma c_{2,\gamma}^{(2)}(n)~,
\end{equation}
where $c_{2,\gamma}^{(2)}(n)$ is obtained from $c_{2,g}^{(2)}(n)$ in (\ref{GluonCoeffi2}) by replacing 
$C_A \rightarrow 0$ and $\frac{n_f}{2} \rightarrow 1$~\cite{MVV}.

\subsection{Anomalous dimensions}

The one-loop  anomalous dimensions for the hadronic sector were 
calculated long time ago~\cite{GrossWilczek,GeorgiPolitzer}. The expressions of 
$\gamma^{(0),n}_{\psi\psi}=\gamma^{(0),n}_{NS}$, 
$\gamma^{(0),n}_{\psi G}$, $\gamma^{(0),n}_{G\psi}$ and $\gamma^{(0),n}_{GG}$ 
are given, for example, 
in Eqs.(4.1), (4.2), (4.3) and (4.4) of Ref.\cite{BB}, respectively, with $f$ being replaced 
by $n_f$.
As for the one-loop anomalous dimension row vector $\bm{K}_n^{(0)}=(K_\psi^{{(0),n}}, 
K_G^{{(0),n}}, K_{NS}^{{(0),n}} )$, we have $K_G^{{(0),n}}=0$,  and $K_\psi^{{(0),n}}$
and $K_{NS}^{{(0),n}}$ are given, respectively, in Eqs.(4.5) and (4.6) of Ref.\cite{BB} 
with $f$ being replaced by $n_f$ again.

The two-loop  anomalous dimensions for the hadronic sector  
were calculated in Ref.\cite{FRS} and recalculated 
using a different method and a different gauge in Ref.\cite{CFP}. The results by the two groups 
agreed with each other except in the part of $\gamma^{(1),n}_{GG}$ proportional to $C_G^2$, but 
this discrepancy was  solved later~\cite{HvN}. They are given by
\bea
\gamma^{(1),n}_{NS}&=&2\gamma^{(1)+}_{\rm ns}(n)~, \label{TwoLoopNS}\\
\gamma^{(1),n}_{\psi\psi}&=&2(\gamma^{(1)+}_{\rm ns}(n)+\gamma^{(1)}_{\rm ps}(n))~, \\
\gamma^{(1),n}_{\psi G}&=&2\gamma^{(1)}_{\rm qg}(n)~,\\
\gamma^{(1),n}_{G \psi}&=&2\gamma^{(1)}_{\rm gq}(n)~,\\
\gamma^{(1),n}_{G G}&=&2\gamma^{(1)}_{\rm gg}(n)~,\label{TwoLoopGG}
\eea
where $\gamma^{(1)+}_{\rm ns}(n)$ is given in Eq.(3.5) of Ref.\cite{MVV1},  and 
$\gamma^{(1)}_{\rm ps}(n)$, $\gamma^{(1)}_{\rm qg}(n)$, $\gamma^{(1)}_{\rm gq}(n)$ and 
$\gamma^{(1)}_{\rm gg}(n)$ are given, respectively, in Eqs.(3.6), (3.7), (3.8) and (3.9) of 
Ref.\cite{MVV2},  with $N$ being replaced by $n$. The factor of 2 in (\ref{TwoLoopNS}) - 
(\ref{TwoLoopGG}) appears since, in Refs.\cite{MVV1} and  \cite{MVV2}, the anomalous
dimension $\gamma$ of the renormalized  operator $O$ is defined as $d O/{d {\rm ln}
\mu^2}=-\gamma O$ instead of $d O/{d {\rm ln}\mu}=-\gamma O$. 

The two-loop anomalous dimensions $K_\psi^{{(1),n}}$, $K_{NS}^{{(1),n}}$ and $ K_G^{{(1),n}}$
can be obtained from $\gamma^{(1),n}_{\psi G}$ and $\gamma^{(1),n}_{G G}$ by replacing 
color factors with relevant charge factors~\cite{BB}. Moreover we need  an additional 
procedure for $ K_G^{{(1),n}}$. 
They are given by
\bea
K_\psi^{{(1),n}}&=&-3n_f\langle e^2 \rangle C_F D_{\psi G}(n)~, \\
K_{NS}^{{(1),n}}&=&-3n_f(\langle e^4 \rangle-\langle e^2 \rangle^2)C_F D_{\psi G}(n)~,\\
 K_G^{{(1),n}}&=&-3n_f\langle e^2 \rangle C_F( D_{G G}(n)-8)~, \label{TwoLoopKG}
\eea
where $D_{\psi G}(n)$ and $D_{G G}(n)$ are obtained from  $\gamma^{(1),n}_{\psi G}$ and
$\gamma^{(1),n}_{G G}$, respectively, by replacing 
$C_A \rightarrow 0$ and $C_F n_f \rightarrow 2$. The number 8 in (\ref{TwoLoopKG}) is due to 
the gluon self-energy contribution to $\gamma^{(1),n}_{G G}$, which should be dropped for 
$ K_G^{{(1),n}}$~\cite{FP,GRV}.

The three-loop  anomalous dimensions for the hadronic sector  have been
calculated recently in Refs.\cite{MVV1} and \cite{MVV2}. They are expressed as
\bea
\gamma^{(2),n}_{NS}&=&2\gamma^{(2)+}_{\rm ns}(n)~, \label{ThreeLoopNS}\\
\gamma^{(2),n}_{\psi\psi}&=&2(\gamma^{(2)+}_{\rm ns}(n)+\gamma^{(2)}_{\rm ps}(n))~, \\
\gamma^{(2),n}_{\psi G}&=&2\gamma^{(2)}_{\rm qg}(n)~,\\
\gamma^{(2),n}_{G \psi}&=&2\gamma^{(2)}_{\rm gq}(n)~,\\
\gamma^{(2),n}_{G G}&=&2\gamma^{(2)}_{\rm gg}(n)~,\label{ThreeLoopGG}
\eea
where $\gamma^{(2)+}_{\rm ns}(n)$ is given in Eq.(3.7) of Ref.\cite{MVV1},  and 
$\gamma^{(2)}_{\rm ps}(n)$, $\gamma^{(2)}_{\rm qg}(n)$, $\gamma^{(2)}_{\rm gq}(n)$ and 
$\gamma^{(2)}_{\rm gg}(n)$ are given, respectively, in Eqs.(3.10), (3.11), (3.12) and (3.13) of 
Ref.\cite{MVV2},  with $N$ being replaced by $n$.

Concerning 
the three-loop anomalous dimensions $K_\psi^{{(2),n}}$, $K_{NS}^{{(2),n}}$ and $ K_G^{{(2),n}}$, 
the exact expressions have not been in literature yet. In fact, the lowest six even-integer 
Mellin moments, $n=2, \cdots, 12$, of these anomalous dimensions were calculated and given in Ref.\cite{MVV}.
Quite recently, the authors of Ref.\cite{MVV} have presented compact parameterizations of the three-loop
photon-non-singlet quark and photon-gluon splitting functions, $P^{(2)}_{{\rm ns}\gamma}(x)$ and
$P^{(2)}_{{\rm g}\gamma}(x)$, instead of providing the exact analytic results~\cite{MVV3}.  It is remarked
there that their parameterizations deviate from the lengthy full expressions by about 0.1\% or less.
They also gave  in Ref.\cite{MVV3} the  analytic expression of the three-loop photon-pure-singlet quark
splitting function $P^{(2)}_{{\rm ps}\gamma}(x)$.
It is true that we can infer the analytic expressions for some parts of $K_\psi^{{(2),n}}$, $K_{NS}^{{(2),n}}$
and
$ K_G^{{(2),n}}$ from the known three-loop results of $\gamma^{(2),n}_{\psi G}$ and
$\gamma^{(2),n}_{GG}$. For instance, the expressions of $K_\psi^{{(2),n}}$ and $K_{NS}^{{(2),n}}$ 
which have the color factor $C_F^2$ are obtained from $\gamma^{(2),n}_{\psi G}$ by taking the terms
which are proportional to the color factor $n_f C_F^2$. Also the terms of $ K_G^{{(2),n}}$ 
which have the color factors $n_f C_F$ and $C_F^2$ are related to the ones of 
$\gamma^{(2),n}_{GG}$ with the color factors $n_f^2 C_F$ and $n_f C_F^2$, respectively. 
But at present we do not have the exact analytic expressions of $K_\psi^{{(2),n}}$, $K_{NS}^{{(2),n}}$ and $
K_G^{{(2),n}}$ as a whole.

Under these circumstances we are reconciled to use of
approximate  expressions for $K_\psi^{{(2),n}}$, $K_{NS}^{{(2),n}}$ and $ K_G^{{(2),n}}$.  
They are obtained by taking the Mellin moments of 
the parameterizations for $P^{(2)}_{{\rm ns}\gamma}(x)$ and  $P^{(2)}_{{\rm
g}\gamma}(x)$, and of the exact result for $P^{(2)}_{{\rm ps}\gamma}(x)$ which are presented in
Ref.\cite{MVV3}.  
Then we have
\bea
K_{NS}^{{(2),n}}&\approx&K_{NS\ \rm approx}^{{(2),n}} \equiv -3n_f(\langle e^4 \rangle-\langle e^2 \rangle^2)~
2E_{{\rm ns}\gamma}^{\rm approx}(n)~,\label{EnsApprox}\\
K_\psi^{{(2),n}}&\approx&K_{\psi\ \rm approx}^{{(2),n}} \equiv-3n_f\langle e^2 \rangle 2\Bigl\{E_{{\rm
ns}\gamma}^{\rm approx}(n) + E_{{\rm ps}\gamma}(n)\Bigr\}~, \label{Eps}\\
 K_G^{{(2),n}}&\approx&K_{G\ \rm approx}^{{(2),n}} \equiv-3n_f\langle e^2 \rangle 2E_{G\gamma}^{\rm
approx}(n)~, \label{EgApprox}
\eea
where the explicit expressions of $E_{{\rm ns}\gamma}^{\rm approx}(n)$, $ E_{{\rm ps}\gamma}(n)$, 
and $E_{G\gamma}^{\rm approx}(n)$ are given in Appendix \ref{Ensgamma}. 
Again the appearance of the factor of 2 in (\ref{EnsApprox}) - (\ref{EgApprox})
is due to the difference in definition of  the anomalous dimensions.
As mentioned earlier, the lowest six even-integer 
Mellin moments, $n=2, \cdots, 12$, of $K_{NS}^{{(2),n}}$, $K_\psi^{{(2),n}}$ and $ K_G^{{(2),n}}$  
were given in Ref.\cite{MVV}. When we write $K_{NS}^{{(2),n}}$ and $ K_G^{{(2),n}}$ as 
\bea
K_{NS}^{{(2),n}}&\equiv& -3n_f(\langle e^4 \rangle-\langle e^2 \rangle^2)~2E_{{\rm ns}\gamma}(n)~, \\
 K_G^{{(2),n}}&\equiv&-3n_f\langle e^2 \rangle 2E_{G\gamma}(n)~,
\eea
then we get the exact results of $E_{{\rm ns}\gamma}(n)$ and $E_{G\gamma}(n)$ for even 
$n=2,  \cdots, 12$. We give in Table\ref{ApproxVsExact} the results of $E_{{\rm ns}\gamma}(n)$, 
$E_{{\rm ns}\gamma}^{\rm approx}(n)$, $E_{{\rm G}\gamma}(n)$ and $E_{{\rm G}\gamma}^{\rm approx}(n)$
in numerical form for the lowest six even-integer values of $n$. 
We see the deviations of $E_{{\rm ns}\gamma}^{\rm approx}(n)$   from
$E_{{\rm ns}\gamma}(n)$  and $E_{{\rm G}\gamma}^{\rm approx}(n)$  from $E_{{\rm G}\gamma}(n)$
are both far less than 0.1\% for these values of $n$.

\begin{table}
   \caption{\ Numerical values of $E_{{\rm ns}\gamma}(n)$, 
$E_{{\rm ns}\gamma}^{\rm approx}(n)$, $E_{{\rm G}\gamma}(n)$ and $E_{{\rm G}\gamma}^{\rm approx}(n)$ for 
the lowest six even-integer values of $n$. The values for $E_{{\rm ns}\gamma}(n)$ ($E_{{\rm G}\gamma}(n)$) 
are found in Eq.(3.1) (Eq.(3.3)) or obtained by evaluating Eqs.(A.1)-(A.6) (Eqs.(A.7)-(A.12)) 
of Ref.\cite{MVV}. The values of $E_{{\rm ns}\gamma}^{\rm approx}(n)$ and $E_{{\rm G}\gamma}^{\rm approx}(n)$
are obtained from the expressions given in (\ref{EnsgammaApprox}) and 
(\ref{EGgammaApprox}) in Appendix B, respectively. }
    \begin{center}
      \small
      \catcode`?=\active \def?{\phantom{0}}
      \begin{ruledtabular}
      \begin{tabular}{l|c@{\hspace{8pt}}c|c@{\hspace{8pt}}c}
        & $E_{{\rm ns}\gamma}(n)$ & $E_{{\rm ns}\gamma}^{\rm approx}(n)$
          & $E_{G\gamma}(n)$ & $E_{G\gamma}^{\rm approx}(n)$ \\
        \hline
        $n=2$
          & -?86.9753+1.47051 $n_f$
          & -?86.9844+1.47104 $n_f$
          &  31.4197?+5.15775?$n_f$
          &  31.4155?+5.15803?$n_f$ \\
        $n=4$
          & -102.831?+1.47737 $n_f$
          & -102.848?+1.47787 $n_f$
          &  23.9427?+1.10886?$n_f$
          &  23.9419?+1.10888?$n_f$ \\
        $n=6$
          & -109.278?+1.65653 $n_f$
          & -109.299?+1.65699 $n_f$
          &  15.6517?+0.695953$n_f$
          &  15.6507?+0.695944$n_f$ \\
        $n=8$
          & -111.167?+1.69550 $n_f$
          & -111.192?+1.69592 $n_f$
          &  10.9661?+0.498196$n_f$
          &  10.9651?+0.498178$n_f$ \\
        $n=10$
          & -111.035?+1.67061 $n_f$
          & -111.062?+1.67099 $n_f$
          &  ?8.16031+0.379060$n_f$
          &  ?8.15953+0.379038$n_f$ \\
        $n=12$
          & -109.943?+1.61908 $n_f$
          & -109.972?+1.61943 $n_f$
          &  ?6.34829+0.300274$n_f$
          &  ?6.34777+0.300250$n_f$ \\
      \end{tabular}
      \end{ruledtabular}
      \label{ApproxVsExact}
    \end{center}
  \end{table}

\subsection{Photon matrix elements}

The two-loop operator matrix elements have been calculated up to the finite terms 
by Matiounine, Smith and van Neerven (MSvN)~\cite{MSvN}. Using their results and changing color-group 
factors, we obtain the photon matrix elements of hadronic operators  
up to the two-loop level. 
 
First we clear up a subtle issue which appears in the calculation of the 
photon matrix elements of the hadronic operators. The one-loop gluon coefficient 
function ${\overline B}^n_G$ in (\ref{OneLoopCoefficient}) was calculated by two 
groups, BBDM and FRS (we have taken initials of the authors of Refs.\cite{BBDM} and \cite{FRS}, 
respectively). Both groups evaluated  one-loop diagrams contributing to the forward
virtual  photon-gluon scattering as well as those contributing to the matrix element of the quark operator
between gluon  states, and they took a difference between the two to obtain ${\overline B}^n_G$. 
But actually BBDM calculated the gluon spin averaged contributions, i.e., multiplying 
$g_{\rho\tau}$ and contracting pairs of Lorentz indices $\rho$ and $\tau$, 
whereas FRS picked up the parts which are proportional to $g_{\rho\tau}$. 
Thus the BBDM results on the contributions to
the forward virtual photon-gluon  scattering and 
the gluon matrix element of quark operator are different from those by FRS, 
but the difference between the two contributions, i.e., ${\overline B}^n_G$, 
is the same as it should be. 

We have defined the photon structure functions $F_2^\gamma$ and $F_L^\gamma$ in (\ref{Wmunu}) and 
(\ref{DefF2FL}), taking a spin average of the target photon for the structure tensor 
$W_{\mu\nu\rho\tau}(p,q)$. We, therefore,  adopt the BBDM result rather than that of  FRS 
and convert it to the photon case.
Then for the 
photon matrix elements of the hadronic operators at one-loop level we get , 
\bea
{\widetilde A}_n^{(1)\psi}&=&3n_f\langle e^2 \rangle H_q^{(1)}(n)~, \nonumber \\
{\widetilde A}_n^{(1)G}&=&0~, \\
{\widetilde A}_n^{(1)NS}&=&3n_f \Bigl(\langle e^4 \rangle-\langle e^2 \rangle^2 \Bigr)  H_q^{(1)}(n)~,
\nonumber
\label{PhotonHadronicMatOneLoop}
\eea
where
\begin{equation}
H_q^{(1)}(n)=4\biggl[ -\frac{1}{n}+\frac{1}{n^2}-\frac{4}{(n+1)^2}+\frac{4}{(n+2)^2} 
+\biggl(\frac{1}{n}-\frac{2}{n+1}+\frac{2}{n+2}    \biggr) S_1(n)\biggr]~,   \label{Hq1} 
\end{equation}
with $S_1(n)=\sum^n_{j=1}\frac{1}{j}$~.
Actually, 
 $H_q^{(1)}(n)$ is related to the BBDM result on the one-loop gluon matrix element of quark operator 
$A_{n G}^{(2)\psi}$ given in Eq.(6.2) of Ref.\cite{BBDM} as 
$A_{n G}^{(2)\psi}=\frac{\alpha_s}{4\pi}\frac{n_f}{2}H_q^{(1)}(n)$.

MSvN have presented in Appendix A of Ref.\cite{MSvN} full expressions for the 
two-loop corrected operator matrix elements which are unrenormalized and include external 
self-energy corrections. The expressions are given in parton momentum fraction  space, 
i.e., in $z$-space. Taking the
moments,  the unrenormalized matrix elements of the (flavor-singlet) quark operators between gluon states are
written  in the form as (see Eq.(2.18) of Ref.\cite{MSvN}), 
\begin{equation}
{\widehat A}_{qg, \rho\tau}(n, \frac{-p^2}{\mu^2}, \frac{1}{\epsilon} )
={\widehat A}_{qg}^{\rm PHYS}(n) T^{(1)}_{\rho\tau} +{\widehat A}_{qg}^{\rm EOM}(n)
T^{(2)}_{\rho\tau}+{\widehat A}_{qg}^{\rm NGI}(n) T^{(3)}_{\rho\tau}~,
\end{equation} 
where
\begin{equation}
{\widehat A}_{qg}^{k}(n)=\int_0^1 dz z^{n-1}{\widehat A}_{qg}^{k}(z, \frac{-p^2}{\mu^2},
\frac{1}{\epsilon} )~, \qquad k={\rm PHYS~, EOM\ and\ NGI} ~,
\end{equation}
and the expressions of ${\widehat A}_{qg}^{\rm PHYS}(z, \frac{-p^2}{\mu^2},\frac{1}{\epsilon} )$, 
${\widehat A}_{qg}^{\rm EOM}(z, \frac{-p^2}{\mu^2},\frac{1}{\epsilon} )$ and ${\widehat A}_{qg}^{\rm NGI}(z,
\frac{-p^2}{\mu^2},\frac{1}{\epsilon} )$  are given in Eqs.(A7), (A8) and (A9) of Ref.\cite{MSvN},
respectively. Refer to Ref.\cite{MSvN} for the explanation of the ``PHYS", ``EOM" and ``NGI" parts. 
The tensors $T^{(i)}_{\rho\tau}\ (i=1,2,3)$ are given by 
(see Eqs.(2.19)-(2.21) of Ref.\cite{MSvN} and note that we have changed the Lorentz indices of 
gluon fields from $\mu\nu$ to $\rho\tau$), 
\bea
T^{(1)}_{\rho\tau}&=&\Bigl[g_{\rho\tau}-\frac{p_\rho\Delta_\tau+\Delta_\rho p_\tau}{\Delta\cdot p}
+\frac{\Delta_\rho\Delta_\tau p^2}{(\Delta\cdot p)^2}   \Bigr](\Delta\cdot p)^n~, \\
T^{(2)}_{\rho\tau}&=&\Bigl[\frac{p_\rho p_\tau}{p^2}-\frac{p_\rho\Delta_\tau+\Delta_\rho p_\tau}{\Delta\cdot p}
+\frac{\Delta_\rho\Delta_\tau p^2}{(\Delta\cdot p)^2}   \Bigr](\Delta\cdot p)^n~, \\
T^{(3)}_{\rho\tau}&=&\Bigl[-\frac{p_\rho\Delta_\tau+\Delta_\rho p_\tau}{2\Delta\cdot p}
+\frac{\Delta_\rho\Delta_\tau p^2}{(\Delta\cdot p)^2}   \Bigr](\Delta\cdot p)^n~, 
\eea
where $\Delta_\mu$ is a lightlike vector ($\Delta^2=0$).
The renormalization of ${\widehat A}_{qg, \rho\tau}(n, \frac{-p^2}{\mu^2}, \frac{1}{\epsilon} )$ proceeds as
follows: First the coupling constant and gauge constant renormalizations are performed. Then the remaining 
ultraviolet divergences are removed by multiplication of the operator 
renormalization constants. We get the finite expression at $\mu^2=-p^2$ as
\bea
A_{qg, \rho\tau}(n) |_{\mu^2=-p^2}&=&\ \Bigl\{\frac{\alpha_s}{4\pi} ~ a_{qg}^{(1)}(n) 
+ \Bigl(\frac{\alpha_s}{4\pi} \Bigr)^2 ~ a_{qg}^{(2)}(n) \Bigr\} 
T^{(1)}_{\rho\tau} \nonumber\\
&&+\Bigl\{\frac{\alpha_s}{4\pi} ~ b_{qg}^{(1)}(n)  
 + \Bigl(\frac{\alpha_s}{4\pi} \Bigr)^2 ~ b_{qg}^{(2)}(n)  \Bigr\} 
T^{(2)}_{\rho\tau} \nonumber\\
&&+\Bigl(\frac{\alpha_s}{4\pi} \Bigr)^2 ~ a_{qA}^{(2)}(n) ~T^{(3)}_{\rho\tau}~.
\eea
The expressions of $a_{qg}^{(i)}(n)$ and $b_{qg}^{(i)}(n)\ (i=1,2)$ are given in Appendix C, while 
$a_{qA}^{(2)}(n)$ is made up of the terms proportional to $C_A\frac{n_f}{2}$ and is, therefore, irrelevant
to the photon matrix element of the quark operator.  Now  multiplying 
$g^{\rho\tau}$ and contracting pairs of indices $\rho$ and $\tau$, we get  
\bea
\frac{1}{2}g^{\rho\tau}\frac{1}{(\Delta\cdot p)^n}A_{qg, \rho\tau}(n) |_{\mu^2=-p^2}&=&~
\frac{\alpha_s}{4\pi} ~\Bigl\{a_{qg}^{(1)}(n) -\frac{1}{2} b_{qg}^{(1)}(n)    \Bigr\} \nonumber\\
&&+\Bigl(\frac{\alpha_s}{4\pi} \Bigr)^2 ~\Bigl\{a_{qg}^{(2)}(n) -\frac{1}{2} b_{qg}^{(2)}(n)   
-\frac{1}{2} a_{qA}^{(2)}(n)\Bigr\}~.  \label{SpinAveragedAqg}
\eea
We can see from the expressions of $a_{qg}^{(1)}(n)$ and  $b_{qg}^{(1)}(n)$ in (\ref{aqg1}) and
(\ref{bqg1}), respectively, that the FRS result for the one-loop gluon matrix element of 
quark operator corresponds to $a_{qg}^{(1)}(n)$, while the BBDM result 
corresponds to the combination $\Bigl\{a_{qg}^{(1)}(n) -\frac{1}{2} b_{qg}^{(1)}(n) \Bigr\}$.
Indeed we find that $H_q^{(1)}(n)$ in (\ref{Hq1}) is written as
$\frac{n_f}{2}H_q^{(1)}(n)=\Bigl\{a_{qg}^{(1)}(n) -\frac{1}{2} b_{qg}^{(1)}(n) \Bigr\}$. 

The two-loop photon matrix elements of the quark operators are derived from the combination
$\Bigl\{a_{qg}^{(2)}(n) -\frac{1}{2} b_{qg}^{(2)}(n)   
-\frac{1}{2} a_{qA}^{(2)}(n)\Bigr\}$ in (\ref{SpinAveragedAqg}) with the following 
replacements: $C_A\rightarrow 0$, $\Bigl(\frac{n_f}{2}\Bigr)^2 \rightarrow 0$ and  $C_F\frac{n_f}{2}
\rightarrow \Bigl[ C_F\times {\rm charge\ factor}\Bigr]$.  The terms proportional to
$\Bigl(\frac{n_f}{2}\Bigr)^2$ in $a_{qg}^{(2)}(n)$ and $ b_{qg}^{(2)}(n)$ come from the 
external gluon self-energy corrections and should be discarded for the photon case. 
Thus we obtain
\bea
{\widetilde A}_n^{(2)\psi}&=&3n_f\langle e^2 \rangle H_q^{(2)}(n)~, \nonumber \\
{\widetilde A}_n^{(2)NS}&=&3n_f \Bigl(\langle e^4 \rangle-\langle e^2 \rangle^2 \Bigr)  H_q^{(2)}(n)~,
\label{PhotonQuarkMatTwoLoop}
\eea
where
\bea
H_q^{(2)}(n)&=&C_F \biggl\{ \Bigl(\frac{1}{n}- \frac{2}{n+1}+\frac{2}{n+2}  \Bigr) 
\Bigl( -\frac{4}{3}S_1(n)^3 -4S_2(n)S_1(n)+\frac{64}{3}S_3(n)\nonumber \\
&&\hspace{8cm}-16S_{2,1}(n) -48\zeta_3 \Bigr) \nonumber \\
&&\hspace{1.5cm}+S_1(n)^2\Bigl( \frac{6}{n}- \frac{8}{n+1}+\frac{16}{n+2}  
-\frac{16}{n^2}+ \frac{40}{(n+1)^2}-\frac{32}{(n+2)^2}  \Bigr) \nonumber \\
&&\hspace{1.5cm}+S_1(n)\Bigl( \frac{4}{n}- \frac{80}{n+1}+\frac{56}{n+2}  
+\frac{16}{n^2}- \frac{48}{(n+1)^2}+\frac{64}{(n+2)^2} \nonumber \\
&&\hspace{6.5cm}-\frac{32}{n^3}+ \frac{176}{(n+1)^3}-\frac{128}{(n+2)^3}   \Bigr) \nonumber \\
&&\hspace{1.5cm}+S_2(n)\Bigl( \frac{6}{n}+ \frac{8}{n+1} 
-\frac{16}{n^2}+ \frac{40}{(n+1)^2}-\frac{32}{(n+2)^2}  \Bigr) \nonumber \\
&&\hspace{1.5cm}+\frac{38}{n}- \frac{70}{n+1}+\frac{56}{n+2}
+\frac{56}{n^2}- \frac{198}{(n+1)^2}+\frac{144}{(n+2)^2}
-\frac{22}{n^3}\nonumber \\
&&\hspace{1.5cm}- \frac{40}{(n+1)^3}+\frac{128}{(n+2)^3} 
+\frac{20}{n^4}+ \frac{88}{(n+1)^4}
\biggr\}~. \label{Hq2}
\eea

Similarly the renormalized matrix elements of the gluon operators between gluon states 
at $\mu^2=-p^2$ are written as (the unrenormalized version is given in Eq.(2.33) of Ref.\cite{MSvN}),
\bea
A_{gg, \rho\tau}(n) |_{\mu^2=-p^2}&=&\ \Bigl\{\frac{\alpha_s}{4\pi} ~ a_{gg}^{(1)}(n) 
+ \Bigl(\frac{\alpha_s}{4\pi} \Bigr)^2 ~ a_{gg}^{(2)}(n) \Bigr\} 
T^{(1)}_{\rho\tau} \nonumber\\
&&+\Bigl\{\frac{\alpha_s}{4\pi} ~ b_{gg}^{(1)}(n)  
 + \Bigl(\frac{\alpha_s}{4\pi} \Bigr)^2 ~ b_{gg}^{(2)}(n)  \Bigr\} 
T^{(2)}_{\rho\tau} \nonumber\\
&&+\Bigl\{\frac{\alpha_s}{4\pi} ~ a_{gA}^{(1)}(n)  
 + \Bigl(\frac{\alpha_s}{4\pi} \Bigr)^2 ~ a_{gA}^{(2)}(n)  \Bigr\}
 ~T^{(3)}_{\rho\tau}~.
\eea
The one-loop results $a_{gg}^{(1)}(n)$, $b_{gg}^{(1)}(n) $ and $a_{gA}^{(1)}(n) $ are all proportional 
to the color factor $C_A$ and thus they are irrelevant to the photon matrix elements. 
Also the two-loop result $a_{gA}^{(2)}(n)$ is made up of the terms proportional to 
$C_A^2$ or $C_A\frac{n_f}{2}$ and is irrelevant.  
The expressions of $a_{gg}^{(2)}(n)$ and $b_{gg}^{(2)}(n)$ are given by (\ref{agg2}) and (\ref{bgg2}),
respectively, in Appendix \ref{AppenC}. Then, we take the  combination
$\Bigl\{a_{gg}^{(2)}(n) -\frac{1}{2} b_{gg}^{(2)}(n) \Bigr\}$ and make  replacements, 
$C_A\rightarrow 0$, $\Bigl(\frac{n_f}{2}\Bigr)^2 \rightarrow 0$ and  $C_F\frac{n_f}{2}
\rightarrow \Bigl[ C_F\times {\rm charge\ factor}\Bigr]$. Furthermore, we realize that 
the last two terms in parentheses of (\ref{agg2}) have also resulted from the external gluon self-energy 
corrections and are thus irrelevant for the photon case. In the end we obtain for 
the photon matrix elements of the gluon operators, 
\begin{equation}
{\widetilde A}_n^{(2)G}=3n_f\langle e^2 \rangle H_G^{(2)}(n)~, 
\label{PhotonGluonMatTwoLoop}
\end{equation} 
where
\bea
H_G^{(2)}(n)&=&C_F \biggl\{ \Bigl(S_1(n)^2 +S_2(n) \Bigr)
\Bigl(\frac{16}{3(n-1)}+\frac{4}{n}- \frac{4}{n+1}-\frac{16}{3(n+2)}  
-\frac{8}{n^2}- \frac{8}{(n+1)^2}  \Bigr) \nonumber \\
&&\hspace{1.2cm}+S_1(n)\Bigl(-\frac{32}{9(n-1)}- \frac{32}{n}+ \frac{32}{n+1}+\frac{32}{9(n+2)}  
+\frac{32}{n^2}+ \frac{8}{(n+1)^2} \nonumber \\
&&\hspace{7cm} -\frac{64}{3(n+2)^2}-\frac{32}{n^3}- \frac{48}{(n+1)^3}   \Bigr) \nonumber \\
&&\hspace{1.2cm}+S_{-2}(n)\Bigl( \frac{32}{3(n-1)}- \frac{32}{n}+ \frac{32}{n+1}-\frac{32}{3(n+2)}  
  \Bigr) \nonumber \\
&&\hspace{1.2cm}+\frac{872}{27(n-1)}-\frac{80}{n}+ \frac{16}{n+1}+\frac{856}{27(n+2)}
-\frac{40}{n^2}+ \frac{104}{(n+1)^2}+\frac{64}{9(n+2)^2}
+\frac{44}{n^3}\nonumber \\
&&\hspace{1.2cm}+ \frac{28}{(n+1)^3}-\frac{128}{3(n+2)^3} 
-\frac{40}{n^4}- \frac{88}{(n+1)^4}
\biggr\}~.\label{HG2}
\eea

\bigskip

With all these necessary parameters at hand, we are now ready to analyze the moments of 
$F_2^\gamma(x,Q^2,P^2)$ up to the NNLO. First we evaluate the coefficients 
${\cal L}^n_i$, ${\cal A}^n_i$, ${\cal B}^n_i$, ${\cal C}^n$, ${\cal D}^n_i$, 
${\cal E}^n_i$, ${\cal F}^n_i$ and ${\cal G}^n$ with $i=+, -, NS$, the expressions of which are  given
in Eqs.(\ref{ExpressionL})-(\ref{CoeffiG}),  for $n\!=\!2,4,\cdots,12$ in the cases of $n_f\!=\!3$ and
$n_f\!=\!4$. The results are listed in Tables \ref{ABCDEFGNF3} (for $n_f\!=\!3$) and \ref{ABCDEFGNF4} (for
$n_f\!=\!4$). In Table 1 of Ref.\cite{UW2}, the numerical values of the seven NLO
coefficients,  ${\cal A}^n_i$, ${\cal B}^n_i$, and ${\cal C}^n$ with $i=+, -, NS$ for 
$n\!=\!2,4,\cdots,20$ in the case of $n_f=4$ were already given. Our results on ${\cal A}^n_i$, ${\cal
B}^n_i$, and
${\cal C}^n$ in Table \ref{ABCDEFGNF4} are consistent with those in Ref.\cite{UW2} except for the values 
of ${\cal A}^n_+$ and  ${\cal A}^n_-$. The discrepancy in the values of ${\cal A}^n_+$ and  ${\cal
A}^n_-$ arises from a term $-8$ in the parentheses of Eq.(\ref{TwoLoopKG}). See the discussion below 
Eq.(\ref{TwoLoopKG}). The numerical calculation of the NNLO coefficients ${\cal D}^n_+$,${\cal D}^n_-$
and ${\cal D}^n_{NS}$  for $n\!=\!2,4,\cdots,12$
 in Tables \ref{ABCDEFGNF3} and \ref{ABCDEFGNF4} was performed
by using the ``exact" values of the three-loop  anomalous dimensions, $K_{NS}^{{(2),n}}$,
$K_\psi^{{(2),n}}$ and
$ K_G^{{(2),n}}$, for  
$n\!=\!2, \cdots, 12$ given in Ref.\cite{MVV} (at upper levels) and also by using the approximate 
expressions, $K_{NS\ \rm approx}^{{(2),n}}$, $K_{\psi\ \rm approx}^{{(2),n}}$ and $K_{G\ \rm
approx}^{{(2),n}}$  defined in Eqs.(\ref{EnsApprox})-(\ref{EgApprox}) (in parentheses at lower levels).
The coefficients  ${\cal A}^n_-$ and ${\cal E}^n_-$ cannot be evaluated at
$n\!=\!2$ since they become singular there. More details concerning this singularity will be discussed 
in the next section. 

The coefficients ${\cal D}^n_-$ and ${\cal D}^n_{NS}$ in Table \ref{ABCDEFGNF3} take extremely large values 
at $n=6$. The values of ${\cal D}^{n=6}_-$ and ${\cal D}^{n=6}_{NS}$ in Table \ref{ABCDEFGNF4} are also
large.  This is due to the fact that 
${\cal D}^n_-$ and ${\cal D}^n_{NS}$ have terms with the factors $\frac{1}{1-d_-^n}$ and 
$\frac{1}{1-d_{NS}^n}$, respectively, and that $d_-^n$ and $d_{NS}^n$ happen to be very close to one 
at $n=6$. Actually we obtain $d_-^{n=6}=0.995846$ and $d_{NS}^{n=6}\!=\!1.00035$ for $n_f\!=\!3$ (Table
\ref{ABCDEFGNF3}),   and  $d_-^{n=6}\!=\!1.07427$ and $d_{NS}^{n=6}\!=\!1.08038$ for $n_f\!=\!4$ (Table
\ref{ABCDEFGNF4}).  But we see from (\ref{master1}) that  ${\cal D}_-^n$ and ${\cal D}_{NS}^n$ are
multiplied, respectively,  by  the factors
$\Bigl[1-\left(\frac{\alpha_s(Q^2)}{\alpha_s(P^2)}\right)^{d_-^n-1}\Bigr]$   and
$\Bigl[1-\left(\frac{\alpha_s(Q^2)}{\alpha_s(P^2)}\right)^{d_{NS}^n-1}\Bigr]$ which become very small
when  
$d_-^n$ and $d_{NS}^n$ are close to one. Thus the contributions of the parts with 
${\cal D}_-^{n=6}$ and ${\cal D}_{NS}^{n=6}$ to the 6-th moment of $F_2^\gamma(x,Q^2,P^2)$ do not stand out from the
others. 


\begin{table}
  \begin{center}
    \caption{%
      Numerical values of ${\cal L}^n_i, {\cal A}^n_i, {\cal B}^n_i, {\cal D}^n_i,
      {\cal E}^n_i, {\cal F}^n_i (i=+, -, NS)$, and ${\cal C}^n$ and  ${\cal G}^n$ 
      for $n=2, 4, \cdots, 12$ in the case of $n_f=3$.
     The calculation of ${\cal D}^n_+$,${\cal D}^n_-$ and ${\cal D}^n_{NS}$ 
 was performed by using
the ``exact" values of  $K_{NS}^{{(2),n}}$, $K_\psi^{{(2),n}}$ and
$ K_G^{{(2),n}}$ given in Ref.\cite{MVV} (at upper levels) and also by using the approximate 
expressions, $K_{NS\ \rm approx}^{{(2),n}}$, $K_{\psi\ \rm approx}^{{(2),n}}$ and $K_{G\ \rm
approx}^{{(2),n}}$  defined in Eqs.(\ref{EnsApprox})-(\ref{EgApprox}) (in parentheses at lower levels).}
    \label{ABCDEFGNF3}
    \makeatletter
    \ifnum \@pointsize=11 
      \footnotesize
    \else\ifnum \@pointsize=12 
      \scriptsize
    \fi\fi
    \makeatother
    \catcode`?=\active \def?{\phantom{0}} 
    \catcode`*=\active \def*{\phantom{-}} 
    \def\dd#1#2{$\genfrac{}{}{0pt}{}{\mbox{#1}}{\mbox{(#2)}}$} 
    \begin{ruledtabular}
      \begin{tabular}{c|cccccccccc}
        n & ${\cal L}^n_+$ & ${\cal L}^n_-$ & ${\cal L}^n_{NS}$
          & ${\cal A}^n_+$ & ${\cal A}^n_-$ & ${\cal A}^n_{NS}$
          & ${\cal B}^n_+$ & ${\cal B}^n_-$ & ${\cal B}^n_{NS}$ & ${\cal C}^n$ \\
        \hline
        $2$
          & 0.4690???? & 0.4267  & 0.4248?
          & -2.8403??  & ---     & -5.5940
          & 1.7481??   & -1.8535 & 0.8290
          & ?-9.3333 \\
        $4$
          & 0.004336?? & 0.3639  & 0.1836?
          & -0.5543??  & -2.6267 & -1.3299
          & 0.07353?   & *3.3149 & 1.4607
          & -10.7467 \\
        $6$
          & 0.0005428? & 0.2324  & 0.1164?
          & *0.06133?  & -1.8806 & -0.9403
          & 0.01652?   & *2.9783 & 1.5349
          & ?-9.1088 \\
        $8$
          & 0.0001493? & 0.1689  & 0.08451
          & *0.009544  & -1.6566 & -0.8277
          & 0.006245   & *2.9612 & 1.4906
          & ?-7.7504 \\
        $10$
          & 0.00005803 & 0.1318  & 0.06591
          & *0.002817  & -1.5336 & -0.7664
          & 0.002993   & *2.8263 & 1.4169
          & ?-6.7116 \\
        $12$
          & 0.00002748 & 0.1075  & 0.05375
          & *0.001087  & -1.4425 & -0.7210
          & 0.001652   & *2.6744 & 1.3390
          & ?-5.9074
      \end{tabular}
      \bigskip \\
      \begin{tabular}{c|cccccccccc}
        n & ${\cal D}^n_+$ & ${\cal D}^n_-$ & ${\cal D}^n_{NS}$
          & ${\cal E}^n_+$ & ${\cal E}^n_-$ & ${\cal E}^n_{NS}$
          & ${\cal F}^n_+$ & ${\cal F}^n_-$ & ${\cal F}^n_{NS}$ & ${\cal G}^n$ \\
        \hline
        $2$
          & \dd{*60.5098??}{*60.5014??} & \dd{*??32.9286}{*??32.9251} & \dd{???63.1965}{???63.1909}
          & -10.5867? & ---      & -10.9168
          & 6.9729?   & -13.7973 & ?3.7817
          & -251.3619 \\
        $4$
          & \dd{*?7.9871??}{*?7.9873??} & \dd{*??25.9791}{*??25.9222} & \dd{???11.6147}{???11.5840}
          & ?-9.3990? & -23.9288 & -10.5807
          & 1.3106?   & *48.8620 & 20.6599
          & -204.5836 \\
        $6$
          & \dd{*?0.01877?}{*?0.01877?} & \dd{-4007.0415}{-4011.3304} & \dd{24025.6303}{24050.8845}
          & ?*1.8667? & -24.0991 & -12.4017
          & 0.4596?   & *56.4575 & 29.3881
          & -176.9466 \\
        $8$
          & \dd{*?0.03222?}{*?0.03221?} & \dd{*?165.7976}{*?165.9277} & \dd{???82.2116}{???82.2758}
          & ?*0.3993? & -29.0367 & -14.5993
          & 0.2217?   & *67.5380 & 34.0579
          & -157.4181 \\
        $10$
          & \dd{?-0.001732}{?-0.001738} & \dd{*?109.3285}{*?109.4090} & \dd{???54.5447}{???54.5847}
          & ?*0.1453? & -32.8877 & -16.4753
          & 0.1249?   & *73.0111 & 36.6197
          & -142.6108 \\
        $12$
          & \dd{?-0.01825?}{?-0.01826?} & \dd{*??86.8381}{*??86.9019} & \dd{???43.3780}{???43.4098}
          & ?*0.06532 & -35.8891 & -17.9598
          & 0.07764   & *75.9024 & 38.0024
          & -130.8717
      \end{tabular}
    \end{ruledtabular}
  \end{center}
\end{table}

\begin{table}
  \begin{center}
    \caption{%
      Numerical values of ${\cal L}^n_i, {\cal A}^n_i, {\cal B}^n_i, {\cal D}^n_i,
      {\cal E}^n_i, {\cal F}^n_i (i=+, -, NS)$, and ${\cal C}^n$ and  ${\cal G}^n$ 
      for $n=2, 4, \cdots, 12$ in the case of $n_f=4$.
      The calculation of ${\cal D}^n_+$,${\cal D}^n_-$ and ${\cal D}^n_{NS}$ 
 was performed by using
the ``exact" values of  $K_{NS}^{{(2),n}}$, $K_\psi^{{(2),n}}$ and
$ K_G^{{(2),n}}$ given in Ref.\cite{MVV} (at upper levels) and also by using the approximate 
expressions, $K_{NS\ \rm approx}^{{(2),n}}$, $K_{\psi\ \rm approx}^{{(2),n}}$ and $K_{G\ \rm
approx}^{{(2),n}}$  defined in Eqs.(\ref{EnsApprox})-(\ref{EgApprox}) (in parentheses at lower levels). }
    \label{ABCDEFGNF4}
    \makeatletter
    \ifnum \@pointsize=11 
      \footnotesize
    \else\ifnum \@pointsize=12 
      \scriptsize
    \fi\fi
    \makeatother
    \catcode`?=\active \def?{\phantom{0}} 
    \catcode`*=\active \def*{\phantom{-}} 
    \def\dd#1#2{$\genfrac{}{}{0pt}{}{\mbox{#1}}{\mbox{(#2)}}$} 
    \begin{ruledtabular}
      \begin{tabular}{c|cccccccccc}
        n & ${\cal L}^n_+$ & ${\cal L}^n_-$ & ${\cal L}^n_{NS}$
          & ${\cal A}^n_+$ & ${\cal A}^n_-$ & ${\cal A}^n_{NS}$
          & ${\cal B}^n_+$ & ${\cal B}^n_-$ & ${\cal B}^n_{NS}$ & ${\cal C}^n$ \\
        \hline
        $2$
          & 0.8078???? & 1.0582  & 0.6231?
          & 2.7608??   & ---     & -6.0944
          & 3.8774??   & -8.5894 & 1.3076
          & -16.3237 \\
        $4$
          & 0.009356?? & 0.7327  & 0.2661?
          & 5.1244??   & -3.7321 & -1.3858
          & 0.1688??   & *0.4820 & 2.1599
          & -18.7956 \\
        $6$
          & 0.001235?? & 0.4656  & 0.1679?
          & 0.09529?   & -2.9038 & -1.0480
          & 0.03909?   & *6.0485 & 2.2395
          & -15.9311 \\
        $8$
          & 0.0003465? & 0.3374  & 0.1215?
          & 0.01953?   & -2.7046 & -0.9735
          & 0.01495?   & *5.9573 & 2.1613
          & -13.5552 \\
        $10$
          & 0.0001362? & 0.2627  & 0.09461
          & 0.006354   & -2.5904 & -0.9322
          & 0.007216   & *5.6671 & 2.0468
          & -11.7384 \\
        $12$
          & 0.00006497 & 0.2140  & 0.07704
          & 0.002598   & -2.4906 & -0.8963
          & 0.003999   & *5.3501 & 1.9293
          & -10.3319
      \end{tabular}
      \bigskip \\
      \begin{tabular}{c|cccccccccc}
        n & ${\cal D}^n_+$ & ${\cal D}^n_-$ & ${\cal D}^n_{NS}$
          & ${\cal E}^n_+$ & ${\cal E}^n_-$ & ${\cal E}^n_{NS}$
          & ${\cal F}^n_+$ & ${\cal F}^n_-$ & ${\cal F}^n_{NS}$ & ${\cal G}^n$ \\
        \hline
        $2$
          & \dd{-84.4549?}{-84.4748?} & \dd{*?64.6182}{*?64.6102} & \dd{*?63.5804}{*?63.5722}
          & 13.2519 & ---       & -12.7900
          & 7.0067  & ?-68.4928 & ?0.8275
          & -439.6247 \\
        $4$
          & \dd{-17.3048?}{-17.3044?} & \dd{-140.7574}{-140.9078} & \dd{?-64.7231}{?-64.7847}
          & 92.4633 & ?-2.4550  & -11.2489
          & 2.6666  & ?-28.0786 & 24.5807
          & -357.8108 \\
        $6$
          & \dd{*?0.4163?}{*?0.4163?} & \dd{*894.6070}{*895.0955} & \dd{*301.7867}{*301.9492}
          & ?3.0154 & -37.7241  & -13.9828
          & 1.0159  & *?99.2476 & 37.1197
          & -309.4744 \\
        $8$
          & \dd{?-0.04747}{?-0.04749} & \dd{*326.5791}{*326.7480} & \dd{*116.7791}{*116.8392}
          & ?0.8428 & -47.7463  & -17.3117
          & 0.5046  & *121.0094 & 43.9510
          & -275.3197 \\
        $10$
          & \dd{?-0.2306?}{?-0.2306?} & \dd{*228.9242}{*229.0460} & \dd{*?82.2373}{*?82.2809}
          & ?0.3366 & -55.8735  & -20.1683
          & 0.2888  & *132.4677 & 47.7946
          & -249.4221 \\
        $12$
          & \dd{?-0.7548?}{?-0.7548?} & \dd{*183.5002}{*183.6024} & \dd{*?65.9952}{*?66.0319}
          & ?0.1600 & -62.2711  & -22.4456
          & 0.1813  & *139.2236 & 49.9533
          & -228.8909
      \end{tabular}
    \end{ruledtabular}
  \end{center}
\end{table}


\section{Sum rule of $F_2^\gamma(x,Q^2,P^2)$\label{SectionSecondMoment}}
The sum rule of the structure function $F_2^\gamma$,
\begin{equation}
\int_0^1dx F_2^\gamma(x,Q^2,P^2)~,
\end{equation}
can be studied by taking the  $n\to 2$ limit of Eq.(\ref{master1}). 
At $n\!=\!2$ one of the eigenvalues of ${\widehat \gamma}_{n=2}(g)$, 
the anomalous dimension matrix in the hadronic sector given in (\ref{HadronicAnomalousDim}),   
vanishes, due to the conservation of energy momentum tensor.
Thus we have a zero eigenvalue, $\lambda^{n=2}_- \!=\!0$, for 
the one-loop anomalous dimension matrix $ {\widehat \gamma}^{(0)}_{n=2}$ and, 
therefore, we get $d_-^{n=2}\!=\!\frac{\lambda^{n=2}_-}{2\beta_0}\!=\!0$. Among the 
coefficients which appeared in (\ref{master1}), two of them, namely, ${\cal A}_-^n$ and ${\cal E}_-^n$  
would develop singularities at $n\!=\!2$, since those coefficients have  terms with 
the factor $\frac{1}{d_-^n}$. However, as we see from (\ref{master1}), both ${\cal A}_-^n$ and ${\cal
E}_-^n$ are multiplied by  a factor
$\Bigl[1-\left(\frac{\alpha_s(Q^2)}{\alpha_s(P^2)}\right)^{d_-^n}\Bigr]$ which also vanishes at $n=2$.
Provided that we regard the expression
$\frac{1}{\epsilon}(1-x^\epsilon)$ as its limiting value for $\epsilon\to 0$, $-{\rm ln} x$~, then the
${\cal A}_-^n$ and ${\cal E}_-^n$ parts of (\ref{master1}) give  finite contribution as
\bea
\lim_{n\to 2}~{\cal A}_-^n
\left[1-\left(\frac{\alpha_s(Q^2)}{\alpha_s(P^2)}\right)^{d_-^n}\right] &=& -
{\overline {\cal A}}_-^{n=2}\ln \frac{\alpha_s(Q^2)}{\alpha_s(P^2)}~,\\
\lim_{n\to 2}~ {\cal E}_-^n\left[1-\left(\frac{\alpha_s(Q^2)}{\alpha_s(P^2)}\right)^{d_-^n}\right] &=& -
{\overline {\cal E}}_-^{n=2}\ln \frac{\alpha_s(Q^2)}{\alpha_s(P^2)}~,
\eea
where
\bea
{\overline {\cal A}}_-^{n=2}&=&\Bigl[-\bm{K}_n^{(0)} \sum_j \frac{P^n_j \hat \gamma_n^{(1)}
P^n_-} {\lambda^n_j +2\beta_0 } {\bm{C}}_{2,n}^{(0)}
            -\bm{K}_n^{(0)}P^n_- {\bm{C}}_{2,n}^{(0)}\frac{\beta_1}{\beta_0}
    + \bm{K}_n^{(1)} P^n_- {\bm{C}}_{2,n}^{(0)}~\Bigr]_{n=2},\\
&&\nonumber \\
{\overline{\cal E}}^{n=2}_-&=&\Bigl[-\bm{K}_n^{(0)}P^n_- {\bm{C}}_{2,n}^{(1)}
\frac{\beta_1}{\beta_0}
     -\bm{K}_n^{(0)} \sum_j\frac{P^n_j\hat \gamma_n^{(1)} P^n_-}
             {\lambda^n_j  +2\beta_0} {\bm{C}}_{2,n}^{(1)}
  +\bm{K}_n^{(1)}P^n_- {\bm{C}}_{2,n}^{(1)} \nonumber\\
&&
-\bm{K}_n^{(0)} \sum_j\frac{P^n_-\hat \gamma_n^{(1)} P^n_j}
             { - \lambda^n_j +2\beta_0} {\bm{C}}_{2,n}^{(0)}
    \frac{\beta_1}{\beta_0} -\bm{K}_n^{(0)} \sum_{j, k}
\frac{P^n_j\hat \gamma_n^{(1)} P^n_-\hat \gamma_n^{(1)} P^n_k}
             {( - \lambda^n_k +2\beta_0)(\lambda^n_j 
+2\beta_0)} {\bm{C}}_{2,n}^{(0)}
   \nonumber\\
&&
+\bm{K}_n^{(1)} \sum_j\frac{P^n_-\hat \gamma_n^{(1)} P^n_j}
             { - \lambda^n_j +2\beta_0} {\bm{C}}_{2,n}^{(0)}\Bigr]_{n=2}~.
\eea
The coefficient functions, anomalous dimensions and photon matrix elements at $n=2$ 
are given in Appendix D. Using these values we obtain ${\overline {\cal A}}_-^{n=2}\!=\!-1.3274~
(-2.2857)$  and ${\overline{\cal E}}^{n=2}_-\!=\! 5.7664~ (18.553)$ for $n_f\!=\!3~ (4)$.
The numerical values of 
${\cal L}^{n=2}_i$, ${\cal A}^{n=2}_i$, ${\cal B}^{n=2}_i$, ${\cal D}^{n=2}_i$,
${\cal E}^{n=2}_i$, ${\cal F}^{n=2}_i$ $(i=+, -, NS)$ and ${\cal C}^{n=2}$ and  ${\cal G}^{n=2}$, 
except for ${\cal A}_-^{n=2}$ and ${\cal E}_-^{n\!=\!2}$, were already given in Table \ref{ABCDEFGNF3} 
(for $n_f\!=\!3$) and \ref{ABCDEFGNF4} (for $n_f\!=\!4$).

Let us express the sum rule  in the following form as 
\bea
\int_0^1dx F_2^\gamma(x,Q^2,P^2)=\frac{\alpha}{4\pi}\frac{1}{2\beta_0}
\Bigl\{ \frac{4\pi}{\alpha_s(Q^2)} c_{LO} +c_{NLO}  +\frac{\alpha_s(Q^2)}{4\pi}c_{NNLO}
+{\cal O}(\alpha_s^2)
\Bigr\}~, \label{SecondMoment}
\eea
where the first, second and third terms in the curly brackets correspond 
to the LO, NLO and NNLO contributions, respectively. 
The coefficients $c_{LO}$, $c_{NLO}$ and $c_{NNLO}$ depend on the number of the active quark flavors 
$n_f$, and also on $\alpha_s(Q^2)$ and $\alpha_s(P^2)$. 
For the QCD running coupling constant  $\alpha_s(Q^2)$, we use the 
following formula which takes into account the $\beta$ function parameters 
up to the three-loop level~\cite{ParticleData}, 
\begin{equation}
    \frac{\alpha_s(Q^2)}{4\pi}
    = \frac{1}{\beta_0 L}
      - \frac{1}{(\beta_0 L)^2} \frac{\beta_1}{\beta_0} \ln L
      + \frac{1}{(\beta_0 L)^3} \left(\frac{\beta_1}{\beta_0}\right)^2
            \left[ \left(\ln L - \frac{1}{2} \right)^2
            + \frac{\beta_0 \beta_2}{\beta_1^2} - \frac{5}{4} \right]
      + {\cal O} \left(\frac{1}{L^4}\right) , \label{runningAlpha}
\end{equation}
where $L\!=\!\ln(Q^2/\Lambda^2)$, and $\beta_0$, $\beta_1$ and $\beta_2$ are given in 
Eqs.(\ref{beta0})-(\ref{beta2}). Taking $\Lambda\!=\!0.2$ GeV, we get, for example, 
$\alpha_s(Q^2\!=\!100{\rm GeV}^2)\!=\!0.1461~(0.1595)$ and 
$\alpha_s(Q^2\!=\!3{\rm GeV}^2)\!=\!0.2487~(0.2717)$ for the case $n_f\!=\!3$ (4).

We list in Table \ref{NLONNLOvsLO} the numerical values of the  coefficients $c_{LO}$, 
$c_{NLO}$ and $c_{NNLO}$ for the cases $n_f\!=\!3$ and 4. We have studied three cases:
$(Q^2, P^2)=$(30${\rm GeV}^2$, 1${\rm GeV}^2$), (100${\rm GeV}^2$, 1${\rm GeV}^2$) and 
(100${\rm GeV}^2$, 3${\rm GeV}^2$).
We already know that  $c_{NLO}$ takes 
negative values~\cite{UW2}. We find that the coefficient $c_{NNLO}$ also 
takes negative values which are rather large in magnitude compared with 
those of $c_{LO}$ and $c_{NLO}$. Also listed in Table \ref{NLONNLOvsLO} 
are the NLO ($\alpha\alpha_s$) and NNLO ($\alpha\alpha_s^2$) corrections relative to LO 
($\alpha$)  and the ratios of the NNLO to the sum of the LO and NLO contributions
for the sum rule of $F_2^\gamma(x,Q^2,P^2)$. We see that the NNLO corrections 
give negative contribution to the sum rule. In fact, we will see in the next section that 
the NNLO corrections reduce $F_2^\gamma(x,Q^2,P^2)$ at larger $x$. 
For the kinematical region of $Q^2$ and $P^2$ which we have studied, 
the NNLO corrections are found to be rather large. When 
 $P^2\!=\!1 {\rm GeV}^2$ and  $Q^2\!=\!30\sim 100{\rm GeV}^2$ or 
$P^2\!=\!3 {\rm GeV}^2$ and  $Q^2\!=\!100{\rm GeV}^2$,  and $n_f$ is  three or four,  
the NNLO corrections are  $7\% \sim10\%$ of the sum of the LO and NLO contributions.

\begin{table}
      \caption{ The numerical values of coefficients $c_{LO}$, $c_{NLO}$ and $c_{NNLO}$ in
                Eq.(\ref{SecondMoment}), and 
                the NLO and NNLO corrections relative to LO for the sum rule of $F_2^\gamma(x,Q^2,P^2)$
                in several cases of $Q^2$ and $P^2$.
                The ratios of the NNLO to the sum of
                the LO and NLO contributions are also listed.
                 For the QCD running coupling constant $\alpha_s$, we have used the 
                formula given in Eq.(\ref{runningAlpha}) with $\Lambda\!=\!0.2$ GeV.}
 \begin{center}
      \small
      \catcode`?=\active \def?{\phantom{0}}
      \begin{ruledtabular}
      \begin{tabular}{c|c c c c c | c c c c }
         & $Q^2({\rm GeV}^2)$ & $P^2({\rm GeV}^2)$
          & $c_{LO}$ & $c_{NLO}$ & $c_{NNLO}$
          & LO & NLO & NNLO & NNLO/(LO+NLO)  \\
        \hline
        $n_f=3$
          & 30
          & 1
          & 0.7631
          &  -11.66 
          & -331.2
          &  1
          &  -0.2063
          &  -0.0791
          &  -0.0997 \\
          & 100
          & 1
          & 0.8613
          &  -12.21 
          &  -355.3
          &  1
          &  -0.1649
          &  -0.0558
          &  -0.0668\\
           & 100
          & 3
          & 0.6690
          &  -11.22 
          &  -313.8
          &  1
          &  -0.1949
          &  -0.0634
          &  -0.0787 \\
        $n_f=4$
          & 30
          & 1
          & 1.429
          &  -18.90 
          &  -525.7
          &  1
          &  -0.1950
          &  -0.0800
          &  -0.0993\\
           & 100
          & 1
          & 1.614
          &  -19.59 
          &  -551.4
          &  1
          &  -0.1541
          &  -0.0551
          &  -0.0651 \\
          & 100
          & 3
          & 1.257
          &  -18.38 
          &  -507.5
          &  1
          &  -0.1855
          &  -0.0650
          &  -0.0798 \\
      \end{tabular}
      \end{ruledtabular}
      \label{NLONNLOvsLO}
    \end{center}
  \end{table}

\section{Numerical analysis of $F_2^\gamma(x,Q^2,P^2)$ \label{NumericalAnalysis}}

 We now perform the inverse Mellin transform of 
 (\ref{master1}) to obtain $F_2^\gamma$ as a function of $x$.
 The $n$-th moment is denoted as
 \begin{equation}
 M_2^\gamma(n,Q^2,P^2)=\int_0^1 dx~x^{n-1}\frac{F_2^\gamma(x,Q^2,P^2)}{x}~.
 \label{mom}
 \end{equation}
 Then by inverting the moments (\ref{mom}) we get
 \begin{equation}
 \frac{F_2^\gamma(x,Q^2,P^2)}{x}=\frac{1}{2\pi i}
 \int_{C-i\infty}^{C+i\infty}dn~ x^{-n} M_2^\gamma(n,Q^2,P^2)~,
 \end{equation}
where the integration contour runs to the right of all singularities
of $ M_2^\gamma(n,Q^2,P^2)$ in the complex $n$-plane. 
 In order to have better convergence of the numerical integration,
we change the contour  from the vertical line connecting 
$C-i\infty$ with $C+i\infty$ ($C$ is an appropriate positive constant),
 introducing a small positive constant $\varepsilon$, to
  \begin{equation}
 n=C-\varepsilon|y|+iy, \qquad\qquad -\infty <y< \infty~.
 \end{equation}
Hence we have
 \bea
\frac{F_2^\gamma(x,Q^2,P^2)}{x}
 &=&\frac{1}{\pi}\int_0^\infty
 \left[\mbox{Re}\{ M_2^\gamma(z,Q^2,P^2)e^{-z\ln(x)}
 \} \right.\nonumber\\
 &&\qquad\qquad-\left.\varepsilon\mbox{Im}\{ M_2^\gamma(z,Q^2,P^2)
 e^{-z\ln(x)}\}\right]dy~, \label{invmom}
 \eea
where $z=C-\varepsilon y+iy$.

As we see from Eqs.(\ref{master1})-(\ref{CoeffiG}), the $n$-th moment $M_2^\gamma(n,Q^2,P^2)$ 
is  written in terms of coefficient functions, anomalous dimensions and photon matrix elements, which 
in turn are  expressed by the rational functions of  integer $n$ and also by the various
harmonic  sums~\cite{Vermaseren1}. Thus we need to make an analytic continuation 
of these harmonic sums from integer $n$ to complex $n$. 
There are several proposals for this continuation~\cite{Blumlein1,Kotikov1}. 
The method we adopted here is to use the asymptotic  expansions of the harmonic sums and 
their  translation relations. The details are explained in Appendix \ref{HarmonicSum}.

\begin{figure}
\begin{center}
\includegraphics[scale=0.7]{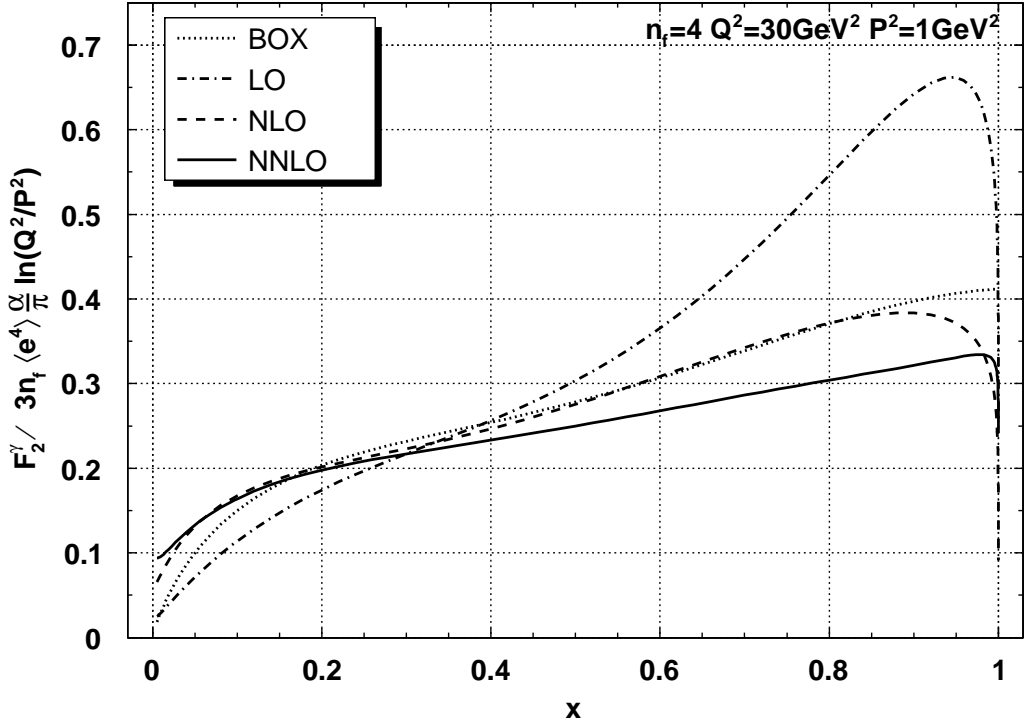}
\vspace{-0.5cm}
\caption{\label{F2_4_30_1_200} Virtual photon structure function $F_2^\gamma(x,Q^2,P^2)$ in units of 
$(3\alpha n_f\langle e^4\rangle/\pi)\ln (Q^2/P^2)$ for $Q^2\!=\!30$ GeV$^2$ and $P^2\!=\!1$ GeV$^2$ with 
 $n_f\!=\!4$ and the QCD scale parameter $\Lambda\!=\!0.2$ GeV. We plot 
the Box (tree) diagram contribution including non-leading corrections (short-dashed line), the QCD leading
order (LO) (dash-dotted line),  the next-to-leading order (NLO) (long-dashed line) and the
next-to-next-to-leading order (NNLO) (solid line) results.}
\end{center}
\end{figure}
In Fig.\ref{F2_4_30_1_200} we  plot the virtual photon structure function  $F_2^\gamma(x,Q^2,P^2)$
predicted by pQCD for the case of $n_f\!=\!4$, $Q^2\!=\!30$ GeV$^2$ and $P^2\!=\!1$ GeV$^2$ with the QCD
scale parameter
$\Lambda\!=\!0.2$ GeV. The vertical axis corresponds to
 \begin{equation}
F_2^\gamma(x,Q^2,P^2)/\frac{3\alpha}{\pi}n_f\langle e^4 \rangle\ln\frac{Q^2}{P^2}~.
\label{normalized}
\end{equation}
Here we show four curves; the LO, NLO  and NNLO QCD results and
the Box (tree) diagram contribution including non-leading corrections. 
The box contribution is expressed by~\cite{UW2}
\bea
F_2^{\gamma({\rm Box})}(x,Q^2,P^2)&=&\frac{3\alpha}{\pi}n_f\langle e^4 \rangle
\biggl\{x\Bigl[x^2+(1-x)^2\Bigr]\ln\frac{Q^2}{P^2}\nonumber \\
&&\qquad \qquad -2x\Bigl[1-3x+3x^2+(1-2x+2x^2)\ln x\Bigr]\biggr\}~,
\eea
where power corrections $P^2/Q^2$ and  quark mass effects are ignored. 
It is noted that, in these analyses, even for the LO and NLO  QCD curves,
 we have used the QCD running coupling constant
$\alpha_s(Q^2)$ which is  valid up to the three-loop level and is governed by 
the formula (\ref{runningAlpha}) and we have put  $\Lambda\!=0.2$GeV. 

The LO and NLO QCD results with the same values of $n_f$, $Q^2$ and $P^2$
as well as the Box contribution   
were already given in Fig.6 of Ref.\cite{UW2}. But in Ref.\cite{UW2} 
the formula for $\alpha_s(Q^2)$ which is valid in the one-loop level was used 
to obtain the LO curve,  while the two-loop level formula for $\alpha_s(Q^2)$ was applied for the NLO
graph,  and the QCD scale  parameter $\Lambda$ was set to be 0.1 GeV in both cases.
The LO result in Fig.\ref{F2_4_30_1_200} has a similar shape as the corresponding one 
in Ref.\cite{UW2} but is different in magnitude; the former is slightly larger than the latter for 
almost the whole $x$ region.  
This is  due to the fact that the one-loop-level formula for $\alpha_s(Q^2)$  
was used for the LO curve in Ref.\cite{UW2} while we applied the three-loop-level 
formula even for the LO result. 
On the other hand, the NLO curve in Fig.\ref{F2_4_30_1_200} is similar to the corresponding one
in  Ref.\cite{UW2} in shape and magnitude. 

Now we observe in Fig.\ref{F2_4_30_1_200} that there exist  notable NNLO QCD corrections at larger $x$.
The corrections are negative and the NNLO curve comes below the NLO one 
in the region  $0.3\lesssim x<1$. This is expected from the $n\!=\!2$ moment analysis 
in Sec.\ref{SectionSecondMoment}. From Table \ref{NLONNLOvsLO} we see that 
the ratio of the NNLO to the sum of the LO and NLO contributions
for the sum rule of $F_2^\gamma(x,Q^2,P^2)$ is~ $-0.099$~ for the case of 
$n_f\!=\!4$, $Q^2\!=\!30$ GeV$^2$ and $P^2\!=\!1$ GeV$^2$. 
At lower $x$ region, $0.05\lesssim x \lesssim 0.3$, the NNLO corrections to the NLO results are found to 
be negligibly small.

We have also studied the QCD corrections to $F_2^\gamma(x,Q^2,P^2)$
with different $Q^2$ and $P^2$ but with $n_f\!=\!4$. 
\begin{figure}
\begin{center}
\includegraphics[scale=0.7]{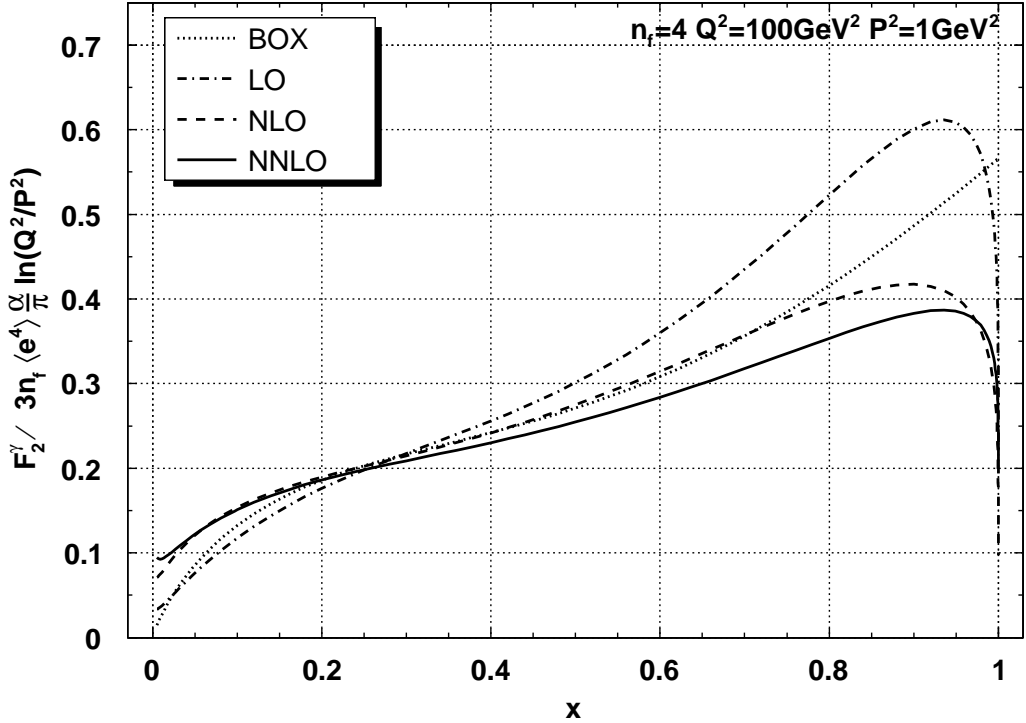}
\vspace{-0.5cm}
\caption{\label{F2_4_100_1_200} Virtual photon structure function $F_2^\gamma(x,Q^2,P^2)$ 
for $Q^2\!=\!100$ GeV$^2$ and $P^2\!=\!1$ GeV$^2$ with 
 $n_f\!=\!4$ and $\Lambda\!=\!0.2$ GeV. }
\end{center}
\end{figure}
\begin{figure}
\begin{center}
\includegraphics[scale=0.7]{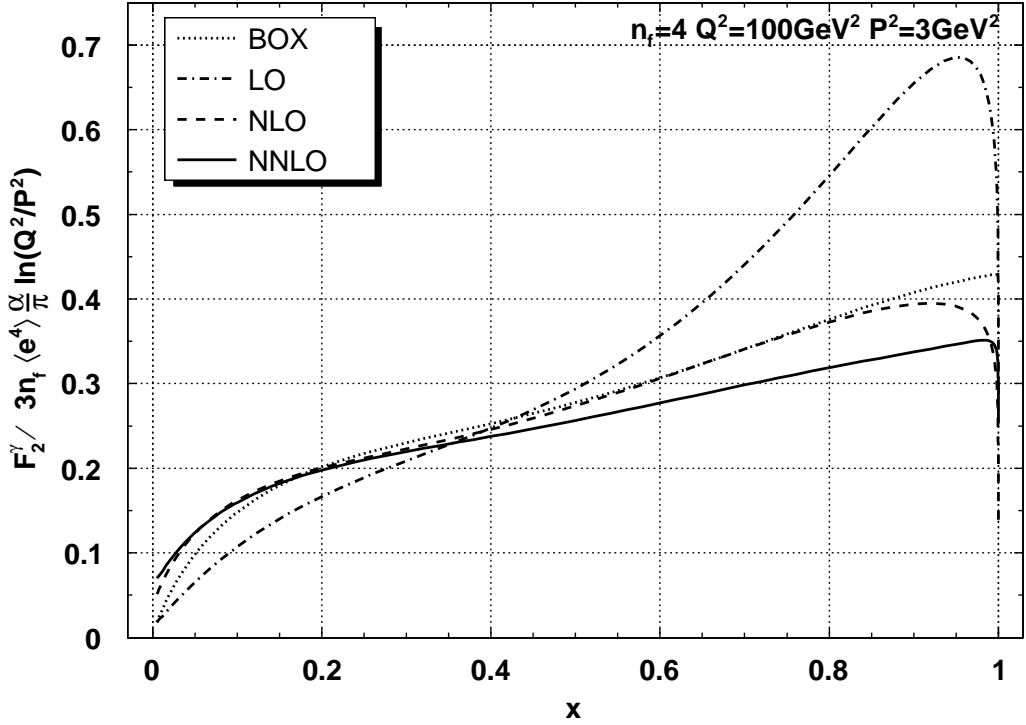}
\vspace{-0.5cm}
\caption{\label{F2_4_100_3_200} Virtual photon structure function $F_2^\gamma(x,Q^2,P^2)$ 
for $Q^2\!=\!100$ GeV$^2$ and $P^2\!=\!3$ GeV$^2$ with 
 $n_f\!=\!4$ and $\Lambda\!=\!0.2$ GeV. }
\end{center}
\end{figure}
In Fig.\ref{F2_4_100_1_200} we plot the case for $Q^2\!=\!100$ GeV$^2$ and $P^2\!=\!1$ GeV$^2$.
Another  case for $Q^2\!=\!100$ GeV$^2$ and $P^2\!=\!3$ GeV$^2$ is shown in Fig.\ref{F2_4_100_3_200}.
We have not seen any sizable change for the normalized structure
function (\ref{normalized}) for these different values of $Q^2$ and $P^2$. 
In both cases the NNLO corrections reduce $F_2^\gamma(x,Q^2,P^2)$ at larger $x$.
We have examined the $n_f=3$ case as well. It is observed that the normalized
structure function (\ref{normalized}) is insensitive to the number of active flavors. 


Finally we should 
note that the spin-averaged structure function directly accessible
in the experiment is not $F_2^\gamma$, but the so-called 
effective structure function
$F_{\rm eff}^\gamma \simeq F_2^\gamma+(3/2)F_L^\gamma$ as discussed in 
ref. \cite{Krawczyk,GRStratmann,GRSchienbein}


\section{Longitudinal structure function $F_L^\gamma(x,Q^2,P^2)$\label{Longitudinal}}

We have considered the structure function  $F_2^\gamma$ so far. 
Regarding another structure function $F_L^\gamma$, its LO 
 contribution, which is of order $\alpha$, was calculated in QCD for the real photon ($P^2\!=\!0$) target  
in Refs.\cite{Witten,BB}. The analysis was extended  to the case of the 
virtual photon ($\Lambda^2\ll P^2 \ll Q^2$) target~\cite{UW2}.
We will now derive a formula for the moment sum rule of $F_L^\gamma(x,Q^2,P^2)$ 
up to the NLO (${\cal O}(\alpha \alpha_s$)) corrections. 
Comparing Eq.(\ref{FLmoment}) with (\ref{F2moment}) and examining the form of (\ref{RGequation}), 
we see that the formula  for $F_L^\gamma(x,Q^2,P^2)$ is obtained from
(\ref{F2momb}) only by replacing $\bm{C}_{2,n}(1,{\bar g}(Q^2)) $ and 
$C_{2,n}^\gamma(1,{\bar g}(Q^2),\alpha) $ with  $\bm{C}_{L,n}(1,{\bar g}(Q^2))$ and
$C_{L,n}^\gamma(1,{\bar g}(Q^2),\alpha) $, respectively.   
An expansion is made for $\bm{C}_{L,n}(1,{\bar g}(Q^2))$ and $C_{L,n}^\gamma(1,{\bar g}(Q^2),\alpha) $
up to the two-loop level as
\bea
{\bm C}_{L,n}(1,{\bar g}(Q^2))&=&{\bm C}_{L,n}^{(0)}+\frac{{\bar
g}^2(Q^2)}{16\pi^2}{\bm C}_{L,n}^{(1)}+
\frac{{\bar g}^4(Q^2)}{(16\pi^2)^2}{\bm C}_{L,n}^{(2)}+\cdots~, \label{ExpandCLhadronic}\\
C_{L,n}^\gamma(1,{\bar g}(Q^2),\alpha)&=&\frac{e^2}{16\pi^2}C_{L,n}^{\gamma(1)}+\frac{e^2{\bar
g}^2(Q^2)}{(16\pi^2)^2}C_{L,n}^{\gamma(2)}+\cdots~.  \label{ExpandCLgamma}
\eea
Here we note that there is no contribution of the tree diagrams to the longitudinal 
coefficient functions and thus we have ${\bm C}_{L,n}^{(0)}={\bm 0}$. 

The moments of 
$F_L^\gamma(x,Q^2,P^2)$ are then given as follows (see Eqs.(\ref{master1})-(\ref{CoeffiG}) for
comparison):

\bea
&&\hspace{-1.5cm}\int_0^1 dx x^{n-2}F_L^\gamma(x,Q^2,P^2)  \nonumber\\
&=&\frac{\alpha}{4\pi}\frac{1}{2\beta_0}
\Biggl\{\sum_{i}{\cal
B}_{(L),i}^n\left[1-\left(\frac{\alpha_s(Q^2)}{\alpha_s(P^2)}\right)^{d_i^n+1}\right]
+{\cal C}_{(L)}^n \nonumber\\
&&\qquad+\frac{\alpha_s(Q^2)}{4\pi}\biggl(\sum_{i}{\cal
E}_{(L),i}^n\left[1-\left(\frac{\alpha_s(Q^2)}{\alpha_s(P^2)}\right)^{d_i^n}\right]
+\sum_{i}{\cal
F}_{(L),i}^n\left[1-\left(\frac{\alpha_s(Q^2)}{\alpha_s(P^2)}\right)^{d_i^n+1}\right]
\nonumber\\
&&\hspace{3.5cm}+{\cal G}_{(L)}^n \biggr) +{\cal O}(\alpha_s^2) 
 \Biggr\}~,\hspace{2cm} {\rm with}\quad i=+, -, NS~,
\label{masterL}
\eea
where 
the coefficients  ${\cal B}^n_{(L),i}$, ${\cal C}^n_{(L)}$, 
${\cal E}^n_{(L),i}$, ${\cal F}^n_{(L),i}$ and ${\cal G}^n_{(L)}$ are 
\bea
{\cal B}^n_{(L),i}&=& \bm{K}_n^{(0)} P^n_i{\bm{C}}_{L,n}^{(1)}
    \frac{1}{1+d_i^n}~,  \\
{\cal C}^n_{(L)}&=&2\beta_0 C_{L,n}^{\gamma (1)}~, \\
{\cal E}^n_{(L),i}&=&-\bm{K}_n^{(0)}P^n_i {\bm{C}}_{L,n}^{(1)}
\frac{\beta_1}{\beta_0}\frac{1-d_i^n}{d_i^n}
     -\bm{K}_n^{(0)} \sum_j\frac{P^n_j\hat \gamma_n^{(1)} P^n_i}
             {\lambda^n_j - \lambda^n_i +2\beta_0} {\bm{C}}_{L,n}^{(1)}
   \frac{1}{d_i^n} \nonumber\\
&&+\bm{K}_n^{(1)}P^n_i {\bm{C}}_{L,n}^{(1)}\frac{1}{d_i^n}
 -2\beta_0{\widetilde{\bm{A}}}_n^{(1)}P^n_i
{\bm{C}}_{L,n}^{(1)}~,\\
{\cal F}^n_{(L),i}&=&\bm{K}_n^{(0)}P^n_i {\bm{C}}_{L,n}^{(2)}
\frac{1}{1+d_i^n}-\bm{K}_n^{(0)}P^n_i {\bm{C}}_{L,n}^{(1)}
\frac{\beta_1}{\beta_0}\frac{d_i^n}{1+d_i^n}\nonumber\\
&&    +\bm{K}_n^{(0)} \sum_j\frac{P^n_i\hat \gamma_n^{(1)} P^n_j}
             {\lambda^n_i - \lambda^n_j +2\beta_0} {\bm{C}}_{L,n}^{(1)}
   \frac{1}{1+d_i^n}~, \\
{\cal G}^n_{(L)}&=&2\beta_0 (C_{L,n}^{\gamma (2)} +
         {\widetilde{\bm{A}}}_n^{(1)} \cdot{\bm{C}}_{L,n}^{(1)} )~,
\eea
with $i,j=+, -, NS$. The coefficients ${\cal B}^n_{(L),i}$ and ${\cal C}^n_{(L)}$ 
represent the LO terms~\cite{Witten,BB,UW2}, while the terms with 
${\cal E}^n_{(L),i}$, ${\cal F}^n_{(L),i}$ and ${\cal G}^n_{(L)}$ are the NLO 
($\alpha\alpha_s$) corrections and new. 
It is noted that, among these coefficients, ${\cal E}^n_{(L),-}$ becomes singular at $n\!=\!2$ 
since it has terms with the factor 
$\frac{1}{d_-^n}$~and $d_-^{n}$ vanishes as $n\!\rightarrow\!2$. But again, as in the
case of the moments of $F_2^\gamma(x,Q^2,P^2)$, this coefficient is 
multiplied by a factor $\Bigl[1-\Bigl(\alpha_s(Q^2)/\alpha_s(P^2)\Bigr)^{d_-^{n}}\Bigr]$, 
and thus the product remain finite at $n\!=\!2$.

The one-loop longitudinal coefficient functions were well known~\cite{ZWT,HL,BBDM}.
They are written in the form as
\begin{equation}
C_{L,n}^{\psi(1)}=\delta_\psi B^n_{\psi,L}, \quad C_{L,n}^{G(1)}=\delta_\psi B^n_{G,L}~, \quad
C_{L,n}^{NS(1)}=\delta_{NS}B^n_{NS,L} ~, \quad C_{L,n}^{\gamma(1)}=\delta_{\gamma}B^n_{\gamma,L}~,
\label{OneLoopCoefficientL}
\end{equation}
where $B^n_{\psi,L}=B^n_{NS,L}$, $B^n_{G,L}$ and $B^n_{\gamma,L}$ are given, 
for example, in Eqs.(6.2) - (6.4) of Ref.\cite{BB}.
The two-loop longitudinal coefficient functions corresponding to the hadronic operators 
were calculated in the $\overline{\rm MS}$ scheme in Ref.\cite{vNZ,ZvN1}
\footnote{The earlier calculations~\cite{EarlierCL,KazakovKotikov,KKPSS,SanchezGuillen} were found
partly incorrect. For  quark coefficient functions $c_{L,q}^{(2),{\rm ns}}$ and $c_{L,q}^{(2),{\rm
ps}}$ in Eq.(\ref{QuarkCoeffiL}), there is a
complete agreement between  Ref.~\cite{SanchezGuillen} and ~\cite{vNZ,ZvN1,LarinVermaseren} while 
for gluon coefficient $c_{L,g}^{(2)}$  in Eq.(\ref{GluonCoeffiL})  the result of 
Ref.\cite{KazakovKotikov} was corrected in Ref.\cite{KazakovKotikov2}.}. 
The results in Mellin space 
as functions of $n$ are found, for example, in Ref.~\cite{MochVermaseren}:
\bea
C_{L,n}^{\psi(2)}&=&\delta_\psi~ \Bigl\{c_{L,q}^{(2),{\rm ns}}(n) + c_{L,q}^{(2),{\rm ps}}(n)
\Bigr\}, \label{QuarkCoeffiL}\\
C_{L,n}^{G(2)}&=&\delta_\psi~ c_{L,g}^{(2)}(n), \label{GluonCoeffiL}\\
\quad C_{L,n}^{NS(2)}&=&\delta_{NS}~c_{L,q}^{(2),{\rm ns}}(n)~. ~
\eea
where $c_{L,q}^{(2),{\rm ns}}(n)$, $c_{L,q}^{(2),{\rm
ps}}(n)$ and $c_{L,g}^{(2)}(n)$ are given in Eqs.(203), (204), and (205)  
in Appendix B of Ref.\cite{MochVermaseren}, respectively,  with $N$ being replaced by $n$.
The two-loop photon longitudinal coefficient function
$C_{L,n}^{\gamma(2)}$ is expressed as 
\begin{equation}
C_{L,n}^{\gamma(2)}=\delta_\gamma c_{L,\gamma}^{(2)}(n)~,
\end{equation}
and $c_{L,\gamma}^{(2)}(n)$ is obtained from $c_{L,g}^{(2)}(n)$ in (\ref{GluonCoeffiL}) by replacing 
$C_A \rightarrow 0$ and $\frac{n_f}{2} \rightarrow 1$.

\begin{figure}
\begin{center}
\includegraphics[scale=0.7]{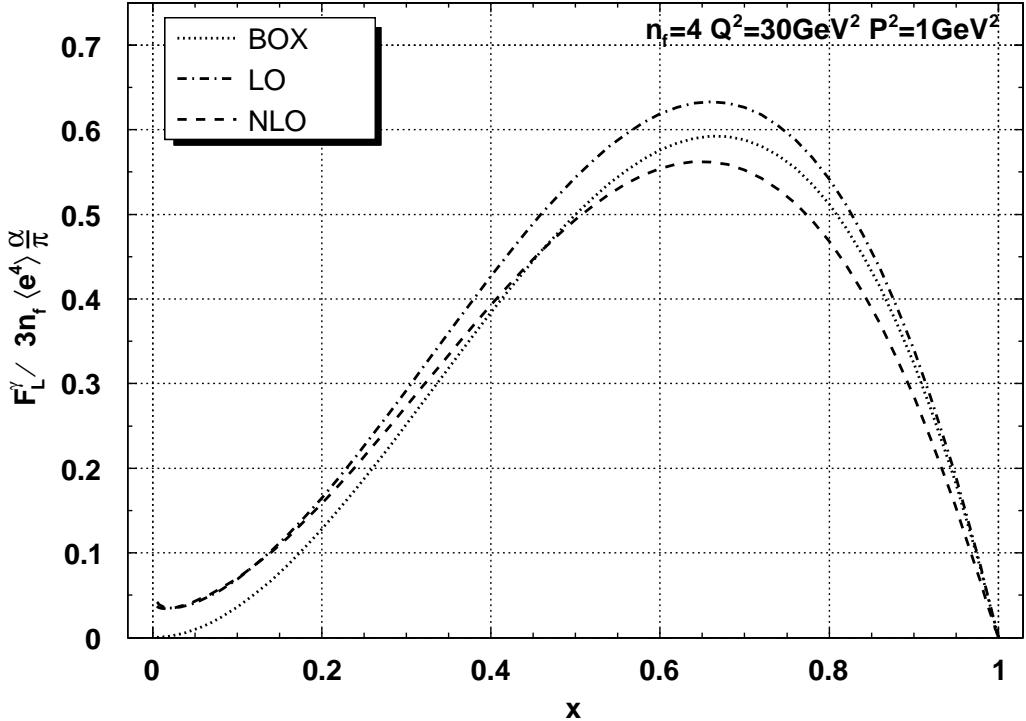}
\vspace{-0.5cm}
\caption{\label{FL_4_30_1_200} Longitudinal photon structure function $F_L^\gamma(x,Q^2,P^2)$ in units of 
$(3\alpha n_f\langle e^4\rangle/\pi)$ for $Q^2\!=\!30$ GeV$^2$ and $P^2\!=\!1$ GeV$^2$ with 
 $n_f\!=\!4$ and the QCD scale parameter $\Lambda\!=\!0.2$ GeV. We plot 
the Box (tree) diagram contribution (short-dashed line), the QCD leading
order (LO) (dash-dotted line) and  the next-to-leading order (NLO) (long-dashed line) results.}
\end{center}
\end{figure}

Inverting the moments (\ref{masterL}), we plot in Fig.\ref{FL_4_30_1_200} 
the longitudinal virtual photon structure function  $F_L^\gamma(x,Q^2,P^2)$
predicted by pQCD for the case of $n_f\!=\!4$, $Q^2\!=\!30$ GeV$^2$ and $P^2\!=\!1$ GeV$^2$ with the QCD
scale parameter
$\Lambda\!=\!0.2$ GeV. The vertical axis is in units of
$F_L^\gamma(x,Q^2,P^2)/\frac{3\alpha}{\pi}n_f\langle e^4 \rangle$.
Here we show three curves; the LO and NLO QCD results and
the Box (tree) diagram contribution, which is expressed by
\begin{equation}
F_L^{\gamma({\rm Box})}(x,Q^2,P^2)=\frac{3\alpha}{\pi}n_f\langle e^4 \rangle
\Bigl\{4x^2(1-x)\Bigr\}~.
\end{equation}
The LO result in Fig.\ref{FL_4_30_1_200} is consistent with the corresponding one 
in Fig.5 of Ref.\cite{UW2}, although the formulae used for $\alpha_s(Q^2)$ differ in detail. 
We see from Fig.\ref{FL_4_30_1_200} that the NLO QCD corrections are negative and the NLO curve comes below
the LO one  in the region  $0.2\lesssim x<1$.

\begin{figure}
\begin{center}
\includegraphics[scale=0.7]{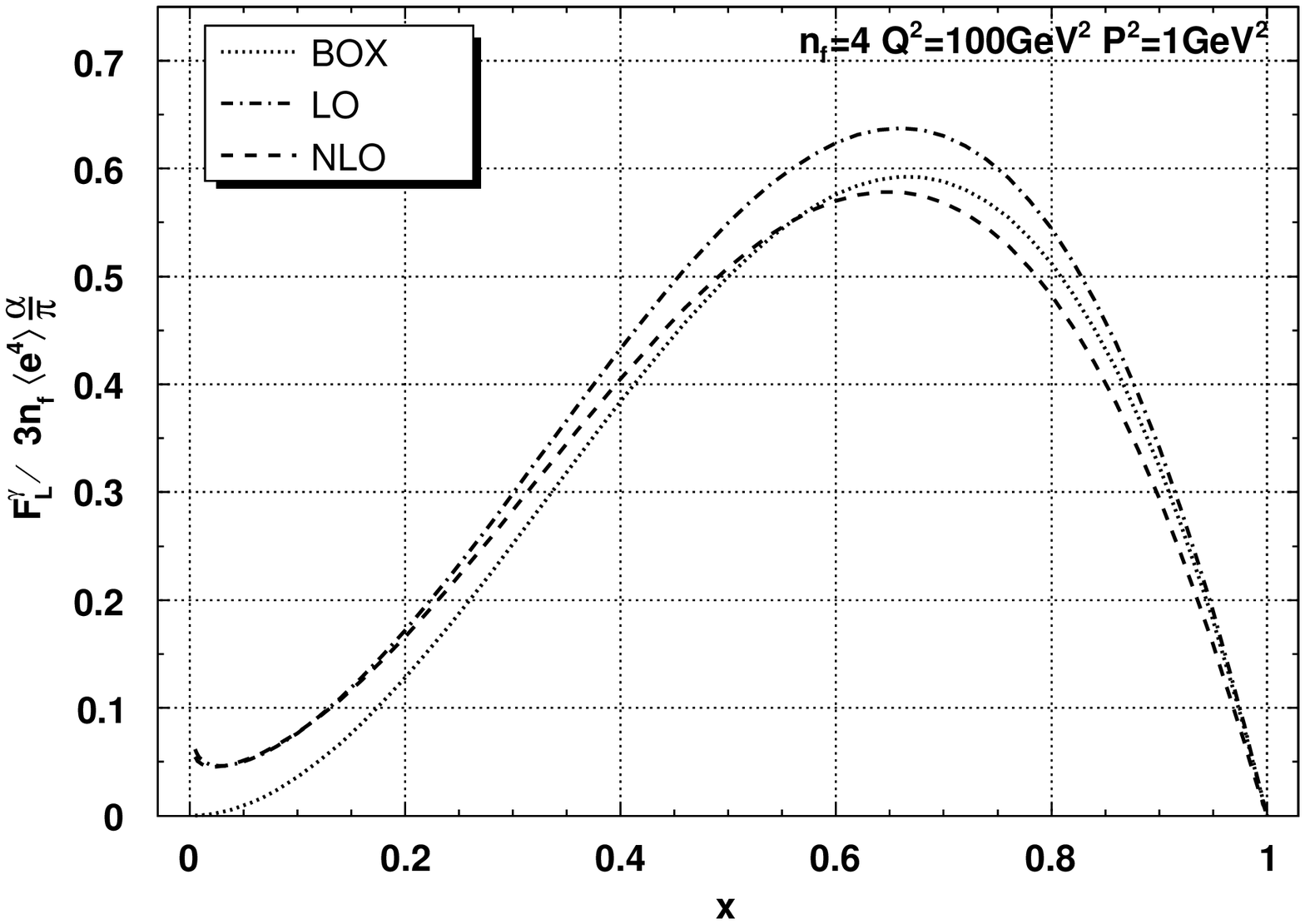}
\vspace{-0.5cm}
\caption{\label{FL_4_100_1_200} Longitudinal photon structure function $F_L^\gamma(x,Q^2,P^2)$ for
$Q^2\!=\!100$ GeV$^2$ and $P^2\!=\!1$ GeV$^2$ with 
 $n_f\!=\!4$ and $\Lambda\!=\!0.2$ GeV.}
\end{center}
\end{figure}
The QCD corrections to $F_L^\gamma(x,Q^2,P^2)$
for different values of $Q^2$, $P^2$ and $n_f$ are also studied. 
The case for  $Q^2\!=\!100$ GeV$^2$ and $P^2\!=\!1$ GeV$^2$ with $n_f\!=\!4$ is shown 
in Fig.\ref{FL_4_100_1_200}. The LO curve has hardly changed from the one for 
$Q^2\!=\!30$ GeV$^2$ and $P^2\!=\!1$ GeV$^2$. The NLO corrections get smaller. 
The LO and NLO QCD curves for $Q^2\!=\!100$ GeV$^2$ and $P^2\!=\!3$ GeV$^2$ with $n_f\!=\!4$
appear to be almost the same as those in the case of $Q^2\!=\!30$ GeV$^2$ and $P^2\!=\!1$.
The cases for $n_f\!=\!3$ are examined as well and 
we find that the normalized
function $F_L^\gamma(x,Q^2,P^2)/\frac{3\alpha}{\pi}n_f\langle e^4 \rangle$ is insensitive to the number of
active flavors.

\section{Conclusions}

We have investigated the unpolarized virtual photon structure
functions $F_2^\gamma(x,Q^2,P^2)$ and $F_L^\gamma(x,Q^2,P^2)$ 
for the kinematical region $\Lambda^2 \ll P^2 \ll Q^2    $ in QCD.
In the framework of the OPE supplemented by the RG method, 
we gave the definite predictions for the moments of $F_2^\gamma(x,Q^2,P^2)$
up to the NNLO (the order $\alpha\alpha_s$) and for the 
moments of $F_L^\gamma(x,Q^2,P^2)$  up to the NLO  (the order $\alpha\alpha_s$).
In the course of  our evaluation,
we utilized
the recently calculated results of
 the three-loop anomalous dimensions for the quark and gluon 
operators. Also we derived the photon matrix elements of hadronic operators  
up to the two-loop level. 

The sum rule of  $F_2^\gamma(x,Q^2,P^2)$, i.e., the second moment, was numerically
examined. The NNLO corrections are found to be  
$7\% \sim10\%$ of the sum of the LO and NLO contributions, when
 $P^2\!=\!1 {\rm GeV}^2$ and  $Q^2\!=\!30\sim 100{\rm GeV}^2$ or 
$P^2\!=\!3 {\rm GeV}^2$ and  $Q^2\!=\!100{\rm GeV}^2$,  and $n_f$ is  three or four. 

The inverse Mellin transform of the moments was performed to 
express the structure functions $F_2^\gamma(x,Q^2,P^2)$ and $F_L^\gamma(x,Q^2,P^2)$
as  functions of $x$.
We found that there exist sizable NNLO  contributions for $F_2^\gamma$ at larger $x$.
The corrections are negative and the NNLO curve comes below the NLO one 
in the region  $0.3\lesssim x<1$.
 At lower $x$ region, $0.05\lesssim x \lesssim 0.3$, the NNLO corrections to the NLO
results are found to  be negligibly small. Concerning  $F_L^\gamma$, 
the NLO corrections  reduce the magnitude in the region 
 $0.2\lesssim x<1$.

The comparison of the present NNLO theoretical prediction for the virtual
photon structure functions with the existing experimental data will be
discussed elsewhere.


\begin{acknowledgments}
This research is supported in part by Grant-in-Aid
for Scientific Research  from the Ministry of Education, Culture, Sports, Science and Technology,
Japan No.18540267. This work is dedicated to the memory of Jiro Kodaira, who passed away in September 
2006. 
\end{acknowledgments}

\newpage
\appendix

\section{\label{M_nX_n}Evaluation of $M_n(Q^2/P^2,{\bar g}(P^2))$  and 
$\bm{X}_n(Q^2/P^2,{\bar g}(P^2),\alpha)$}

In order to evaluate the integrals for $M_n(Q^2/P^2,{\bar g}(P^2))$ given in 
(\ref{M_nExpansion}), we employ the same method that was used by Bardeen and Buras in Ref.\cite{BB} 
and make a full use of the projection operators obtained from 
the one-loop anomalous dimension matrix ${\widehat \gamma}^{0}_{n}$:
\begin{equation}
  {\widehat \gamma}^{(0)}_{n}=\sum_{i=+, -, NS} \lambda^n_i~P^n_i~, \label{ProjectionAppendix}
\end{equation}
where $\lambda^n_i~ (i=+, -, NS)$ are eigenvalues of ${\widehat \gamma}^{0}_{n}$ 
and are expressed as
\begin{eqnarray}
     \lambda^n_{\pm}&=&\frac{1}{2}\Bigl\{\gamma^{(0),n}_{\psi\psi}+
 \gamma^{(0),n}_{GG} \pm \Bigl[(\gamma^{(0),n}_{\psi\psi}-\gamma^{(0),n}_{GG})^2+4
\gamma^{(0),n}_{\psi G}\gamma^{(0),n}_{G\psi} \Bigr]^{1/2}\Bigr\} ~, \\
 \lambda^n_{NS}&=&\gamma^{(0),n}_{NS}~,
\end{eqnarray}
and $P^n_i$ are the corresponding projection operators,
\begin{eqnarray}
  P^n_{\pm}&=&\frac{1}{\lambda^n_{\pm}-\lambda^n_{\mp}}
\begin{pmatrix}\gamma^{(0),n}_{\psi\psi}-\lambda^n_{\mp}&\gamma^{(0),n}_{G\psi}&0 \\
\gamma^{(0),n}_{\psi G}& \gamma^{(0),n}_{GG}-\lambda^n_{\mp}&0 \\0&0&0
\end{pmatrix}~, \\
    P^n_{NS}&=&
\begin{pmatrix} 0&0&0\\ 0&0&0\\ 0&0&1
\end{pmatrix}~.
\end{eqnarray}
With an expansion of $\beta(g)$ up to the three-loop level in (\ref{Beta}), 
we get its inverse as follows:
\begin{equation}
\frac{1}{\beta(g)}=-\frac{16\pi^2}{\beta_0}\frac{1}{g^3}\left\{1-\frac{g^2}{16\pi^2}\frac{\beta_1}{\beta_0}+ \frac{g^4}{(16\pi^2)^2}
\left(\frac{\beta_1^2}{\beta_0^2}-\frac{\beta_2}{\beta_0}  \right) +\cdots
\right\}~. \label{betaInverse}
\end{equation}
Then using (\ref{ProjectionAppendix}) and (\ref{betaInverse}), we perform integration in
(\ref{M_nExpansion}). 

\newpage
The result is 
\begin{eqnarray}
&&\hspace{-1cm}M_n(Q^2/P^2,{\bar g}(P^2))\nonumber\\
&=&\sum_i P_i^n\left(\frac{{\bar g}_2^2}{{\bar g}_1^2}  
\right)^{d_i^n}
+\frac{1}{16\pi^2}\sum_i  P_i^n\left(\frac{{\bar g}_2^2}{{\bar g}_1^2}  
\right)^{d_i^n}d_i^n~\frac{\beta_1}{\beta_0}({\bar g}_1^2-{\bar g}_2^2)\nonumber\\
&-&\frac{1}{16\pi^2}\sum_{i,j}P_i^n{\hat{\bm{\gamma}}}_{n}^{(1)}P_j^n 
\frac{1}{\lambda_i^n-\lambda_j^n+2\beta_0}
\biggl[ {\bar g}_1^2\Bigl(\frac{{\bar g}_2^2}{{\bar g}_1^2}  
\Bigr)^{d_j^n}-{\bar g}_2^2\Bigl(\frac{{\bar g}_2^2}{{\bar g}_1^2}  
\Bigr)^{d_i^n }  \biggr]\nonumber\\
&+&\frac{1}{(16\pi^2)^2}\sum_i P_i^n\left(\frac{{\bar g}_2^2}{{\bar g}_1^2}  
\right)^{d_i^n}\frac{1}{2}\biggl\{ -d_i^n\Bigl[\frac{\beta_1^2}{\beta_0^2}
-\frac{\beta_2}{\beta_0}   \Bigr]( {\bar g}_1^4-{\bar g}_2^4  )
+\Bigl(d_i^n~\frac{\beta_1}{\beta_0} \Bigr)^2
({\bar g}_1^2-{\bar g}_2^2)^2
\biggr\}\nonumber\\
&-&\frac{1}{(16\pi^2)^2}\sum_{i,j}P_i^n{\hat{\bm{\gamma}}}_{n}^{(1)}P_j^n 
\frac{1}{\lambda_i^n-\lambda_j^n+2\beta_0}~\frac{\beta_1}{\beta_0}
\Bigl(d_i^n{\bar g}_1^2 - d_j^n{\bar g}_2^2 \Bigr)
\biggl[ {\bar g}_1^2\Bigl(\frac{{\bar g}_2^2}{{\bar g}_1^2}  
\Bigr)^{d_j^n}-{\bar g}_2^2\Bigl(
\frac{{\bar g}_2^2}{{\bar g}_1^2}  
\Bigr)^{d_i^n }  \biggr]\nonumber\\
&+&\frac{1}{(16\pi^2)^2}\sum_{i,j}P_i^n{\hat{\bm{\gamma}}}_{n}^{(1)}P_j^n 
\frac{ \Bigl(1+ d_i^n-d_j^n
\Bigr)}{\lambda_i^n-\lambda_j^n+4\beta_0}~\frac{\beta_1}{\beta_0}
\biggl[ ({\bar g}_1^2)^2\Bigl(\frac{{\bar g}_2^2}{{\bar g}_1^2}  
\Bigr)^{d_j^n}-({\bar g}_2^2)^2\Bigl(\frac{{\bar g}_2^2}{{\bar g}_1^2}  
\Bigr)^{d_i^n }  \biggr]\nonumber\\
&-&\frac{1}{(16\pi^2)^2}\sum_{i,j} 
~\frac{P^n_i{\hat{\bm{\gamma}}}_{n}^{(2)}
P^n_j}{\lambda^n_i-\lambda^n_j+4\beta_0}
\biggl[ ({\bar g}_1^2)^2\Bigl(\frac{{\bar g}_2^2}{{\bar g}_1^2}  
\Bigr)^{d_j^n}-({\bar g}_2^2)^2\Bigl(
\frac{{\bar g}_2^2}{{\bar g}_1^2}  
\Bigr)^{d_i^n}  \biggr]\nonumber\\
&+&\frac{1}{(16\pi^2)^2}\sum_{i,j,k}
P^n_i{\hat{\bm{\gamma}}}_{n}^{(1)}P^n_j {\hat{\bm{\gamma}}}_{n}^{(1)}P^n_k
\frac{1}{(\lambda^n_j-\lambda^n_k+2\beta_0)}\nonumber\\
&&\qquad \qquad\times \Biggl[
\frac{1}{\lambda^n_i-\lambda^n_k+4\beta_0}\biggl\{({\bar g}_1^2)^2
\Bigl(\frac{{\bar g}_2^2}{{\bar g}_1^2}\Bigr)^{d_k^n}
-({\bar g}_2^2)^2\Bigl(\frac{{\bar g}_2^2}{{\bar g}_1^2}\Bigr)^{d_i^n}
\biggr\} \nonumber\\
&& \qquad \qquad \qquad \qquad
-\frac{1}{\lambda^n_i-\lambda^n_j+2\beta_0}\biggl\{{\bar g}_1^2~{\bar g}_2^2~
\Bigl(\frac{{\bar g}_2^2}{{\bar g}_1^2}\Bigr)^{d_j^n}
-({\bar g}_2^2)^2\Bigl(\frac{{\bar g}_2^2}{{\bar g}_1^2}\Bigr)^{d_i^n}
\biggr\} \Biggr]~,   \label{M_nExpression}
\end{eqnarray}
where ${\bar g}_1^2={\bar g}^2(P^2)$ and ${\bar g}_2^2={\bar g}^2(Q^2)$.
The first term is the leading, and the second and third terms are the next-to-leading  
terms. The rest are the next-to-next-to-leading terms. 

Once we get the above expression for  $M_n(Q^2/P^2,{\bar g}(P^2))$ expanded up to the NNLO, we use 
an expansion of ${\bm K}_n(g,\alpha)$ in (\ref{ExpandK}) up to the three-loop level  and can 
evaluate $\bm{X}_n(Q^2/P^2,{\bar g}(P^2),\alpha)$  in (\ref{X_n}) up to the NNLO. The result is

\newpage
\begin{eqnarray}
&&\hspace{-1cm}\bm{X}_n(Q^2/P^2,{\bar g}(P^2),\alpha)\nonumber\\
&=& \frac{e^2}{2\beta_0}\frac{1}{{\bar g}_2^2}\bm{K}_n^{(0)}\sum_i P_i^n
\frac{1}{d_i^n+1}\biggl[ 1-\biggl(\frac{{\bar g}_2^2}
{{\bar g}_1^2} \biggr)^{d_i^n+1}  \biggr]\nonumber\\
&+&\frac{e^2}{2\beta_0}\frac{1}{16\pi^2}\bm{K}_n^{(0)}\frac{\beta_1}{\beta_0}\sum_i P_i^n
\biggl\{-\frac{d_i^n}{d_i^n+1}\biggl[ 1-\biggl(\frac{{\bar g}_2^2}
{{\bar g}_1^2} \biggr)^{d_i^n+1}  \biggr]
+\frac{d_i^n-1}{d_i^n}\biggl[ 1-\biggl(\frac{{\bar g}_2^2}{{\bar g}_1^2} \biggr)^{d_i^n} 
\biggr]
\biggr\}\nonumber\\
&+&\frac{e^2}{2\beta_0}\frac{1}{16\pi^2}\bm{K}_n^{(0)}\sum_{i}
\biggl\{\sum_{j}\frac{P_i^n{\hat{\bm{\gamma}}}_{n}^{(1)}
P_j^n}{\lambda_i^n-\lambda_j^n+2\beta_0}\frac{1}{d_i^n+1}\biggl[ 1-\biggl(\frac{{\bar g}_2^2}
{{\bar g}_1^2} \biggr)^{d_i^n+1}  \biggr]\nonumber\\
&&\hspace{3.5cm}-\sum_{j}\frac{P_j^n{\hat{\bm{\gamma}}}_{n}^{(1)}
P_i^n}{\lambda_j^n-\lambda_i^n+2\beta_0}\frac{1}{d_i^n}\biggl[ 1-\biggl(\frac{{\bar
g}_2^2} {{\bar g}_1^2} \biggr)^{d_i^n}  \biggr]\biggr\}\nonumber\\
&+&\frac{e^2}{2\beta_0}\frac{1}{16\pi^2}~\bm{K}_n^{(1)} 
 \sum_i P_i^n \frac{1}{d_i^n}\biggl[ 1-\biggl(\frac{{\bar g}_2^2}
{{\bar g}_1^2} \biggr)^{d_i^n}  \biggr]\nonumber\\
&+&\frac{e^2}{2\beta_0}\frac{{\bar g}_2^2}{(16\pi^2)^2}\bm{K}_n^{(0)}\sum_i P_i^n~\Biggl\{
\frac{d_i^n}{2}\biggl(\frac{\beta_1^2}{\beta_0^2}-\frac{\beta_2}{\beta_0}
\frac{1}{d_i^n+1}\biggr)
\biggl[ 1-\biggl(\frac{{\bar g}_2^2}{{\bar g}_1^2} \biggr)^{d_i^n+1}  \biggr]
\nonumber\\
&&\  +\frac{\beta_1^2}{\beta_0^2}~(1-d_i^n)~\biggl[ 1-\biggl(\frac{{\bar g}_2^2}{{\bar g}_1^2}
\biggr)^{d_i^n} 
\biggr] +\Bigl(\frac{d_i^n}{2}-1\Bigr)\biggl(\frac{\beta_1^2}{\beta_0^2}+\frac{\beta_2}{\beta_0}
\frac{1}{d_i^n-1}\biggr)
\biggl[ 1-\biggl(\frac{{\bar g}_2^2}{{\bar g}_1^2} \biggr)^{d_i^n-1}  \biggr]
\Biggr\}\nonumber\\
&+&\frac{e^2}{2\beta_0}\frac{{\bar
g}_2^2}{(16\pi^2)^2}\bm{K}_n^{(0)}\frac{\beta_1}{\beta_0}\sum_{i}\nonumber\\
&&\quad\times\Biggl\{
\biggl(-\sum_{j}\frac{P_i^n{\hat{\bm{\gamma}}}_{n}^{(1)}
P_j^n}{\lambda_i^n-\lambda_j^n+2\beta_0}\frac{d_j^n}{d_i^n+1}
-\sum_{j}\frac{P_i^n{\hat{\bm{\gamma}}}_{n}^{(1)}
P_j^n}{\lambda_i^n-\lambda_j^n+4\beta_0}\frac{1+d_i^n-d_j^n}{d_i^n+1}\biggr)\nonumber\\
&&\hspace{8.5cm}\times\biggl[ 1-\biggl(\frac{{\bar
g}_2^2} {{\bar g}_1^2} \biggr)^{d_i^n+1}  \biggr]\nonumber\\
&&\qquad+\biggl(\sum_{j}\frac{P_i^n{\hat{\bm{\gamma}}}_{n}^{(1)}
P_j^n}{\lambda_i^n-\lambda_j^n+2\beta_0}\Bigl( 1-\frac{1}{d_i^n}  \Bigr)
+\sum_{j}\frac{P_j^n{\hat{\bm{\gamma}}}_{n}^{(1)}
P_i^n}{\lambda_j^n-\lambda_i^n+2\beta_0}
\biggr)\biggl[ 1-\biggl(\frac{{\bar
g}_2^2} {{\bar g}_1^2} \biggr)^{d_i^n}  \biggr]\nonumber\\
&&\qquad+\biggl(\sum_{j}\frac{P_j^n{\hat{\bm{\gamma}}}_{n}^{(1)}
P_i^n}{\lambda_j^n-\lambda_i^n+2\beta_0}\frac{1-d_j^n}{d_i^n-1}
+\sum_{j}\frac{P_j^n{\hat{\bm{\gamma}}}_{n}^{(1)}
P_i^n}{\lambda_j^n-\lambda_i^n+4\beta_0}\frac{1+d_j^n-d_i^n}{d_i^n-1}
\biggr)\nonumber\\
&&\hspace{8.5cm}\times\biggl[ 1-\biggl(\frac{{\bar
g}_2^2} {{\bar g}_1^2} \biggr)^{d_i^n-1}  \biggr]\Biggr\}\nonumber\\
&+&\frac{e^2}{2\beta_0}\frac{{\bar g}_2^2}{(16\pi^2)^2}\bm{K}_n^{(0)}
\sum_{i}\biggl\{\sum_{j}\frac{P^n_i{\hat{\bm{\gamma}}}_{n}^{(2)}
P^n_j}{\lambda^n_i-\lambda^n_j+4\beta_0}\frac{1}{d_i^n+1}\biggl[ 1-\biggl(\frac{{\bar g}_2^2}
{{\bar g}_1^2} \biggr)^{d_i^n+1}\biggr]\nonumber\\
&&\hspace{3.5cm} -\sum_{j}\frac{P^n_j{\hat{\bm{\gamma}}}_{n}^{(2)}
P^n_i}{\lambda^n_j-\lambda^n_i+4\beta_0}\frac{1}{d_i^n-1}\biggl[
1-\biggl(\frac{{\bar g}_2^2}{{\bar g}_1^2}
\biggr)^{d_i^n-1} \biggr]  \biggr\}\nonumber
\end{eqnarray}

\begin{eqnarray}
&&\hspace{-2.5cm}+\frac{e^2}{2\beta_0}\frac{{\bar g}_2^2}{(16\pi^2)^2}\bm{K}_n^{(0)}
\sum_{i}\nonumber\\
&&\hspace{-1.5cm}\times\Biggl\{\sum_{j,k}\frac{P^n_i{\hat{\bm{\gamma}}}_{n}^{(1)}P^n_j
{\hat{\bm{\gamma}}}_{n}^{(1)}P^n_k }{\lambda^n_j-\lambda^n_k+2\beta_0}
\biggl(\frac{1}{\lambda^n_i-\lambda^n_j+2\beta_0}-
\frac{1}{\lambda^n_i-\lambda^n_k+4\beta_0}  \biggr)\nonumber\\
&&\hspace{5cm}\times
\frac{1}{d_i^n+1}\biggl[ 1-\biggl(\frac{{\bar g}_2^2}
{{\bar g}_1^2} \biggr)^{d_i^n+1}  \biggr]\nonumber\\
&&\hspace{-1.0cm}-\sum_{j,k}\frac{P^n_j{\hat{\bm{\gamma}}}_{n}^{(1)}P^n_i
{\hat{\bm{\gamma}}}_{n}^{(1)}P^n_k }{\lambda^n_i-\lambda^n_k+2\beta_0}
\frac{1}{\lambda^n_j-\lambda^n_i+2\beta_0}~
\frac{1}{d_i^n}\biggl[ 1-\biggl(\frac{{\bar g}_2^2}
{{\bar g}_1^2} \biggr)^{d_i^n}  \biggr]\nonumber\\
&&\hspace{-1.0cm}+\sum_{j,k}\frac{P^n_k{\hat{\bm{\gamma}}}_{n}^{(1)}P^n_j
{\hat{\bm{\gamma}}}_{n}^{(1)}P^n_i }{\lambda^n_j-\lambda^n_i+2\beta_0}
\frac{1}{\lambda^n_k-\lambda^n_i+4\beta_0}~
\frac{1}{d_i^n-1}\biggl[ 1-\biggl(\frac{{\bar g}_2^2}
{{\bar g}_1^2} \biggr)^{d_i^n-1}  \biggr]\Biggr\}\nonumber\\
&&\hspace{-2.5cm}+\frac{e^2}{2\beta_0}\frac{{\bar
g}^2_2}{(16\pi^2)^2}\bm{K}_n^{(1)}
\frac{\beta_1}{\beta_0}  \sum_i P_i^n \biggl\{-
\biggl[ 1-\biggl(\frac{{\bar g}_2^2}
{{\bar g}_1^2} \biggr)^{d_i^n}\biggr] +
\biggl[ 1-\biggl(\frac{{\bar g}_2^2}
{{\bar g}_1^2} \biggr)^{d_i^n-1}\biggr]\biggr\}\nonumber\\
&&\hspace{-2.5cm}+\frac{e^2}{2\beta_0}\frac{{\bar g}^2_2}{(16\pi^2)^2}
\bm{K}_n^{(1)} \sum_{i}\biggl\{ \sum_{j} \frac{P_i^n{\hat{\bm{\gamma}}}_{n}^{(1)}P_j^n }
{\lambda_i^n-\lambda_j^n+2\beta_0}~\frac{1}{d_i^n}~
\biggl[ 1-\biggl(\frac{{\bar g}_2^2}{{\bar g}_1^2} \biggr)^{d_i^n}\biggr]\nonumber\\
&&\qquad \qquad -\sum_{j} \frac{P_j^n{\hat{\bm{\gamma}}}_{n}^{(1)}P_i^n }
{\lambda_j^n-\lambda_i^n+2\beta_0}~\frac{1}{d_i^n-1}~
\biggl[ 1-\biggl(\frac{{\bar g}_2^2}{{\bar g}_1^2} \biggr)^{d_i^n-1}\biggr\}\nonumber\\
&&\hspace{-2.5cm}+\frac{e^2}{2\beta_0}
\frac{{\bar g}_2^2}{(16\pi^2)^2}~\bm{K}_n^{(2)}
 P_i^n
\biggl\{\frac{1}{d_i^n-1}\biggl[ 1-\biggl(\frac{{\bar g}_2^2}
{{\bar g}_1^2} \biggr)^{d_i^n-1}\biggr]+{\cal O}({\bar g}^2_2) \biggr\}~. \label{X_nExpression}
\end{eqnarray}
where ${\bar g}_1^2={\bar g}^2(P^2)$ and ${\bar g}_2^2={\bar g}^2(Q^2)$.
The first term is the leading, and the second to fourth terms are the next-to-leading  
terms. The rest are the next-to-next-to-leading terms. 

\newpage

\section{$E_{{\rm ns}\gamma}^{\rm approx}(n)$, $E_{G\gamma}^{\rm approx}(n)$ and  
$E_{{\rm ps}\gamma}(n)$ \label{Ensgamma}}

We give the explicit expressions of $E_{{\rm ns}\gamma}^{\rm approx}(n)$, 
$E_{G\gamma}^{\rm approx}(n)$ and $E_{{\rm ps}\gamma}(n)$
which have appeared in (\ref{EnsApprox})-(\ref{EgApprox}).  
They are obtained by taking the Mellin moments of the parameterizations for 
$P^{(2)}_{{\rm ns}\gamma}(x)$ and $P^{(2)}_{{\rm g}\gamma}(x)$  and of the exact result for
$P^{(2)}_{{\rm ps}\gamma}(x)$,  which are presented in Eqs.(6)-(8) of Ref.\cite{MVV3}. Using a 
single harmonic sum $S_m(n)$ defined by
\begin{equation}
S_m(n)=\sum_{j=1}^n \frac{1}{j^m}~,
\end{equation}
they are expressed as 
\bea
E_{{\rm ns}\gamma}^{\rm approx}(n)&\equiv&-\int^1_0 dx x^{n-1}\{ 
\text{the r.h.s. of Eq.(6) in Ref.\cite{MVV3}}  \}\nonumber\\
&=&-\frac{128S_1(n)^4}{27n}+\frac{62.5244S_1(n)^3}{n}-\frac{50.08S_1(n)^3}{n+1}
-\frac{256S_2(n)S_1(n)^2}{9n}\nonumber\\
&&-\frac{175.3S_1(n)^2}{n}-\frac{195.4S_1(n)^2}{n^2}-\frac{150.24S_1(n)^2}{(n+1)^2}
-\frac{203.227S_2(n)S_1(n)}{n}\nonumber\\
&&-\frac{150.24S_2(n)S_1(n)}{n+1}-\frac{1024S_3(n)S_1(n)}{27n}-\frac{128S_2(n)^2}{9n}
+\frac{785.14S_1(n)}{n}\nonumber\\
&&-\frac{325.4S_1(n)}{n^3}-\frac{300.48S_1(n)}{(n+1)^3}-\frac{175.3S_2(n)}{n}
-\frac{520.8S_2(n)}{n^2}-\frac{150.24S_2(n)}{(n+1)^2}\nonumber\\
&&-\frac{591.151S_3(n)}{n}-\frac{100.16S_3(n)}{n+1}-\frac{256S_4(n)}{9n}
-\frac{492.087}{n}+\frac{1262}{n+1}-\frac{449.2}{n+2}
\nonumber\\
&&+\frac{1445}{n+3}+\frac{1279.86}{n^2}+\frac{1169}{(n+1)^2}-\frac{403.2}{n^3}+\frac{160}{n^4}
-\frac{300.48}{(n+1)^4}-\frac{512}{9n^5}\nonumber\\
&+&n_f\biggl\{-\frac{32S_1(n)^3}{27n}-\frac{258.142S_1(n)^2}{n}+\frac{270S_1(n)^2}{n+1}
+\frac{269.4S_1(n)^2}{n^2}\nonumber\\
&&\qquad+\frac{535.244S_2(n)S_1(n)}{n}-\frac{905.06S_1(n)}{n}
+\frac{540S_1(n)}{(n+1)^2}+\frac{17.046S_1(n)}{n^3}\nonumber\\
&&\qquad-\frac{258.142S_2(n)}{n}+\frac{270S_2(n)}{n+1}+\frac{286.446S_2(n)}{n^2}
+\frac{553.476S_3(n)}{n}\nonumber\\
&&\qquad-\frac{628.124}{n}-\frac{114.4}{n+1}+\frac{24.86}{n+2}
+\frac{53.39}{n+3}-\frac{49.5895}{n^2}-\frac{26.63}{(n+1)^2}\nonumber\\
&&\qquad+\frac{21.984}{n^3}+\frac{540}{(n+1)^3}-\frac{64}{9n^4}
\biggr\}~, \label{EnsgammaApprox}
\eea

\bigskip

\bea
\hspace{-1cm}E_{G\gamma}^{\rm approx}(n)&\equiv&-\int^1_0 dx x^{n-1}\{ 
\text{the  r.h.s. of Eq.(7) in Ref.\cite{MVV3}} \}\nonumber\\
&&\hspace{-1.2cm}=\frac{32S_1(n)^3}{27n}-\frac{32S_1(n)^3}{27(n+1)}
+\frac{79.13S_1(n)^2}{n}-\frac{79.13S_1(n)^2}{n+1}-\frac{433.2S_1(n)^2}{n^2}\nonumber\\
&&\hspace{-1cm}+\frac{429.644S_1(n)^2}{(n+1)^2}-\frac{862.844S_2(n)S_1(n)}{n}+
\frac{862.844S_2(n)S_1(n)}{n+1}\nonumber\\
&&\hspace{-1cm}+\frac{1512.39S_1(n)}{n}-\frac{1512.39S_1(n)}{n+1}
+\frac{549.5S_1(n)}{n^2}-\frac{707.76S_1(n)}{(n+1)^2}-\frac{2460S_1(n)}{n^3}\nonumber\\
&&\hspace{-1cm}+\frac{4185.69S_1(n)}{(n+1)^3}
+\frac{628.63S_2(n)}{n}-\frac{628.63S_2(n)}{n+1}-\frac{2893.2S_2(n)}{n^2}\nonumber\\
&&\hspace{-1cm}+\frac{3756.04S_2(n)}{(n+1)^2}
-\frac{3324.03S_3(n)}{n}+\frac{3324.03S_3(n)}{n+1}\nonumber\\
&&\hspace{-1cm}
+\frac{73.1409}{n-1}-\frac{1673.57}{n}+\frac{3180.43}{n+1}-\frac{1420}{n+2}
+\frac{406.7}{n+3}- \frac{566.7}{n+4}+ \frac{128}{3(n-1)^2}    \nonumber\\
&&\hspace{-1cm}+\frac{6400}{3n^2}-\frac{3688.39}{(n+1)^2}-\frac{2247.4}{n^3}+\frac{990.14}{(n+1)^3}
+\frac{1600}{3n^4}+\frac{9438.76}{(n+1)^4}-\frac{3584}{9n^5}\nonumber\\
&&\hspace{-1cm}+\frac{3584}{9(n+1)^5}+\Bigl(\frac{2460}{n}-\frac{2460}{n+1}   \Bigr)\zeta_3+
\Bigl(\frac{2460}{n^2}-\frac{2460}{(n+1)^2}   \Bigr)\zeta_2
\nonumber\\
&&\hspace{-1cm}+n_f\biggl\{\frac{32S_1(n)^2}{9n}-\frac{32S_1(n)^2}{9(n+1)}
-\frac{9.133S_1(n)^2}{n^2}+\frac{9.133S_1(n)^2}{(n+1)^2}-\frac{18.266S_2(n)S_1(n)}{n}\nonumber\\
&&\hspace{-1cm}\qquad+\frac{18.266S_2(n)S_1(n)}{n+1}+\frac{46.4264S_1(n)}{n}
-\frac{46.4264S_1(n)}{n+1}+\frac{16.18S_1(n)}{n^2}
\nonumber\\
&&\hspace{-1cm}\qquad
-\frac{23.2911S_1(n)}{(n+1)^2}-\frac{76.66S_1(n)}{n^3}+\frac{113.192S_1(n)}{(n+1)^3}
+\frac{19.7356S_2(n)}{n}-\frac{19.7356S_2(n)}{n+1}\nonumber\\
&&\hspace{-1cm}\qquad -\frac{85.793S_2(n)}{n^2}
+\frac{104.059S_2(n)}{(n+1)^2}
-\frac{94.926S_3(n)}{n}+\frac{94.926S_3(n)}{n+1}\nonumber \\
&&\hspace{-1cm}\qquad +\frac{40.5597}{n-1}-\frac{21.1683}{n}+\frac{17.0286}{n+1}-\frac{93.37}{n+2}
+\frac{101.05}{n+3}-\frac{44.1}{n+4}+\frac{115.341}{n^2}\nonumber\\
&&\hspace{-1cm}\qquad 
-\frac{161.767}{(n+1)^2}-\frac{52.82}{n^3}+\frac{13.3489}{(n+1)^3}-\frac{128}{9n^4} +\frac{299}{(n+1)^4}
\biggr\}~,\label{EGgammaApprox}
\eea

\bigskip

\bea
E_{{\rm ps}\gamma}(n)&\equiv&-\int^1_0 dx x^{n-1}\{ 
\text{the r.h.s. of Eq.(8) in Ref.\cite{MVV3}}  \}\nonumber\\
&=&
n_f C_F\biggl\{-\frac{2464}{81(n-1)}+\frac{432}{n}+\frac{72}{n+1}-\frac{38360}{81(n+2)}
-\frac{344}{n^2}-\frac{368}{(n+1)^2}\nonumber\\
&&\quad -\frac{3584}{27(n+1)^2}+\frac{288}{n^3}+\frac{208}{(n+1)^3}+\frac{448}{9(n+2)^3}
-\frac{96}{n^4}+\frac{96}{(n+1)^4}\nonumber\\
&&\quad +\frac{256}{3(n+2)^4}+\frac{64}{n^5}-\frac{128}{(n+1)^5}
\biggr\}~.
\eea

Note that $E_{{\rm ps}\gamma}(n)$ is an exact result.

\section{Mellin moments $a_{qg}^{(i)}(n)$ and $b_{qg}^{(i)}(n)$ with $i=1,2$ \\
and $a_{gg}^{(2)}(n)$ and $b_{gg}^{(2)}(n)$ \label{AppenC}}

The expressions of $a_{qg}^{(i)}(n)$ and $b_{qg}^{(i)}(n)$ with $i=$1 and 2 are 
obtained by taking the moments of the functions $a_{qg}^{(i)}(z)$ and $b_{qg}^{(i)}(z)$ as
\begin{equation}
a_{qg}^{(i)}(n)=\int_0^1dz z^{n-1}a_{qg}^{(i)}(z)~, \quad 
b_{qg}^{(i)}(n)=\int_0^1dz z^{n-1}b_{qg}^{(i)}(z)~, \qquad {\rm with}\ i=1,2
\end{equation}
where $a_{qg}^{(i)}(z)$ and $b_{qg}^{(i)}(z)$ are extracted from the $\epsilon$-independent terms of
${\hat A}^{\rm PHYS}_{qg}\Bigl(z, \frac{-p^2}{\mu^2}, \frac{1}{\epsilon} \Bigr)$ given in 
Eq.(A7) and of
${\hat A}^{\rm EOM}_{qg}\Bigl(z, \frac{-p^2}{\mu^2}, \frac{1}{\epsilon} \Bigr)$ in Eq.(A8) 
of Ref.\cite{MSvN}, respectively. See also Eqs.(2.27) and (2.28) of Ref.\cite{MSvN}.

\bigskip

The one-loop results are
\bea
a_{qg}^{(1)}(n)&=&\frac{n_f}{2}4 \biggl\{  \Bigl(\frac{1}{n}- \frac{2}{n+1}+\frac{2}{n+2}  \Bigr) 
\Bigl(S_1(n)-1 \Bigr)
 +\frac{1}{n^2}- \frac{4}{(n+1)^2}+\frac{4}{(n+2)^2}
\biggr\} ~, \label{aqg1}\\
b_{qg}^{(1)}(n)&=&\frac{n_f}{2}16  \Bigl(\frac{1}{n+1}-\frac{1}{n+2}  \Bigr)~.  \label{bqg1}
\eea

\bigskip

The two-loop results are
\bea
a_{qg}^{(2)}(n)&=&C_F \frac{n_f}{2} \biggl\{ \Bigl(\frac{1}{n}- \frac{2}{n+1}+\frac{2}{n+2}  \Bigr) 
\Bigl( -\frac{4}{3}S_1(n)^3 -4S_2(n)S_1(n)+\frac{64}{3}S_3(n)\nonumber \\
&&\hspace{8cm}-16S_{2,1}(n) -48\zeta_3 \Bigr) \nonumber \\
&&\hspace{1.5cm}+S_1(n)^2\Bigl( \frac{6}{n}- \frac{24}{n+1}+\frac{32}{n+2}  
-\frac{16}{n^2}+ \frac{40}{(n+1)^2}-\frac{32}{(n+2)^2}  \Bigr) \nonumber \\
&&\hspace{1.5cm}+S_1(n)\Bigl( \frac{12}{n}+ \frac{8}{n+1}-\frac{40}{n+2}  
+\frac{16}{n^2}- \frac{128}{(n+1)^2}+\frac{128}{(n+2)^2} \nonumber \\
&&\hspace{6.5cm}-\frac{32}{n^3}+ \frac{176}{(n+1)^3}-\frac{128}{(n+2)^3}   \Bigr) \nonumber \\
&&\hspace{1.5cm}+S_2(n)\Bigl( \frac{6}{n}- \frac{8}{n+1}+\frac{16}{n+2}  
-\frac{16}{n^2}+ \frac{40}{(n+1)^2}-\frac{32}{(n+2)^2}  \Bigr) \nonumber \\
&&\hspace{1.5cm}+\frac{14}{n}- \frac{38}{n+1}+\frac{48}{n+2}
+\frac{64}{n^2}- \frac{126}{(n+1)^2}+\frac{80}{(n+2)^2}
-\frac{22}{n^3}\nonumber \\
&&\hspace{1.5cm}- \frac{88}{(n+1)^3}+\frac{128}{(n+2)^3} 
+\frac{20}{n^4}+ \frac{88}{(n+1)^4}
\biggr\}\nonumber\\
&&+\ \  \text{terms proportional to } C_A\frac{n_f}{2}~ \ {\rm or}\ \Bigl(\frac{n_f}{2}\Bigr)^2~ ~,
\label{aqg2}
\eea
\bea
b_{qg}^{(2)}(n)&=&16C_F \frac{n_f}{2} \biggl\{ \Bigl(S_1(n)^2 +S_2(n) \Bigr)\Bigl(-
\frac{2}{n+1}+\frac{2}{n+2}  \Bigr) 
 \nonumber \\
&&\hspace{1.5cm}+S_1(n)\Bigl( \frac{1}{n}+ \frac{11}{n+1}-\frac{12}{n+2}  
- \frac{10}{(n+1)^2}+\frac{8}{(n+2)^2}    \Bigr) \nonumber \\
&&\hspace{1.5cm}-\frac{3}{n}+ \frac{4}{n+1}-\frac{1}{n+2}
+\frac{1}{n^2}+ \frac{9}{(n+1)^2}-\frac{8}{(n+2)^2}
- \frac{6}{(n+1)^3}
\biggr\}\nonumber \\
&&+\ \  \text{terms proportional to } C_A\frac{n_f}{2}~ \ {\rm or}\ \Bigl(\frac{n_f}{2}\Bigr)^2~ ~,
\label{bqg2}
\eea
where $S_{2,1}(n)=\sum^n_{j=1}\frac{1}{j^2}S_1(j)$. The terms proportional to
$\Bigl(\frac{n_f}{2}\Bigr)^2$ in $a_{qg}^{(2)}(n)$ and $ b_{qg}^{(2)}(n)$ come from the 
external gluon self-energy corrections and should be discarded for the photon case. 
\bigskip

Similarly the expressions of $a_{gg}^{(2)}(n)$ and $b_{gg}^{(2)}(n)$  are 
obtained by taking the moments of the functions $a_{gg}^{(2)}(z)$ and $b_{gg}^{(2)}(z)$ 
which  are extracted from the $\epsilon$-independent terms of
${\hat A}^{\rm PHYS}_{gg}\Bigl(z, \frac{-p^2}{\mu^2}, \frac{1}{\epsilon} \Bigr)$ given in 
Eq.(A12) and of 
${\hat A}^{\rm EOM}_{gg}\Bigl(z, \frac{-p^2}{\mu^2}, \frac{1}{\epsilon} \Bigr)$ in Eq.(A13) 
of Ref.\cite{MSvN}, respectively. See also Eqs.(2.34) and (2.35) of Ref.\cite{MSvN} \footnote{Two terms
$\gamma_{gg}^{(0)}b_{gg}^{\epsilon, (1)}$ and 
$\gamma_{gq}^{(0)}b_{qg}^{\epsilon, (1)}$ are missing in the $\epsilon$-independent terms 
of Eq.(2.35) of Ref.\cite{MSvN}. The both are needed in order to extract $b_{gg}^{(2)}(z)$ correctly.}.

The two-loop results for $a_{gg}^{(2)}(n)$ and $b_{gg}^{(2)}(n)$ are
\bea
a_{gg}^{(2)}(n)&=&
C_F \frac{n_f}{2} \biggl\{ \Bigl(S_1(n)^2 +S_2(n) \Bigr)
\Bigl(\frac{16}{3(n-1)}+\frac{4}{n}- \frac{4}{n+1}-\frac{16}{3(n+2)}  
-\frac{8}{n^2}- \frac{8}{(n+1)^2}  \Bigr) \nonumber \\
&&\hspace{1.2cm}+S_1(n)\Bigl(\frac{16}{9(n-1)}- \frac{48}{n}+ \frac{32}{n+1}+\frac{128}{9(n+2)}  
+\frac{32}{n^2}+ \frac{24}{(n+1)^2} \nonumber \\
&&\hspace{7cm} -\frac{64}{3(n+2)^2}-\frac{32}{n^3}- \frac{48}{(n+1)^3}   \Bigr) \nonumber \\
&&\hspace{1.2cm}+S_{-2}(n)\Bigl( \frac{32}{3(n-1)}- \frac{32}{n}+ \frac{32}{n+1}-\frac{32}{3(n+2)}  
  \Bigr) \nonumber \\
&&\hspace{1.2cm}+\frac{680}{27(n-1)}-\frac{48}{n}+ \frac{16}{n+1}+\frac{184}{27(n+2)}
-\frac{56}{n^2}+ \frac{56}{(n+1)^2}+\frac{256}{9(n+2)^2}
+\frac{44}{n^3}\nonumber \\
&&\hspace{1.2cm}+ \frac{76}{(n+1)^3}-\frac{128}{3(n+2)^3} 
-\frac{40}{n^4}- \frac{88}{(n+1)^4}-\frac{55}{3}+16\zeta_3
\biggr\}\nonumber\\
&&+\ \  \text{terms proportional to } C_A^2~ \ {\rm or}\ C_A\frac{n_f}{2}~ \ {\rm or}\
\Bigl(\frac{n_f}{2}\Bigr)^2~~, \label{agg2}
\eea
\bea
b_{gg}^{(2)}(n)&=&
C_F \frac{n_f}{2} \biggl\{ 
S_1(n)\Bigl(\frac{32}{3(n-1)}- \frac{32}{n}+\frac{64}{3(n+2)}  
+ \frac{32}{(n+1)^2} \Bigr) \nonumber \\
&&\hspace{1.2cm}-\frac{128}{9(n-1)}+\frac{64}{n}- \frac{448}{9(n+2)}
-\frac{32}{n^2}- \frac{96}{(n+1)^2}+\frac{128}{3(n+2)^2}
+ \frac{96}{(n+1)^3}
\biggr\}\nonumber \\
&&+\ \  \text{terms proportional to } C_A^2~ \ {\rm or}\ C_A\frac{n_f}{2}~~.\label{bgg2}
\eea
The terms proportional to
$\Bigl(\frac{n_f}{2}\Bigr)^2$ in $a_{gg}^{(2)}(n)$ again come from the 
external gluon self-energy corrections and should be discarded for the photon case. 
Furthermore, the contribution of  the last two terms in the curly brackets of 
(\ref{agg2}), more explicitely, 
$C_F \frac{n_f}{2}\Bigl(-\frac{55}{3}+16\zeta_3\Bigr)$,  is  also 
resulted from the  external gluon self-energy corrections and is thus irrelevant.

\newpage
\section{Values at $n=2$}
\subsection{Coefficient functions}

With $\delta_\psi=\langle e^2 \rangle$, $\delta_{NS}=1$ and 
$\delta_{\gamma}=3n_f \langle e^4 \rangle$, we have

$\bullet$\ At  tree-level
\begin{equation}
C_{2,n=2}^{\psi(0)}=\delta_\psi~, \quad C_{2,n=2}^{G(0)}=0~, \quad C_{2,n=2}^{NS(0)}=\delta_{NS}
~, \quad C_{2,n=2}^{\gamma(0)}=0~.
\end{equation}

$\bullet$\ At one-loop
\bea
C_{2,n=2}^{\psi(1)}&=&\delta_\psi C_F\Bigl(\frac{1}{3}\Bigr), \qquad 
C_{2,n=2}^{G(1)}=\delta_\psi n_f\Bigl(-\frac{1}{2}\Bigr)~, \nonumber \\
C_{2,n=2}^{NS(1)}&=&\delta_{NS}C_F\Bigl(\frac{1}{3}\Bigr) ~, \qquad
C_{2,n=2}^{\gamma(1)}=\delta_{\gamma}(-1)~.
\eea

$\bullet$\ At two-loop
\bea
C_{2,n=2}^{\psi(2)}&=&\delta_\psi \Bigl\{C_AC_F\Bigl(\frac{3677}{135}-\frac{128}{5}\zeta_3\Bigr)
+C_Fn_f\Bigl(-\frac{457}{81}\Bigr)
+C_F^2\Bigl(-\frac{4189}{810}+\frac{96}{5}\zeta_3\Bigr)\Bigr\}, \nonumber\\
C_{2,n=2}^{G(2)}&=&\delta_\psi \Bigl\{
C_Fn_f\Bigl(-\frac{4799}{810}+\frac{16}{5}\zeta_3\Bigr)
+C_An_f\Bigl(\frac{115}{324}-2\zeta_3\Bigr)\Bigr\}, \nonumber\\
\quad C_{2,n=2}^{NS(2)}&=&\delta_{NS}\Bigl\{C_AC_F\Bigl(\frac{3677}{135}-\frac{128}{5}\zeta_3\Bigr)
+C_Fn_f(-4)
+C_F^2\Bigl(-\frac{4189}{810}+\frac{96}{5}\zeta_3\Bigr)\Bigr\}~.\nonumber \\ 
C_{2,n=2}^{\gamma(2)}&=&\delta_\gamma C_F\Bigl(-\frac{4799}{405}+\frac{32}{5}\zeta_3\Bigr)~.
\eea

\subsection{Anomalous dimensions}

$\bullet$\ At  one-loop
\bea
\gamma^{(0),n=2}_{NS}&=&\gamma^{(0),n=2}_{\psi\psi}=C_F\Bigl(\frac{16}{3}\Bigr)~, \qquad 
\gamma^{(0),n=2}_{\psi G}=n_f\Bigl(-\frac{4}{3}\Bigr)~,\nonumber \\
\gamma^{(0),n=2}_{G\psi}&=&C_F\Bigl(-\frac{16}{3}\Bigr) ~, \hspace{2cm}
\gamma^{(0),n=2}_{GG}=n_f\Bigl(\frac{4}{3}\Bigr)~,
\eea

\quad and
\bea
K_\psi^{{(0),n=2}}=3n_f\langle e^2 \rangle \Bigl(\frac{8}{3}\Bigr)~, \quad
K_G^{{(0),n=2}}=0~,\quad  K_{NS}^{(0),n=2}=3n_f(\langle e^4 \rangle -
\langle e^2 \rangle^2) \Bigl(\frac{8}{3}\Bigr)~.
\eea

$\bullet$\ At  two-loop
\bea
\gamma^{(1),n=2}_{NS}&=&C_AC_F \Bigl(\frac{752}{27}\Bigr)   +C_Fn_f \Bigl(-\frac{128}{27}\Bigr)  
  +C_F^2\Bigl(-\frac{224}{27}\Bigr)~,\nonumber \\
\gamma^{(1),n=2}_{\psi\psi}&=&C_AC_F \Bigl(\frac{752}{27}\Bigr)   +C_Fn_f \Bigl(-\frac{208}{27}\Bigr)  
  +C_F^2  \Bigl(-\frac{224}{27}\Bigr)~, \nonumber \\
\gamma^{(1),n=2}_{\psi G}&=&C_An_f\Bigl(-\frac{70}{27}\Bigr)+
C_Fn_f\Bigl(-\frac{148}{27}\Bigr)~, \\
\gamma^{(1),n=2}_{G\psi}&=&C_AC_F\Bigl(-\frac{752}{27}\Bigr)+C_Fn_f \Bigl(\frac{208}{27}\Bigr) 
+ C_F^2\Bigl(\frac{224}{27}\Bigr)~, \nonumber \\
\gamma^{(1),n=2}_{GG}&=&C_An_f\Bigl(\frac{70}{27}\Bigr)+
C_Fn_f\Bigl(\frac{148}{27}\Bigr)~,\nonumber
\eea

\quad and
\bea
K_{NS}^{(1),n=2}&=&3n_f(\langle e^4 \rangle -
\langle e^2 \rangle^2) C_F\Bigl(\frac{296}{27}\Bigr)~,\nonumber \\
K_\psi^{{(1),n=2}}&=&3n_f\langle e^2 \rangle C_F \Bigl(\frac{296}{27}\Bigr)~, \\
K_G^{{(1),n=2}}&=&3n_f\langle e^2 \rangle C_F \Bigl(-\frac{80}{27}\Bigr)~.\nonumber
\eea

$\bullet$\ At  three-loop 

We get from \cite{MVV1} and \cite{MVV2},
\bea
\gamma^{(2),n=2}_{NS}&=&C_AC_Fn_f \Bigl(-\frac{6256}{243}-\frac{128}{3}\zeta_3\Bigr)   
+C_FC_A^2 \Bigl(\frac{41840}{243}+\frac{128}{3}\zeta_3\Bigr)  
+C_Fn_f^2 \Bigl(-\frac{448}{243}\Bigr)  
\nonumber \\
 && +C_F^2C_A \Bigl(-\frac{17056}{243}-128\zeta_3\Bigr)  
+C_F^2n_f\Bigl(-\frac{6824}{243}+\frac{128}{3}\zeta_3\Bigr)
+C_F^3\Bigl(-\frac{1120}{243}+\frac{256}{3}\zeta_3\Bigr)~,\nonumber
\\
\gamma^{(2),n=2}_{\psi\psi}&=&C_AC_Fn_f \Bigl(-\frac{44}{9}-\frac{256}{3}\zeta_3\Bigr) 
+C_FC_A^2 \Bigl(\frac{41840}{243}+\frac{128}{3}\zeta_3\Bigr)  
+C_Fn_f^2 \Bigl(-\frac{568}{81}\Bigr)    \nonumber \\
 &&+C_F^2C_A \Bigl(-\frac{17056}{243}-128\zeta_3\Bigr)  
+C_F^2n_f\Bigl(-\frac{14188}{243}+\frac{256}{3}\zeta_3\Bigr)
+C_F^3\Bigl(-\frac{1120}{243}+\frac{256}{3}\zeta_3\Bigr)~,
\nonumber \\
\gamma^{(2),n=2}_{\psi G}&=&C_AC_Fn_f\Bigl(\frac{278}{9}-\frac{208}{3}\zeta_3\Bigr)+
C_An_f^2\Bigl(\frac{2116}{243}\Bigr)+C_A^2n_f\Bigl(-\frac{3589}{81}+48\zeta_3\Bigr) \nonumber \\
&&+C_Fn_f^2\Bigl(-\frac{346}{243}\Bigr)+C_F^2n_f\Bigl(-\frac{4310}{243}
+\frac{64}{3}\zeta_3\Bigr)~, \\
\gamma^{(2),n=2}_{G\psi}&=&C_AC_Fn_f\Bigl(\frac{44}{9}+\frac{256}{3}\zeta_3\Bigr)  
+C_AC_F^2 \Bigl(\frac{17056}{243}+128\zeta_3\Bigr) 
+C_A^2C_F\Bigl(-\frac{41840}{243}-\frac{128}{3}\zeta_3\Bigr)  
\nonumber \\ 
&& +C_Fn_f^2 \Bigl(\frac{568}{81}\Bigr)+C_F^2n_f\Bigl(\frac{14188}{243}
-\frac{256}{3}\zeta_3\Bigr)+C_F^3 \Bigl(\frac{1120}{243}-\frac{256}{3}\zeta_3\Bigr)   ~, \nonumber \\
\gamma^{(2),n=2}_{GG}&=&C_AC_Fn_f\Bigl(-\frac{278}{9}+\frac{208}{3}\zeta_3\Bigr) 
+C_An_f^2\Bigl(-\frac{2116}{243}\Bigr)+ C_A^2n_f\Bigl(\frac{3589}{81}-48\zeta_3\Bigr) 
\nonumber\\
&&+C_Fn_f^2\Bigl(\frac{346}{243}\Bigr) 
+C_F^2n_f\Bigl(\frac{4310}{243}-\frac{64}{3}\zeta_3\Bigr)~,\nonumber
\eea
and from \cite{MVV}
\bea
K_{NS}^{(2),n=2}&=&-3n_f(\langle e^4 \rangle -\langle e^2 \rangle^2)
       \Bigl\{C_Fn_f\Bigl(\frac{536}{243}\Bigr) 
+C_FC_A\Bigl(-\frac{6044}{243}-\frac{64}{3}\zeta_3\Bigr) \nonumber\\
&&\hspace{7cm}+C_F^2\Bigl(-\frac{8620}{243}
+\frac{128}{3}\zeta_3\Bigr)
\Bigr\}~,\nonumber\\ 
K_\psi^{{(2),n=2}}&=&-3n_f\langle e^2 \rangle  \Bigl\{C_Fn_f\Bigl(-\frac{692}{243}\Bigr) 
 +C_FC_A\Bigl(-\frac{6044}{243}-\frac{64}{3}\zeta_3\Bigr)
+C_F^2\Bigl(-\frac{8620}{243}+\frac{128}{3}\zeta_3\Bigr) \Bigr\}~, \nonumber\\
K_G^{{(2),n=2}}&=&-3n_f\langle e^2 \rangle  \Bigl\{C_Fn_f\Bigl(\frac{1880}{243}\Bigr) 
 +C_FC_A\Bigl(-\frac{1138}{243}+\frac{64}{3}\zeta_3\Bigr)
+C_F^2\Bigl(\frac{9592}{243}-\frac{128}{3}\zeta_3\Bigr) \Bigr\}~.\nonumber\\
\eea

Note that a relation $\gamma^{(i),n=2}_{\psi \psi}\gamma^{(i),n=2}_{G G}-\gamma^{(i),n=2}_{\psi
G}\gamma^{(i),n=2}_{G \psi}=0$ is indeed holds for $i=0, 1, 2$.

\subsection{Photon matrix elements of quark and gluon operators}

$\bullet$\ At  one-loop
\bea
{\widetilde A}_{n=2}^{(1)\psi}&=&3n_f\langle e^2 \rangle \Bigl(\frac{2}{9} \Bigr)~, \nonumber \\
{\widetilde A}_{n=2}^{(1)G}&=&0~, \\
{\widetilde A}_{n=2}^{(1)NS}&=&3n_f \Bigl(\langle e^4 \rangle-\langle e^2 \rangle^2 \Bigr)  \Bigl(\frac{2}{9}
\Bigr)~.
\nonumber
\eea

$\bullet$\ At  two-loop
\bea
{\widetilde A}_{n=2}^{(2)\psi}&=&3n_f\langle e^2 \rangle C_F\Bigl(\frac{616}{81}-16\zeta_3 \Bigr)~, \nonumber
\\ {\widetilde A}_{n=2}^{(2)G}&=&3n_f\langle e^2 \rangle C_F\Bigl(\frac{383}{81} \Bigr)~, \\
{\widetilde A}_{n=2}^{(2)NS}&=&3n_f \Bigl(\langle e^4 \rangle-\langle e^2 \rangle^2 \Bigr)C_F 
\Bigl(\frac{616}{81}-16\zeta_3  \Bigr)~. \nonumber
\eea

\newpage
\section{Analytic continuation of the harmonic sums\label{HarmonicSum}}

The moments of $F_2^\gamma(x,Q^2,P^2)$ given in (\ref{master1}) are expressed 
by the  rational functions of  integer $n$ and the  various harmonic  sums. 
The single harmonic sums are defined by
  \begin{equation}
    S_k(n) = \sum_{j=1}^n \frac{[{\rm sgn}(k)]^j}{j^{|k|}} ,
    \label{eq:061027-01}
  \end{equation}
  where $k=\pm1,\pm2,\cdots$,
  and the higher  harmonic sums are defined recursively as
  \begin{equation}
    S_{k,m_1, \cdots, m_p}(n) = \sum_{j=1}^n\frac{[{\rm sgn}(k)]^j}{j^{|k|}} S_{m_1, \cdots, m_p}(j) ,
    \label{eq:061027-02}
  \end{equation}
where indices $k$ and $m_1, \cdots, m_p$ take nonzero integers.
In order to invert the moments so that we get $F_2^\gamma$ as a function of $x$, 
we need to make an 
analytic continuation  of these harmonic sums from integer $n$ to complex $n$. 
Since the moment sum rules of the $s$-$u$-crossing-even structure function 
$F_2^\gamma$ are defined for even integer $n$,  the  continuation should be performed from even $n$.
Thus whenever a factor $(-1)^n$ appears, it  should be replaced by $(+1)$.
The method we adopted here for the analytic continuation is to use the asymptotic  expansions of the
harmonic sums and  their translation relations. Choosing the following two harmonic sums
\bea
   S_1(n)&=&\sum_{j=1}^n \frac{1}{j}~, \label{S1n}\\
 S_{1,1,-2,1}(n) &=&
      \sum_{i=1}^n \frac{1}{i}  \sum_{j=1}^i \frac{1}{j}\sum_{k=1}^j \frac{(-1)^k}{k^2}
      \sum_{l=1}^k \frac{1}{l} \label{S11-21n}~,
\eea
as examples, we explain how we get approximate analytic formulae for these sums.

The asymptotic expansion of $S_1(n)$ for large $n$ is
  well-known:
\begin{equation}
    S_1(n) = {\rm ln}(n) + \gamma_E + \frac{1}{2n} - \frac{1}{12n^2}
      + \frac{1}{120n^4} + \cdots .
    \label{eq:061127-03}
  \end{equation}
The right-hand side has a simple analytic property.
  On the other hand, $S_1(n)$ satisfies the following translation
  relation:
  \begin{equation}
    S_1(n) = S_1(n+1) - \frac{1}{n+1}~.
    \label{eq:061127-04}
  \end{equation}
  This relation is valid not only for integer $n$, but also for
  complex $n$.
  Therefore, our algorithm to evaluate $S_1(n)$ at  arbitrary complex $n$
  is as follows:
(i) If $|n|\ge n_0$, where $n_0$ is some positive integer  at which
      the asymptotic expansion (\ref{eq:061127-03}) holds 
at a desired accuracy, then we use the expansion (\ref{eq:061127-03}) 
to evaluate $S_1(n)$. 
(ii) For $|n|<n_0$, we apply the translation relation (\ref{eq:061127-04}) 
and shift the argument
$n\rightarrow n+1$ repeatedly,  until the shifted new ${\widetilde n}$ satisfies 
the condition $|{\widetilde n}|\ge n_0$ so that 
the asymptotic  expansion (\ref{eq:061127-03}) for $S_1({\widetilde n})$ 
may be used with a desired accuracy. Then $S_1(n)$ is evaluated by a formula
   \begin{equation}
        S_1(n)
      =S_1({\widetilde n})-\sum_{i=1}^{{\widetilde n}-n}\frac{1}{n+i}~,
   \end{equation}
where the expansion  (\ref{eq:061127-03}) is used for $S_1({\widetilde n})$.

In the case of a more complicated higher harmonic sum $S_{1,1,-2,1}(n)$ with \textit{even} integer $n$, 
its asymptotic expansion for large $n$ is given by
\begin{equation}
        S_{1,1,-2,1}^{\rm even}(n)
      =
        c_{0,2} {\rm ln}^2(n) + c_{0,1} {\rm ln}(n) + c_{0,0}
        + \frac{c_{1,1} {\rm ln}(n)}{n} + \frac{c_{1,0}}{n} 
        +\frac{c_{2,1} \ln(n)}{n^2} + \frac{c_{2,0}}{n^2}+\cdots~,
      \label{AsymExansionS11-21}
\end{equation}
where
 \bea
      c_{0,2}
    &=&
      c_{1,1}
    =
      - \frac{5}{16}\zeta(3) = -0.37564278 \cdots
    , \nonumber\\
      c_{0,1}
    &=&
      - \frac{5}{8} \gamma_E \zeta(3)
      - \frac{3}{40} \zeta^2(2) = -0.63658940 \cdots
    ,\nonumber \\
      c_{0,0}
    &=&
      - \frac{3}{40} \gamma_E \zeta^2(2)
      - \frac{5}{16} \gamma_E^2 \zeta(3)
      - {\rm ln}(2) {\rm Li}_4\left(\frac{1}{2}\right)
      - \frac{7}{16} {\rm ln}^2(2) \zeta(3)
      + \frac{1}{6} {\rm ln}^3(2) \zeta(2)
    \nonumber\\&&\quad
      - \frac{1}{30}{\rm ln}^5(2)
      + \frac{1}{8} \zeta(2) \zeta(3)
      + \frac{1}{8} \zeta(5)
      - {\rm Li}_5\left(\frac{1}{2}\right) = -0.89930722 \cdots
    , \\
      c_{1,0}
    &=&
      - \frac{5}{16} \gamma_E \zeta(3)
      - \frac{3}{80} \zeta^2(2)
      + \frac{5}{16} \zeta(3) = 0.057348080 \cdots
    , \nonumber\\
      c_{2,1}
    &=&
      \frac{5}{96}\zeta(3) = 0.062607130 \cdots
    , \nonumber\\
      c_{2,0}
    &=&
      \frac{5}{96} \gamma_E \zeta(3)
      + \frac{1}{160} \zeta^2(2)
      - \frac{15}{64} \zeta(3) = - 0.22868296 \cdots~.
     \nonumber
      \eea
Also $S_{1,1,-2,1}^{\rm even}(n)$ satisfies the following translation
  relation:
  \begin{equation}
      S_{1,1,-2,1}^{\rm even}(n)
    =
      S_{1,1,-2,1}^{\rm even}(n+2)
      - \left( \frac{1}{n+1} + \frac{1}{n+2}\right) S_{1,-2,1}^{\rm even}(n+2)
      + \frac{1}{(n+1)(n+2)} S_{-2,1}^{\rm even}(n+2)~.
    \label{eq:061127-05}
  \end{equation}
  Note that the right-hand side of \eqref{eq:061127-05} is written
  in terms of the same harmonic sum $S_{1,1,-2,1}^{\rm even}$  and lower harmonic sums
  $S_{1,-2,1}^{\rm even}$ and $S_{-2,1}^{\rm even}$ but with a larger
  argument $n\!+\!2$.
When $|n|\ge n_0$,  the asymptotic expansion 
(\ref{AsymExansionS11-21}) is used to evaluate $S_{1,1,-2,1}^{\rm even}(n)$.
For $|n|<n_0$,
we apply  \eqref{eq:061127-05} and shift the argument
$n\rightarrow n+2$ repeatedly,  until the shifted new ${\widetilde n}$ satisfies 
the condition $|{\widetilde n}|\ge n_0$ so that 
the asymptotic expansion (\ref{AsymExansionS11-21}) of $S_{1,1,-2,1}^{\rm even}({\widetilde n})$ 
may be used with a desired accuracy.  The lower harmonic sums $S_{1,-2,1}^{\rm even}(n)$ and $S_{-2,1}^{\rm
even}(n)$ are evaluated in a similar fashion. 

  In practice, we take $n_0\!=\!16$ for all harmonic sums. Then 
the asymptotic expansion formula for each harmonic sum is derived so as to 
ensure double precision accuracy (15 significant figures) at $n\!=\!n_0$.
For example, $S_{1}(n)$ and $S_{1,1,-2,1}^{\rm even}(n)$ are expanded up to 
the terms with $1/n^{10}$ and $1/n^{18}$, respectively.


\newpage


\end{document}